\documentclass{aa} 

\usepackage[varg]{txfonts}
\usepackage{graphicx,url}
\usepackage{natbib}

\usepackage{lscape}

\usepackage{latexsym}

\usepackage{multirow}

\newcommand{\msun}{\mbox{M$_{\odot}$}}

\begin{document}

\title{Supernova rates from the SUDARE VST-Omegacam search \\ II. Rates in a galaxy sample.
\thanks{Based on observations made with ESO telescopes at the Paranal Observatory under programme ID 
088.D-4006, 088.D-4007, 089.D-0244, 089.D-0248, 090.D-0078, 090.D-0079, 
088.D-4013, 
089.D-0250, 090.D-0081 
}}

\author{M.T. Botticella\inst{\ref{OAC}} \and E. Cappellaro\inst{\ref{OAPD}} \and L. Greggio\inst{\ref{OAPD}},  G. Pignata\inst{\ref{Chile},\ref{millenium}} \and M. Della Valle\inst{\ref{OAC},\ref{ICRA}} \and  A. Grado\inst{\ref{OAC}} \and  L. Limatola\inst{\ref{OAC}}, \\
A. Baruffolo\inst{\ref{OAPD}} \and S. Benetti\inst{\ref{OAPD}} \and F. Bufano\inst{\ref{oact}}
\and M. Capaccioli\inst{\ref{uniNA}} \and E. Cascone\inst{\ref{OAC}} \and G. Covone\inst{\ref{uniNA}} \and D. De Cicco\inst{\ref{uniNA}} \and S. Falocco\inst{\ref{stock}},\\ 
B. Haeussler\inst{\ref{oxford}} \and V. Harutyunyan\inst{\ref{OAC}} \and M. Jarvis\inst{\ref{oxford},\ref{SA}} \and L. Marchetti\inst{\ref{milton}} \and N. R. Napolitano\inst{\ref{OAC}} \and M., Paolillo\inst{\ref{uniNA},\ref{ASDC}} \and A. Pastorello\inst{\ref{OAPD}}\\
M. Radovich\inst{\ref{OAPD}} \and P. Schipani\inst{\ref{OAC}}
 \and L. Tomasella\inst{\ref{OAPD}} \and M. Turatto\inst{\ref{OAPD}} \and  M. Vaccari\inst{\ref{SA},\ref{oabo}}
}
\institute{
  INAF - Osservatorio Astronomico di Capodimonte, Salita Moiariello 16, Napoli, 80131 Italy\label{OAC}
\and
INAF - Osservatorio Astronomico di Padova, vicolo dell'Osservatorio 5, Padova, 35122 Italy\label{OAPD}
 \and   
   Departemento de Ciencias Fisicas, Universidad Andres Bello, Santiago, Chile \label{Chile}
\and  
   Millennium Institute of Astrophysics, Santiago, Chile\label{millenium}
   \and 
   International Center for Relativistic Astrophysics, Piazzale della Repubblica 2, 65122 Pescara, Italy\label{ICRA}
\and
  Department of Physics and Astronomy, University of the Western Cape, 7535 Bellville, Cape Town, South Africa  \label{SA}   
   \and
   INAF - Istituto di Radioastronomia, via Gobetti 101, 40129 Bologna, Italy\label{oabo}
\and
INAF - Osservatorio Astronomico di Catania, via S.Sofia 78, 95123 Catania, Italy \label{oact}
 \and  
  Dipartimento di Fisica, Universit\'a Federico II, Napoli, Italy \label{uniNA}
  \and AlbaNova University Center, KTH Royal Institute of Technology, Roslagstullsbacken 21, SE-106 91 Stockholm, Sweden\label{stock}
 \and
  Astrophysics, University of Oxford, Denys Wilkinson Building, Keble Road, Oxford OX1 3RH, UK
   \label{oxford} 
\and
  Department of Physical Sciences, The Open University, Milton Keynes, MK7 6AA, UK \label{milton}
  \and
  ASI Science Data Center, via del Politecnico snc, 00133 Roma, Italy
  \label{ASDC}
  }

\date{Received: ????; Revised: ??????; Accepted: ????? }
\titlerunning{SUDARE II}
\authorrunning{MTB}


\abstract{} {This is the second paper of a series in which we present measurements of the Supernova (SN) rates from the SUDARE survey. \\
The aim of this survey is to constrain the core collapse (CC) and Type Ia SN progenitors by analysing the dependence of their explosion rate on the properties of the parent stellar population averaging over a population of galaxies with different ages in a cosmic volume and in a galaxy sample.   
In this paper, we study the trend of  the SN rates with the intrinsic colours, the star formation activity  and the mass of the parent galaxies. To constrain the SN  progenitors we compare the observed rates with model predictions assuming four  progenitor models  for SNe~Ia with different distribution functions of the time intervals between the formation of the progenitor and the explosion and a mass range of  $8-40\,  {\rm M}_\odot$ for CC~SN progenitors.}{We have considered a  galaxy sample of about $130000$ galaxies and a SN sample of about 50 events. The wealth of photometric information for our galaxy  sample allows us to apply the spectral energy distribution (SED) fitting technique to estimate the intrinsic rest frame colours, the stellar mass and star formation rate (SFR) for each galaxy of  the sample. The galaxies have been separated in star-forming and quiescent galaxies exploiting both the rest frame $U-V$ vs $V-J$ colour-colour diagram and the best fit values of the specific star formation rate (sSFR) from the SED fitting.}{ We found that  the SN~Ia rate per unit mass is higher  by a factor of six in the star-forming galaxies with respect to the passive galaxies identified as such both on the $U-V$ vs $V-J$ colour-colour diagram and for their sSFR. 
The SN~Ia rate per unit mass is also higher  in the less massive galaxies that are also younger.
These results  suggest a distribution of the delay times (DTD) less populated at long delay times  than at short delays.\\
The CC~SN rate per unit mass is proportional to  both the sSFR  and the galaxy mass, confirming that the CC~SN progenitors explode soon after the end of the star formation activity.
 The trends of the Type Ia and CC SN rates as a function of the sSFR  and  the galaxy mass  that we observed from SUDARE data are in agreement with literature results at  different redshifts suggesting that  the ability of the stellar populations to produce SN events does not vary  with cosmic time. 
\\   The expected  number of SNe Ia is in agreement with the observed one for all four DTD models considered  both in passive and star-forming galaxies so we can not discriminate between different progenitor scenarios. The expected number of CC~SNe  is higher than the observed one, suggesting a higher limit  for the minimum progenitor mass. However, at least  part of this discrepancy between expected and observed number of CC SNe may reflect  a fluctuation due to the relatively poor statistics.
We also compare the expected and observed  trends of  the SN~Ia rate with the  intrinsic $U-J$ colour of the parent galaxy, assumed as a tracer of the age distribution.
While the slope of the relation between the SN~Ia rate and the $U-J$ color in star-forming galaxies can be reproduced well by all four DTD models considered, only the steepest of them is able to account for the rates and colour in star-forming and passive galaxies with the same value of the SN~Ia production
efficiency. The agreement between model predictions and data could be found also for the other DTD models, but with a
productivity of SN~Ia higher in passive galaxies compared to star-forming galaxies.
 }{}


\maketitle

\section{Introduction}\label{intro}

Supernovae (SNe), dramatic and violent end-points of stellar evolution,  are  involved in the formation of neutron stars, black holes and gamma-ray bursts  and are sources of gravitational waves, neutrino emission and high-energy cosmic rays \citep[e.g.,][]{heger:2003xy,woosley:2006lr,fryer:2012vn}.  

SNe  play  also an important role in driving the dynamical and chemical evolution of galaxies, contributing to the feedback processes in galaxies \citep{ceverino:2009ol}, producing  heavy elements and  dust   \citep[e.g.,][]{bianchi:2007eu}.

SNe are formidable cosmological probes  due to their huge intrinsic brightness.
In particular, Type Ia SNe have provided the first evidence for an accelerated expansion of the Universe \citep{perlmutter:1998kq,perlmutter:1999gl,riess:1998qm,schmidt:1998nr} and remain one of the more  promising probes of the nature of dark energy  \citep[e.g.,][]{sullivan:2011zl,salzano:2013gf, DES2016lr}.  


We recognise two main, physically defined, SN classes: core collapse-induced explosions (CC~SNe) of short-lived massive stars  ($M \gtrsim 8\, {\rm M}_{\odot}$) and thermonuclear explosions (SNe~Ia) of long-lived, intermediate mass stars   ($ 3 \lesssim {\rm M} \lesssim 8\, {\rm M}_{\odot}$) in binary systems. 

CC~SNe display a huge range in their physical characteristics, including kinetic energy, radiated energy, ejecta mass and the amount of radioactive elements  created during the explosion.  
Different sub-types of CC~SNe have been identified on the basis of their spectro-photometric evolution (IIP, IIL, IIn, IIb, Ib, Ic, IcBL) and have been associated to a possible sequence of progenitors related to their  mass loss history, with the most massive stars and stars in binary systems losing the largest fraction of their initial mass.
However, this simple  scheme where only mass loss drives the evolution of CC SN progenitors has difficulties to explain the relative fraction of different types, as well as  the variety of properties  exhibited by CC~SNe of the same type \citep[e.g.,][]{smith:2011jy}.  
In recent years the progenitor stars of  several  type IIP  SNe have been detected  on pre-explosion images out to 25\,Mpc \citep[][and reference therein]{smartt:2009mq}, while  there are few progenitors detected for IIb and Ib SNe  \citep{bersten:2012ys,cao:2013jk,bersten:2014yq,eldridge:2015nr,fremling:2016uq} and no progenitors for type Ic SNe.  
The lack of  detection of red supergiant progenitors with initial masses between  $17-30\,\msun$  is a still unsolved issue and several possible explanations  have been suggested  \citep{smartt:2009mq}. 
Moreover, several observational evidences have been suggested that a significant fraction of type Ib/c progenitors are stars of  initial mass lower than $25\,\msun$  in close binary systems stripped of their envelopes  by the interaction with the companion \citep[e.g.,][]{eldridge:2013lr}.

There is a general consensus that SNe~Ia  correspond to thermonuclear explosions of a carbon and oxygen white dwarf (WD) reaching the Chandrasekhar mass due to accretion from a close companion. 
The evolutionary path that leads a WD to mass growth, ignition, and explosion is still unknown. 
The most widely favoured progenitor scenarios are  the single degenerate  scenario (SD) in which a  WD, accreting from a  non degenerate companion (a main sequence star, a sub-giant, a red giant or a helium star), grows in mass until it reaches the Chandrasekhar mass and explodes in a thermonuclear runaway \citep{whelan:1973nr}, and the double degenerate scenario (DD), in which a close double WD system merges after orbital shrinking due to the emission of gravitational waves \citep{tutukov:1981ys,iben:1984rz}.
Unfortunately  to date, there is no conclusive evidence from observations decisively supporting one channel over the other  \citep[e.g.,][]{maoz:2014kt}.
Actually, both channels may well contribute in comparable fractions of events  \citep{greggio:2010pd}, as suggested by some observational facts, but the contribution of each channel to the overall Type Ia population is very hard to assess.

 The modern SN searches discovering transients with no galaxy bias, faint limiting magnitudes, and very fast evolution are challenging the paradigms we hold for SN progenitors suggesting an unexpected large range of explosion  proprieties  and  new  SN classes (e.g. Iax, Ibn or super luminous SNe).

It is then of foremost importance to answer open questions with regard to SN progenitor systems, explosion mechanisms and physical parameters driving the observed SN diversity. \\
\noindent In this framework, the analysis of  the dependence of the SN rates on the age distribution of the parent stellar population  can help  constraining the progenitor scenarios and hence understanding the effects of age on the SN diversity  \citep[e.g.,][]{greggio:2010pd}.
  
  Due to the short lifetime of progenitor stars, the rate of CC~SNe directly traces the current SFR of the host galaxy. Therefore  the mass range for CC~SN progenitors can be probed by comparing  the rate of born stars and the rate of CC~SNe occurring in the host galaxy, assuming the distribution of the masses with which stars are born, i.e. the initial mass function (IMF). 
  On the other hand, the rate of SNe~Ia echoes the whole SF history (SFH) of the host galaxy due to the time elapsed from the birth of the binary system to the final explosion, called delay-time.
SNe~Ia are observed to explode both in young and in old stellar populations and  the age distribution of the SN~Ia progenitors is still considerably debated \citep[e.g.,][]{maoz:2014kt}.
 By comparing the observed SN Ia rate with what is expected for the SFH of the parent stellar population, it is possible to constrain the progenitor scenario and the fraction of the binaries exploding  as SNe~Ia  \citep{greggio:2005ph,greggio:2008er,greggio:2010pd}. \\ 
   Ideally, we should compare SFH and SN rates in each individual galaxy \citep[e.g.,][]{maoz:2012uq} or for each resolved stellar population in a given galaxy.
To minimise  the uncertainties in SN rates and SFHs based on individual galaxies we can  average over a galaxy sample  or in a large cosmic volume.
 In the first case, a characteristic average  age for a sample of galaxies is estimated via morphological types, colours or SED fitting; in the second case, one averages over a population of galaxies with different ages and evolutionary paths, by assuming a cosmic SFH.

With the goal to measure SN rates both at different cosmic epochs and  in a well defined galaxy sample we started the SUpernova Diversity And Rate Evolution (SUDARE) programme ran at the VLT Survey Telescope (VST)  \citep{botticella:2013ys}. 
The survey strategy, the pipeline  and the estimate of  SN rates in a cosmic volume  are described in \citet{cappellaro:2015lr}  (Paper\,$\textrm{I}$).  
In the present paper we illustrate the measurement of  SN rates for the main SN types as a function of  galaxy colours, SFR, sSFR and mass. 

This paper is organised as follows.  In Section~\ref{gal}, we describe the method that we
use to estimate  the properties of the galaxies monitored by SUDARE. 
In Section~\ref{sne}, we outline how we select the SN sample and identify the host
galaxy of each SN.
In Section~\ref{rate}, we illustrate how we measure rates in our galaxy sample. 
Finally, we investigate how SN rates  are dependent on the properties of
the host galaxies: rest frame colours (Section~\ref{ratecolor}), star-formation quenching (Section~\ref{ratesfr}),
 sSFR (Section~\ref{ratessfr}),  stellar
mass (Section~\ref{ratemass}). 
A comparison of the observed rates with predictions  from theoretical models is presented  in Section~\ref{tmodel}, while our conclusions are summarised in Section~\ref{conclusions}.
In the Appendix we present  the analysis of rates for different CC~SN subtypes (Section~\ref{CCsubtypes})  and systematic errors (Section~\ref{systerrors}).
Throughout, we assume a 
standard cosmology with $\rm{H} _{0} = 70$\,km\,s$^{-1}$\,Mpc$^{-1}$, 
$\Omega _{M}= 0.3$ and $\Omega_{\Lambda} = 0.7$ and we adopt a Salpeter IMF from $0.1$ to $100\, {\rm M}_{\odot}$    \citep{salpeter:1955ky} and the AB magnitude system \citep{gunn:1998dq}.  

\section{Galaxy properties}\label{gal}
In order to relate the SN events and their rates to fundamental properties of their progenitors  we need to characterize the stellar populations in the sample galaxies we monitored for SNe, and in particular their SFH. The latter can be derived  from integrated rest-frame colours in various combinations, or, more in detail, from the galaxy SED. In this paper we consider both tracers. In the following we describe in detail  how we determine the redshift  (Sec.~\ref{z}), which is needed to derive the rest frame colours  (Sec.~\ref{sfpass} and Sec.~\ref{restcolours}), and mass, SFR, and age (Sec.~\ref{sed}) of the galaxies in our sample.

\subsection{Photometric Redshifts}\label{z}
The extensive  multi-band coverage of both COSMOS and CDFS  sky-fields make them  excellent samples for performing studies of  stellar populations. 
We exploited the analysis of  recent data from deep multi-band surveys in COSMOS field published by \citet{muzzin:2013fk} retrieving, for the galaxies monitored by SUDARE (1.15\,deg$^2$),  photometric redshifts  from their catalogs\footnote{http://www.strw.leidenuniv.nl/galaxyevolution/ULTRAVISTA/}.
We  performed our own analysis  for the CDFS field (2.05\,deg$^2$ monitored from SUDARE) following the same approach of \cite{muzzin:2013fk}.  A detailed description of the photometric redshift computation and analysis has been reported in Paper\,$\textrm{I}$.

To collect multi-band photometry in CDFS we exploited deep stacked images in $u,g,r,i$  from the SUDARE/VOICE survey, complemented with data
in $J,H,K_{\rm s}$ from the VISTA Deep Extragalactic Observations \citep[VIDEO,][]{Jarvis2013}, in the FUV  from {\it GALEX}, and in IRAC bands from the SERVS survey \citep{mauduit:2012sf}, SWIRE surveys \citep{lonsdale:2003zr}  and Spitzer Data Fusion \footnote{http://www.mattiavaccari.net/df} \citep{vaccari:2015rt}, as reported in detail in Paper\,$\textrm{I}$.

The source catalog for  both CDFS  and COSMOS were obtained with {\sc SExtractor} from the VISTA $K_{\rm s}$ band images reaching  $K_{\rm s}=23.5$\, mag (at 5$\sigma$ in a 2\arcsec diameter aperture).
Star vs. galaxy separation was performed in the  $J  - K_{\rm s}$ versus $u - J $ color space where there is a clear segregation between the two components \citep[Fig.~3 on][]{muzzin:2013fk}.
The photometric redshifts for both catalogs were determined by fitting the multi-band photometry  to SED templates using the {\sc EAZY}\footnote{ http://www.astro.yale.edu/eazy/} code \citep{Brammer2008}. 
The main difference between the COSMOS and the CDFS fields consists in the number of filters available for the analysis, i.e. a maximum of 12 filters for CDFS and 30 for COSMOS,  that results in different quality of the photometric redshifts.
The redshift quality parameter provided by {\sc EAZY}, $Q_z$\footnote{$Q_z = \frac{\chi^2}{N_\mathrm{filt}-3} \, \frac{u^{99} - l^{99}}{p_{\Delta z=0.2}}$ where  $N_\mathrm{filt}$ is the number of photometric measurements used in the fit, $u^{99}- l^{99}$ is the 99\% confidence intervals  and ${p_{\Delta z=0.2}}$ is the fractional probability that the redshift lies within $\pm 0.2$ of the nominal value.}, is good, i.e. $Q_z < 1$,  for 75\% and  for 93\% of the galaxies in CDFS and  COSMOS fields, respectively. 
The normalised median absolute deviation  is $ 0.02$ and   $ 0.005$,  and the fraction of "catastrophic"  redshift estimates\footnote{ The fraction of  "catastrophic"  estimates  is defined as the fraction of  galaxies for which $ \left| z_{\rm phot}-z_{\rm spec}\right| / 1+z_{\rm spec} >5\sigma_{\rm NMAD}$ } is  $\sim 14$\%  and  4\% for CDFS and COSMOS, respectively.

\begin{figure}
\begin{center}
\includegraphics[width=8.5cm]{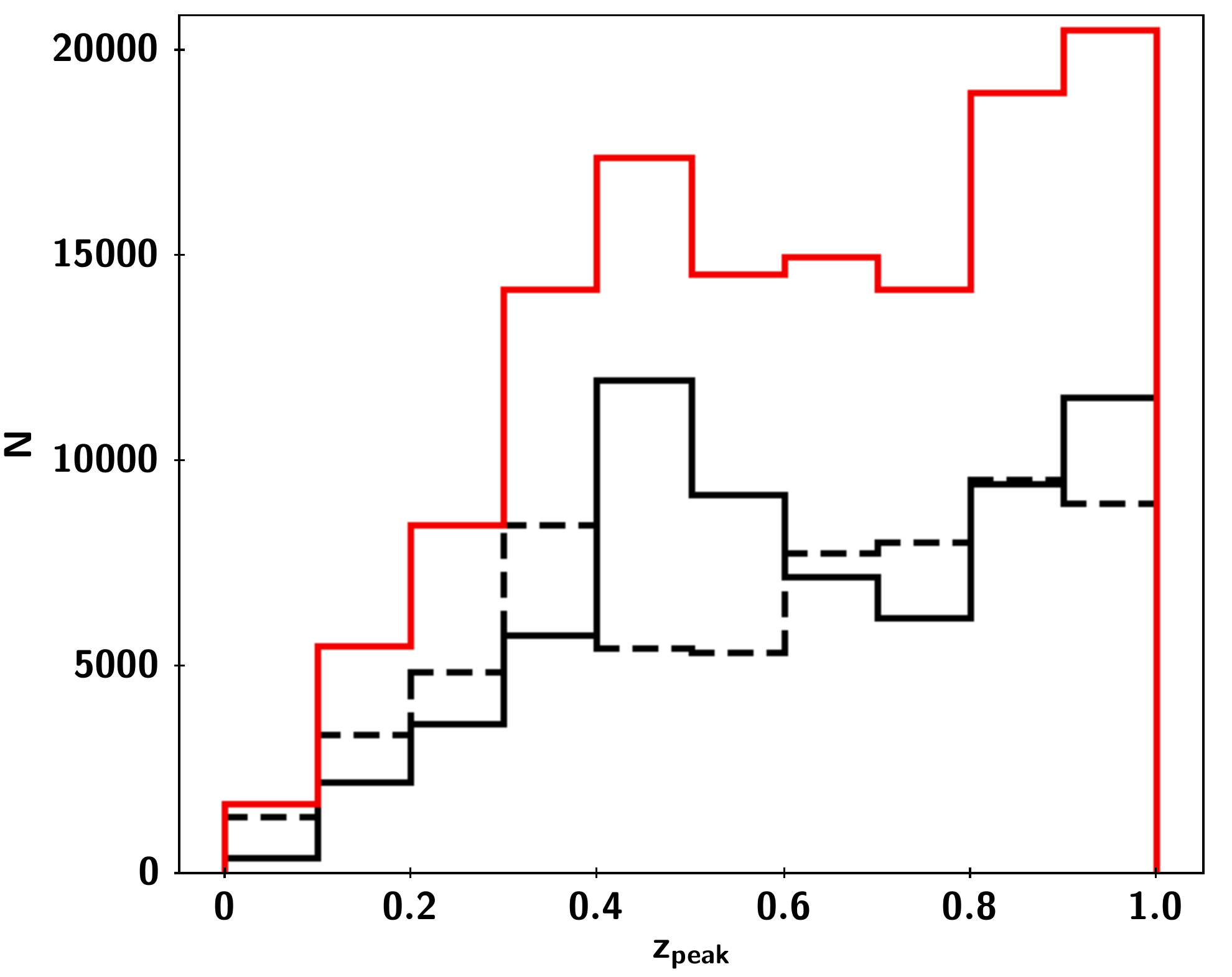}\\
\caption{The distribution of   $z_{\rm peak}$  for  galaxies in the CDFS (black line), COSMOS (dashed line) and in the overall sample (red line).}
\label{zphot}
\end{center}
\end{figure}

For our analysis we selected all galaxies with  
$i)$ $K_{\rm s}$ band magnitude  $ \le 23.5$\, mag, $ii)$ redshift  within the range  $0.15 \le z \le 0.75$  and   $iii)$ $Q_z < 1$,
 which results in $62863$  galaxies  in COSMOS  and $67224$ galaxies in  CDFS, for a final  sample of about 130000 galaxies. 
 Although the area monitored around CDFS is larger than that of  the COSMOS field, the number of galaxies in the two samples  is similar, because the COSMOS sample is more complete (deeper  $K_{\rm s}$ band stack) and  has a  larger fraction of galaxies with $Q_z < 1$. The distributions of the photometric  redshift  of galaxies in the CDFS and COSMOS fields  and in the overall sample are shown  in  Fig.~\ref{zphot}.
 Globally, the vast majority of our galaxies is found at redshifts larger  than $0.3$ and  the distribution does not present major peaks.

\subsection{Passive and star-forming galaxies}\label{sfpass}
The rest frame colours of galaxies show a bimodal distribution  that basically reflects star-formation quenching: in general, "red" galaxies have had their star formation quenched, and "blue" galaxies are still forming stars \citep{Blanton2003,Baldry2004}. 
However, some star-forming galaxies exhibit red colours in spite of their young stars, due to the presence  of dust.
 
\noindent The $U-V$ vs $V-J$ colour-colour diagram allows  for a simple separation of star-forming from passive galaxies: the $U-V$ colour covers the Balmer break and therefore is a good measure of relatively  unobscured recent star formation, while the $V-J$ colour allows to empirically separate  "red" passively evolving galaxies from "red" dusty star-forming galaxies, since dust-free quiescent galaxies
are relatively  blue in $V-J$ \citep[e.g.,][]{Labbe:2005fv,williams:2009fk,Brammer2009}.

\begin{figure}
\begin{center}
$
\begin{array}{c@{\hspace{.1in}}c@{\hspace{.1in}}c}
\includegraphics[width=8.5cm]{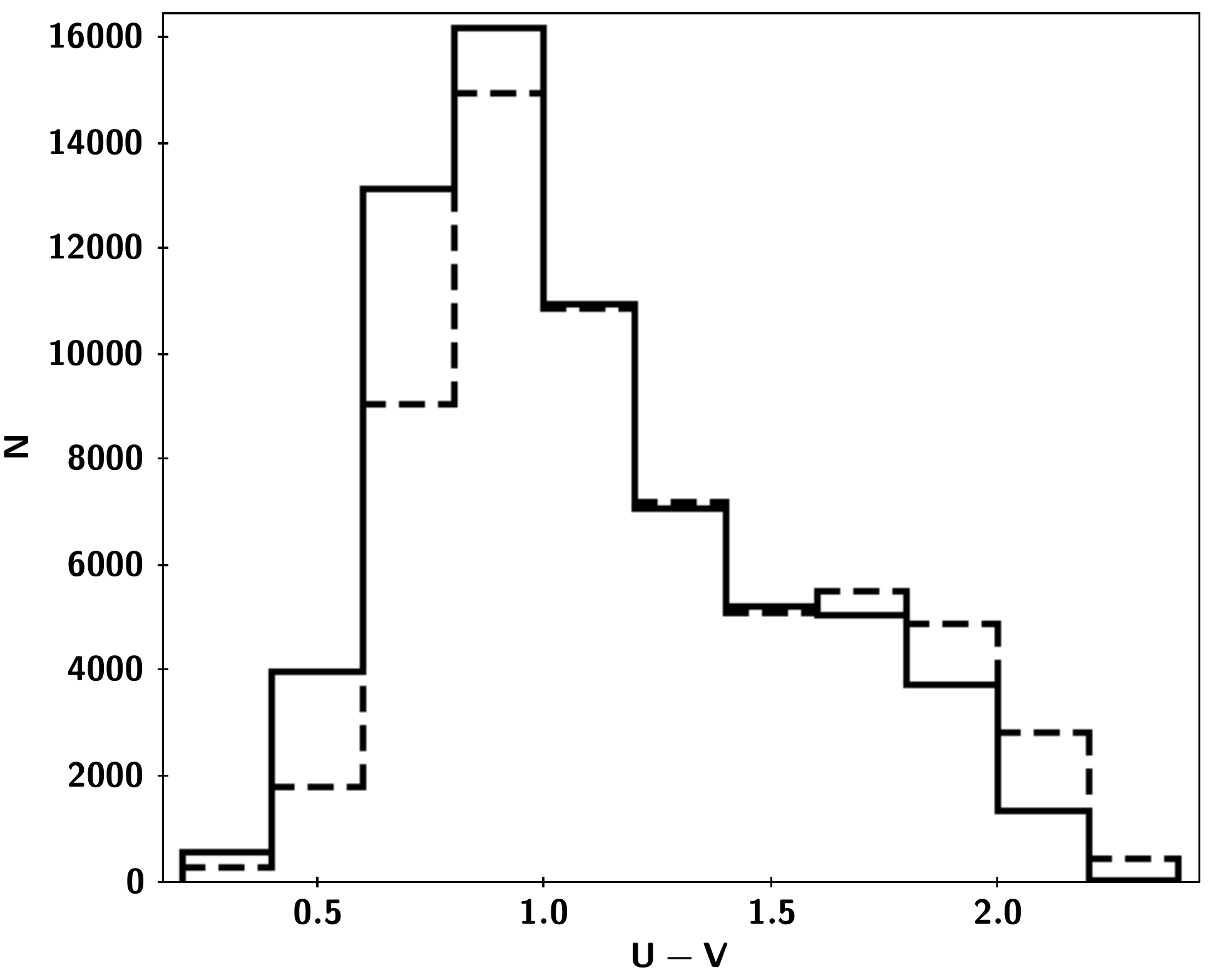} \\
\includegraphics[width=8.5cm]{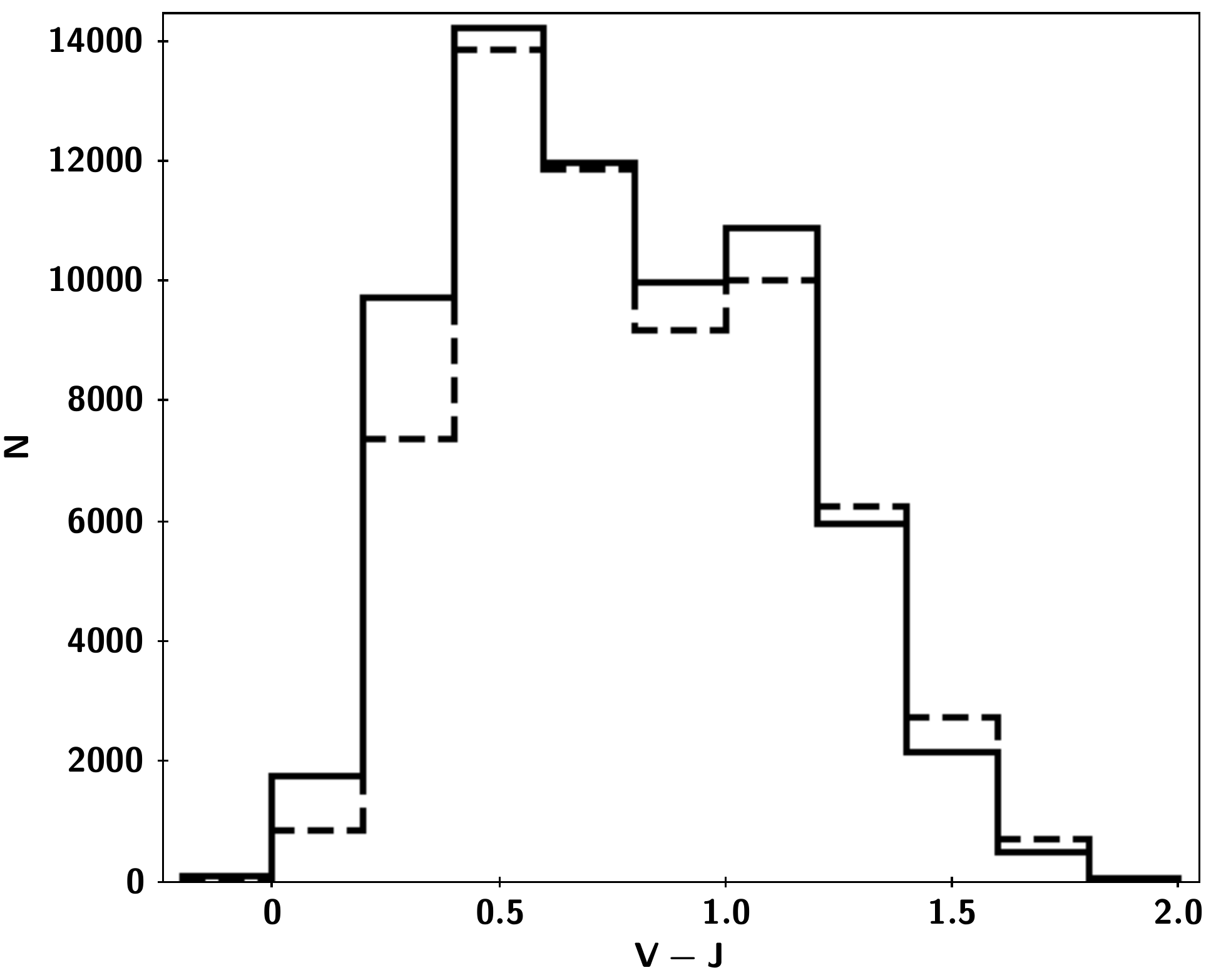}
\end{array}
$
\caption{The distribution of  $U-V$  (top panel) and $V-J$ (bottom panel) rest frame colours for the CDFS  (solid line)  and   the COSMOS (dashed line) galaxy samples.}
\label{restcoldist}
\end{center}
\end{figure}

As in \cite{muzzin:2013fk}, we obtained rest frame  $U-V$ and $V-J$ colours from the  {\sc EAZY} code, which measures the rest-frame fluxes from the best-fitting SED template \citep{Brammer2008}.
The values of  interpolated rest-frame colours primarily depends on the galaxy redshift  and may  be subject to systematic, redshift dependent offsets.
 Moreover, these offsets may be different depending on the SED  templates (e.g., for quiescent and star-forming galaxies).
Therefore, we verified that the distribution on the $U-V$ vs $V-J$ colour-colour diagram for galaxies in the CDFS sample is very similar to that in the COSMOS sample, for which the estimate of photometric redshift is more accurate (Fig.~\ref{restcoldist}).
 
\begin{figure}
\begin{center}
\includegraphics[width=9cm]{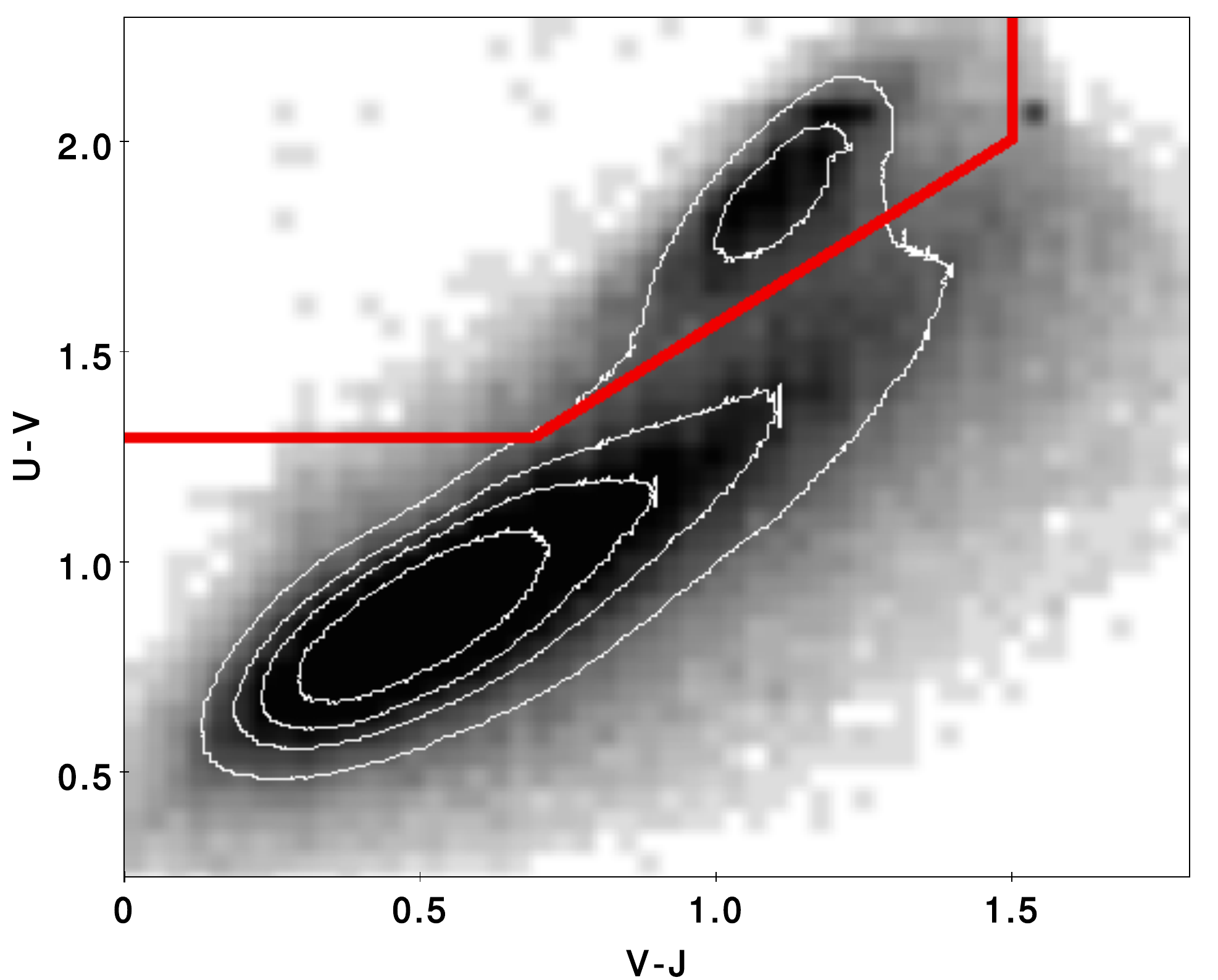}
\caption{The rest frame $U-V$ vs $V-J$ colour-colour diagram for the CDFS galaxy sample. The solid red line separates passive from star-forming galaxies (Eq.~\ref{uvjsep}).}
\label{UmV_VmJ}
\end{center}
\end{figure}

 The rest frame $U-V$ vs $V-J$  colour-colour diagram for galaxies in COSMOS, shown in Fig. 3 of  \cite{muzzin:2013fj}, and in CDFS, shown here in Fig.~\ref{UmV_VmJ}, clearly display an extended sequence of star-forming galaxies (from the lower-left to the upper-right region)  and a localised clump of passive galaxies in the upper region of the figure.
Following \cite{williams:2009fk} and \cite{muzzin:2013fj} we adopted the following criteria  to select passive galaxies:
\begin{eqnarray}\label{uvjsep}
U - V > 1.3   \qquad  V - J < 1.5 \quad    \mbox{at all redshifts}  \\
U - V > (V - J)\times0.88 + 0.69 \qquad \mbox{for}\,  0< z \le 1 \nonumber
\end{eqnarray}
The conditions $U - V > 1.3, V - J < 1.5$ are applied to prevent contamination from unobscured and dusty star-forming galaxies, respectively. 
We found that  the fraction of  passive galaxies (about 12\% and 18\% in CDFS and COSMOS, respectively) and their distribution as a function of redshift are very similar in the two fields.

\subsection{Stellar masses, ages, SFRs and sSFR}\label{sed}
The wealth of photometric information for our samples allows us to apply the SED fitting technique to estimate the stellar mass, and SFH for each galaxy of COSMOS and CDFS samples. We performed this fitting using the  {\sc FAST}\footnote{Fitting and Assessment of Synthetic Templates ( {\sc FAST})  http://astro.berkeley.edu/~mariska/FAST.html} code \citep{Kriek2009}. 
To construct the set of template SEDs, we adopted:  $i)$ the \cite{Bruzual2003} stellar population synthesis (SPS) model library,  $ii)$ a Salpeter IMF,  $iii)$ solar metallicity, $iv)$ an exponentially declining SFH ($\psi(t) \propto \exp(-t/\tau)$ where $t$ is the time since the onset of the star formation  and $\tau$ is the e-folding  star formation timescale, ranging in $10^{7}-10^{10}$\,Gyr) and $v)$ the  \cite{calzetti:2000ht} dust attenuation law. 
For all galaxies we restrict $t$ to be less than the age of the Universe  at the redshift of the galaxy and allowed for a visual attenuation in the range $0-4$\, mag.
Redshifts were fixed to the values derived by  {\sc EAZY}.

\begin{figure}
\begin{center}
$
\begin{array}{c@{\hspace{.1in}}c@{\hspace{.1in}}c}
\includegraphics[width=8.5cm]{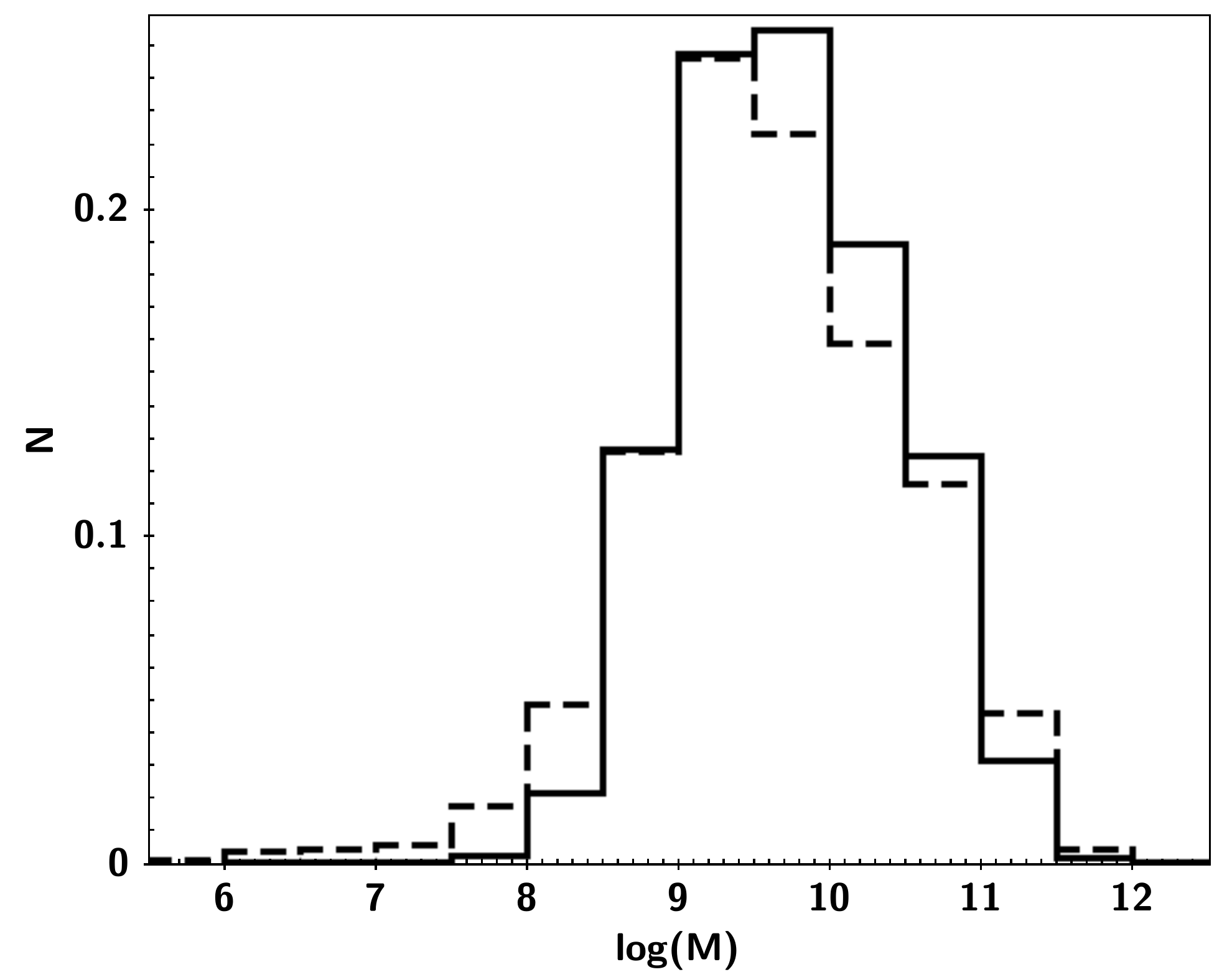} \\
\includegraphics[width=8.5cm]{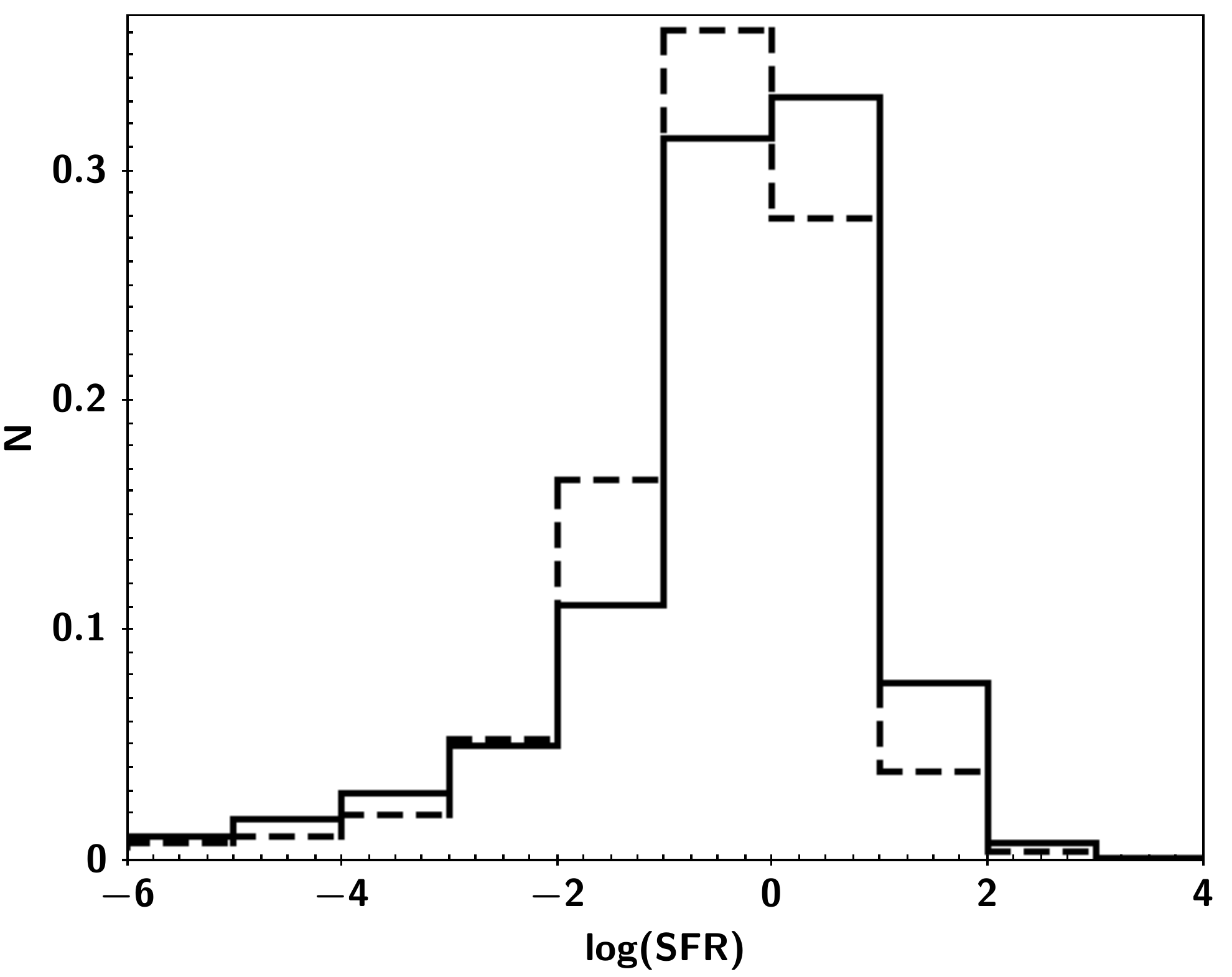} \\
\includegraphics[width=8.5cm]{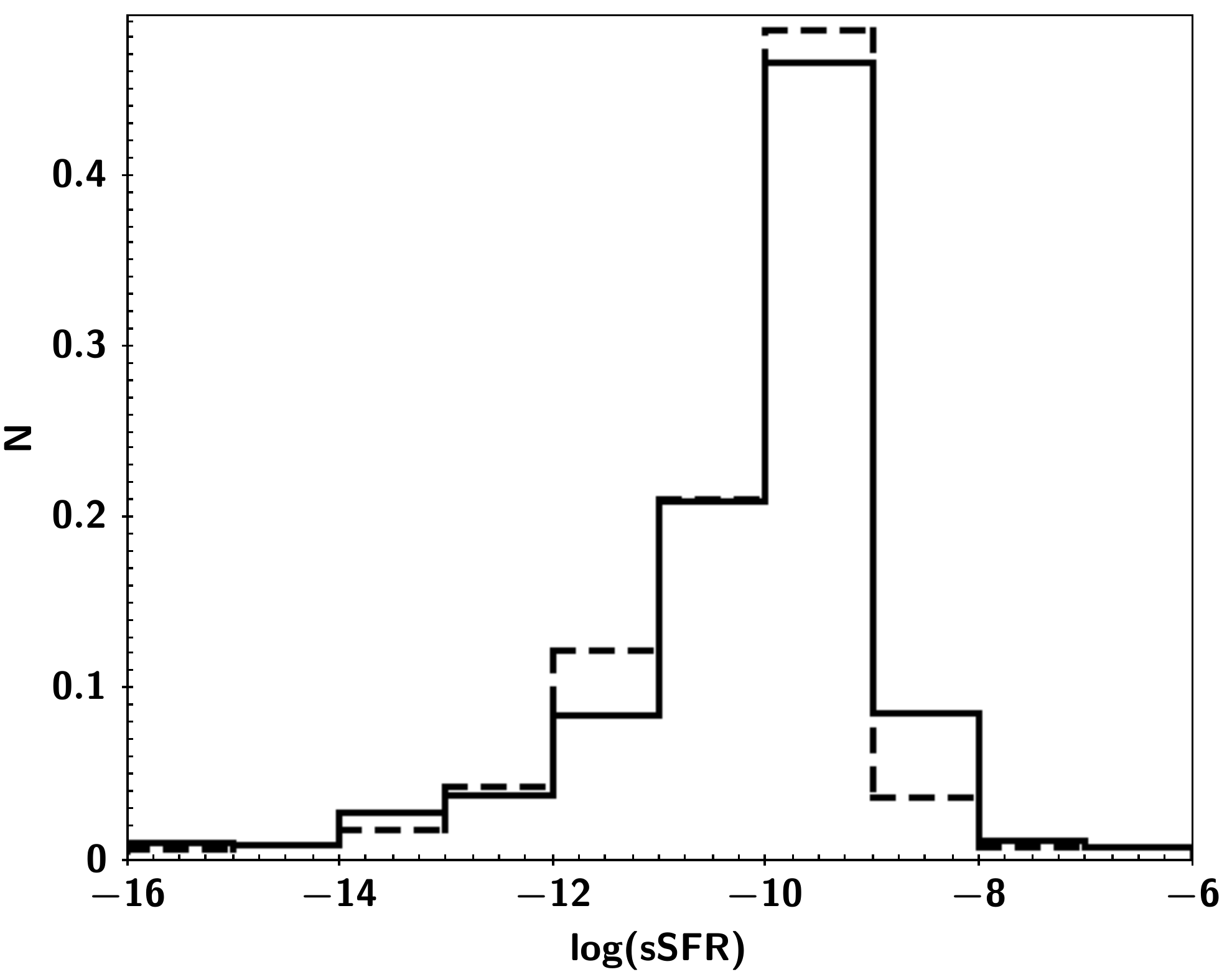}
\end{array}
$
\caption{The distribution of stellar masses (top panel), SFR  (middle panel) and sSFR (bottom panel) estimated with {\sc FAST} for the CDFS  (solid line) and the COSMOS (dashed line) galaxy sample.}
\label{fastdist1}
\end{center}
\end{figure}

In general, stellar masses estimated via the SED fitting technique are relatively well-constrained, however {\sc FAST} does not account for the gas re-cycling and  the "return fraction", which is the  fraction of mass of gas processed by stars and returned to the interstellar medium during their evolution. Therefore {\sc FAST}  overestimates the actual stellar mass by a fraction which amounts to 30\%  in an old stellar population for a Salpeter IMF \citep[see e.g.,][]{greggio:2011lr}. 
In Fig.~\ref{fastdist1}  (top panel) we show the distribution of stellar masses for both COSMOS and CDFS samples.  The COSMOS sample appears to include a larger fraction of low mass galaxies, compared to the CDFS sample.

The SFRs are more uncertain when they are derived solely from optical-NIR photometry, plus the assumption of an exponentially declining SFR.  
The  lack of FUV and MIR data, that probe unobscured and obscured star formation, respectively, has an important  effect on the estimate of SFR for  galaxies experiencing a recent star formation burst.  
It is possible to analyse this effect by comparing the galaxy mass and SFR distribution for the COSMOS and the CDFS samples, since the photometry for galaxies in COSMOS field ranges from FUV to MIR bands. 
This comparison is shown in Fig.~\ref{fastdist1} (middle panel) where we notice that the SFR of the COSMOS galaxies is found systematically lower than that of the CDFS galaxies. Interestingly, the distribution of the sSFR, i.e.,  the ratio between the current SFR and the galaxy mass,  is the same for the two samples, even though the distributions in mass and SFR are slightly different  (Fig.~\ref{fastdist1}  bottom panel).
The sSFR  is less sensitive to the choice of IMF and input  SPS models since the total SFR and mass both exhibit a similar dependence on these parameters \citep{williams:2010ty}.

SED fitting also leads to unrealistically low ages when applied to actively star-forming galaxies since the SED of such galaxies, at all wavelengths, is dominated by their youngest stellar populations,
which outshine the older stellar populations that may inhabit these galaxies \citep{maraston:2010fp}.
Thus, the SED of such galaxies conveys little information on the beginning of star formation, i.e. on the age of their oldest stellar populations.

\begin{figure*}
\begin{center}
$
\begin{array}{c@{\hspace{.1in}}c@{\hspace{.1in}}c}
\includegraphics[width=8.3cm]{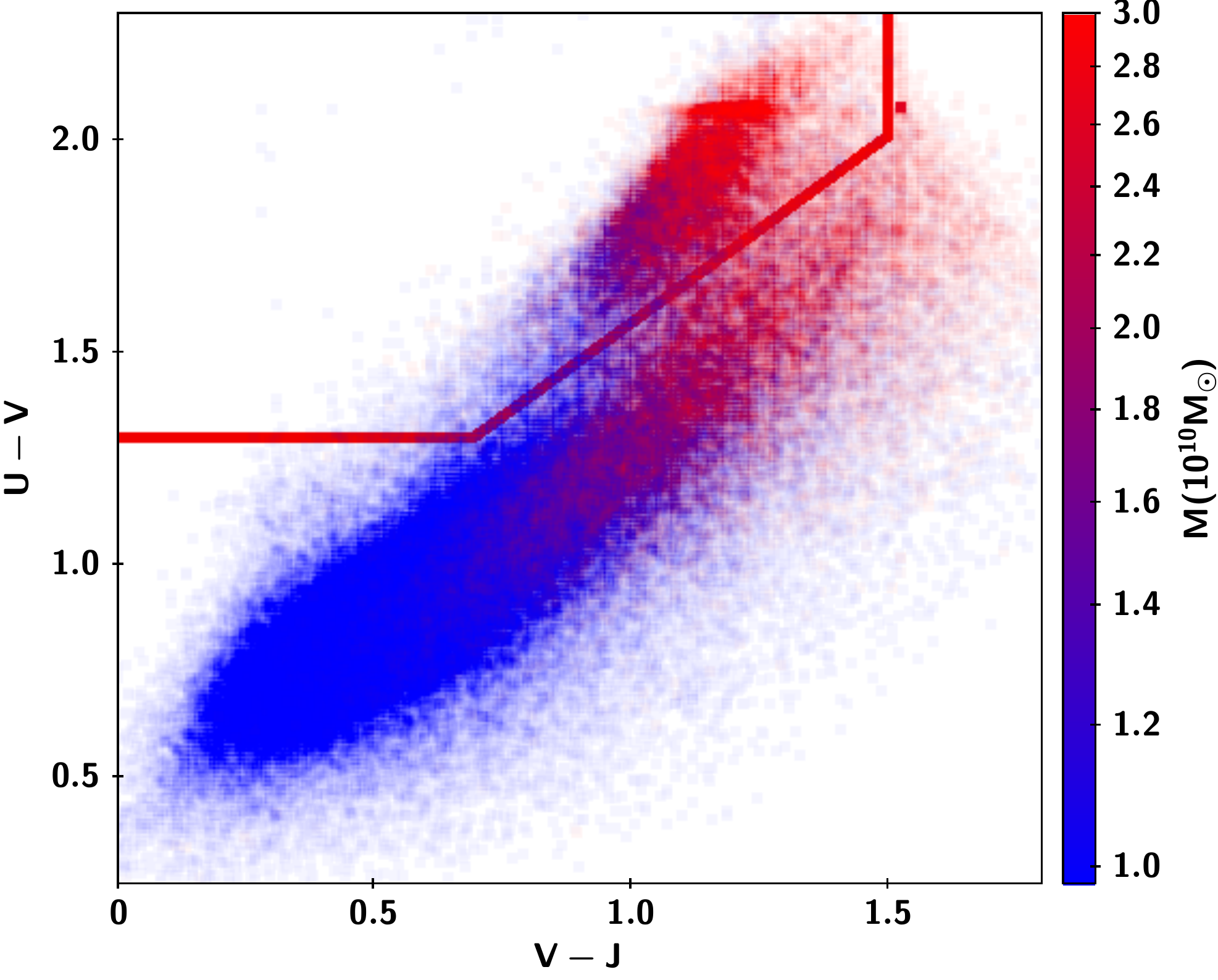} &
\includegraphics[width=8.3cm]{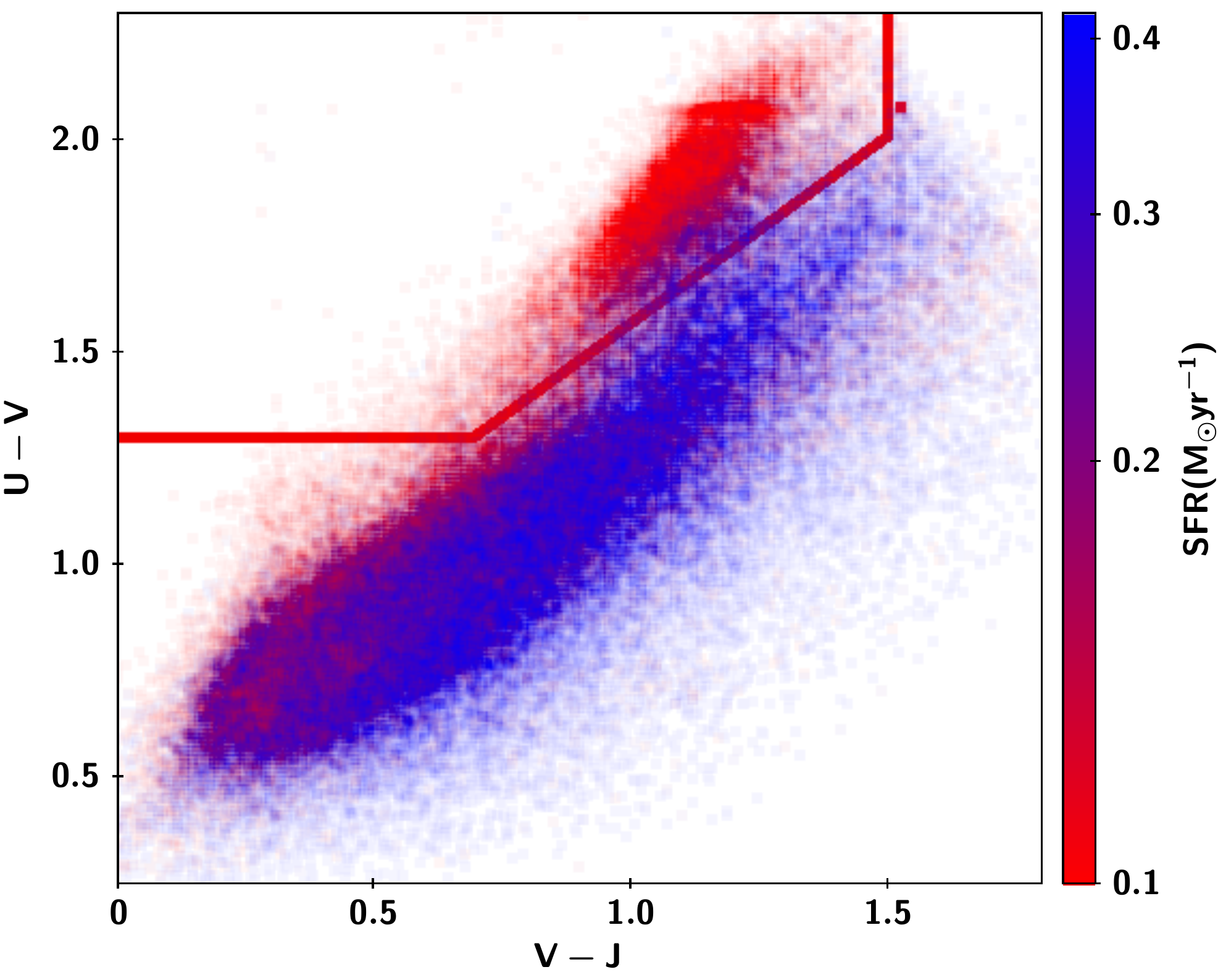}\\
\includegraphics[width=8.7cm]{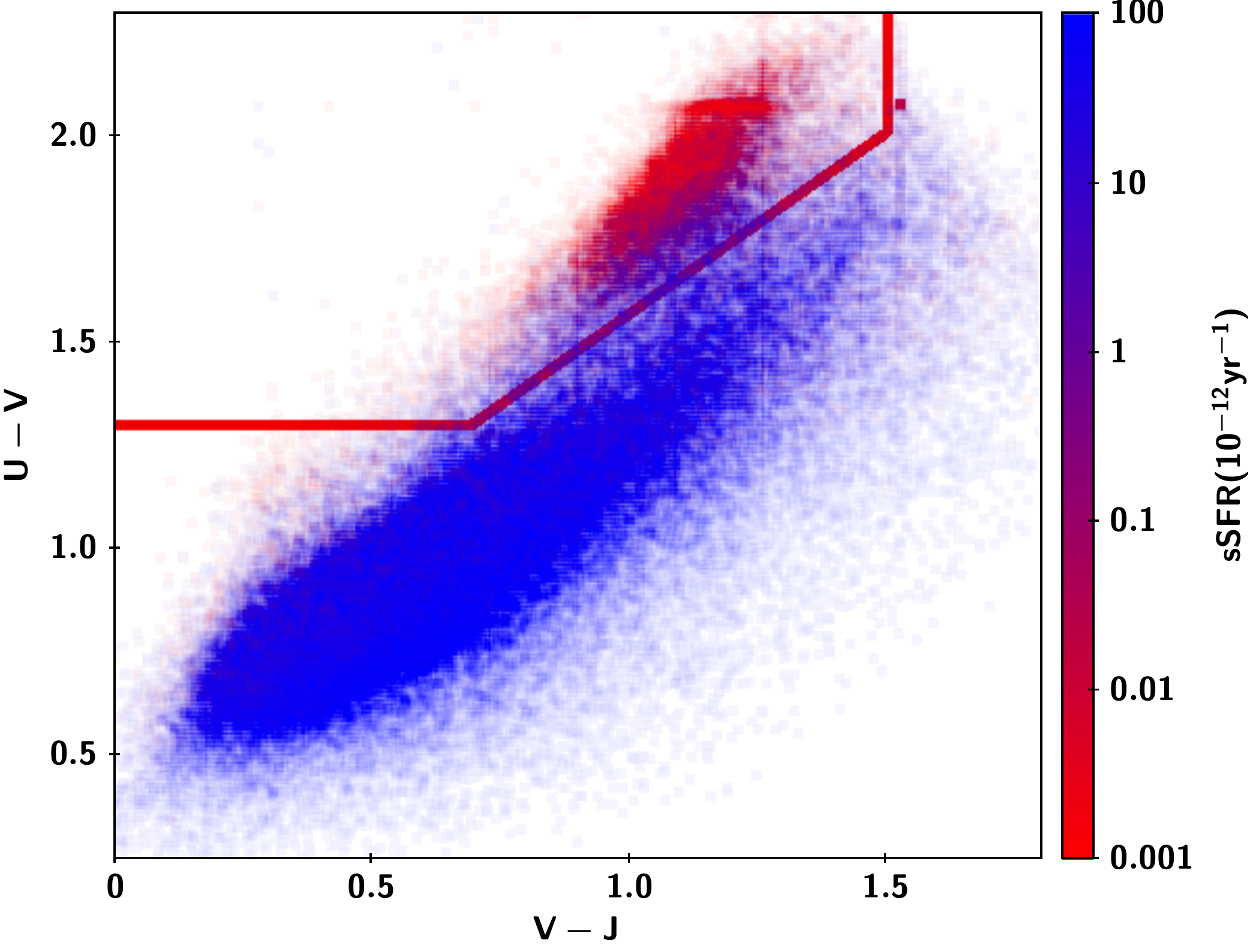}&
\includegraphics[width=8.3cm]{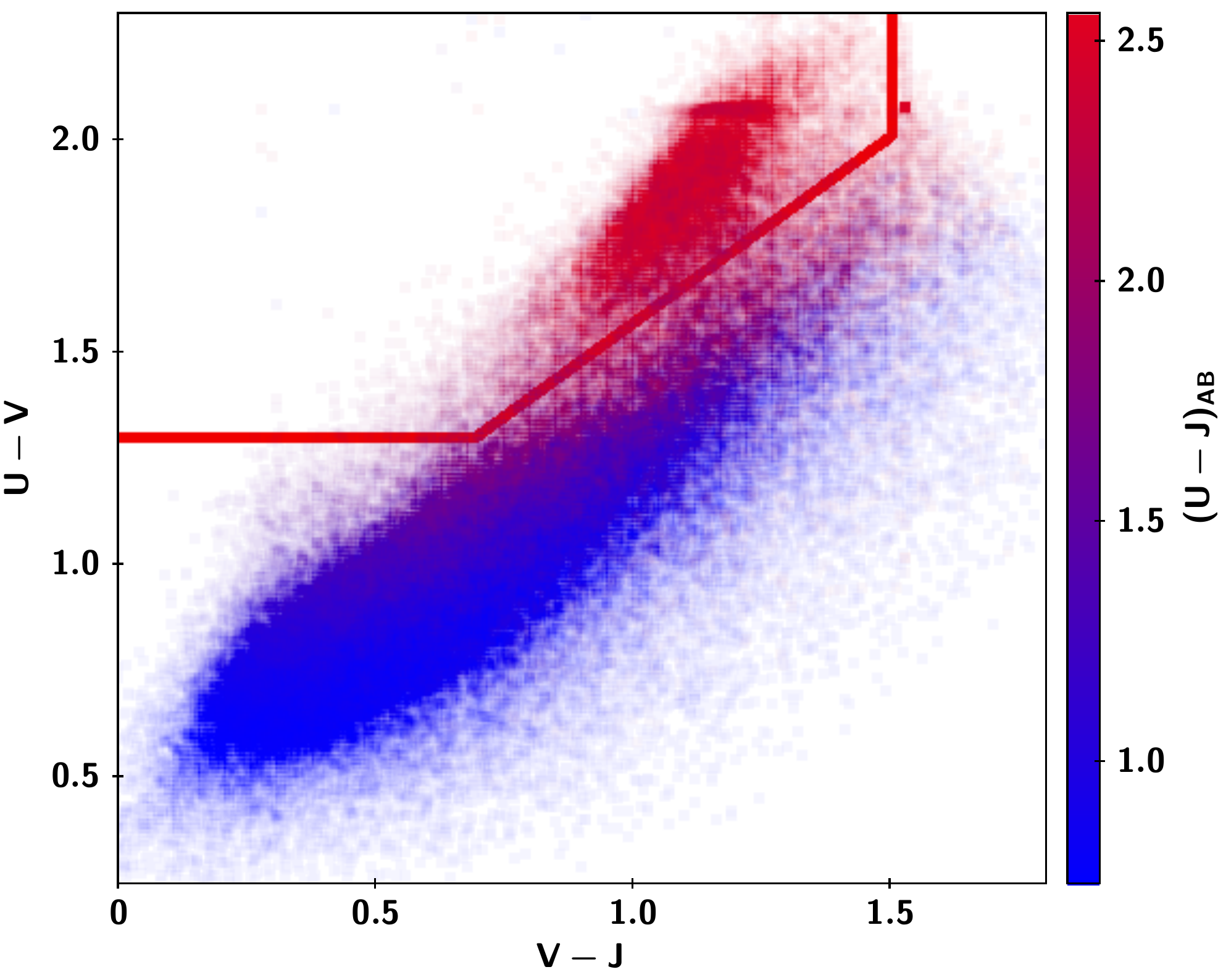}
\end{array}
$
\caption{The rest-frame $U-V$ vs $V-J$ colour-colour plot for galaxies in the overall sample, colour coded by stellar mass (top left), SFR (top right) and sSFR (bottom left), as derived form the SED fitting. In the bottom right 
panel the colour encoding reflects the rest frame de-reddened  $U-J$ colour.} \label{colvsparam}
\end{center}
\end{figure*}

It is instructive to verify whether the separation of star-forming and quiescent galaxies on the $U-V$ vs $V-J$ colour-colour space is well-correlated with separation using SED fitting-determined sSFRs \citep[e.g.][]{williams:2010ty}.
Fig.~\ref{colvsparam}  shows that "blue" star-forming galaxies  exhibit higher sSFR with respect passive "red" galaxies suggesting that the separation of star-forming and passive galaxies based on colours and that based on SED fitting are consistent.  The star-forming galaxies also show lower masses and higher SFRs with respect to the passive galaxies.

In the following we will estimate SN rates in star-forming and passive galaxies exploiting both methods to distinguish between the two classes.

\begin{figure}
\begin{center}
\includegraphics[width=8.5cm]{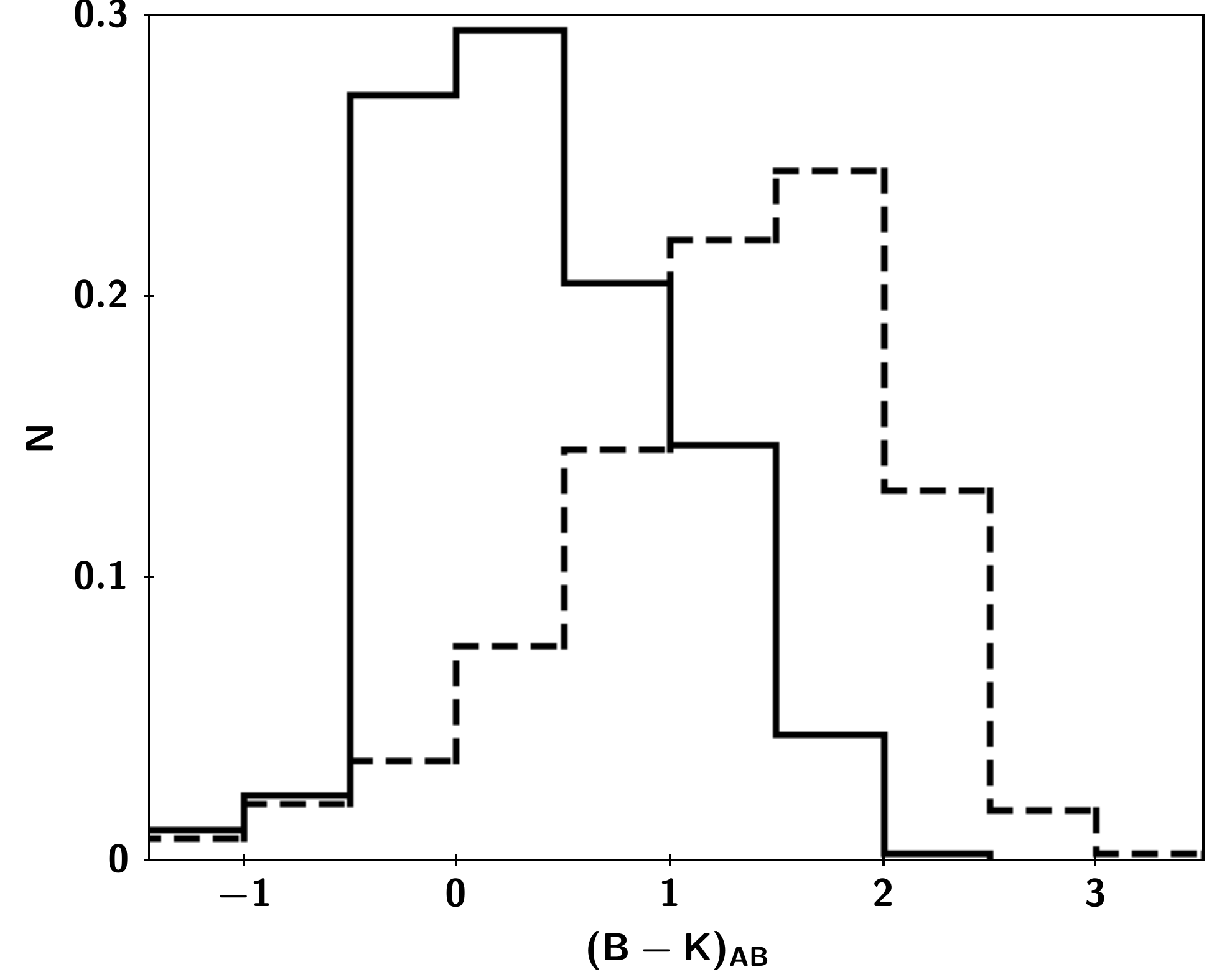} 
\caption{The distribution of the  $B-K$ colour for galaxies in the Local Universe  (dashed line) monitored by the LOSS survey \citep{li:2011qf}, and for galaxies in the SUDARE sample (solid line).
We converted colours  in \citet{li:2011qf} from the Vega  to the AB magnitude system.}
\label{BmK}
\end{center}
\end{figure}

\subsection{Intrinsic dust-corrected colours}\label{restcolours}
Historically,  SN rates have been estimated as a function of  $U-V$ and  $B-K$ colours adopted  as a proxy of the galaxy mean age \citep{cappellaro:1999dg, mannucci:2005mb,li:2011qf}.
To compare the dependence of SN rates on  the $B-K$ colour in the Local Universe and in the redshift range monitored by SUDARE we estimated this rest-frame colour for the galaxies in our sample with {\sc EAZY}.  To correct the colour excess we exploited the extinction coefficients provided by {\sc FAST} for each galaxy assuming the  \cite{calzetti:2000ht} dust attenuation law.
 The distribution of rest-frame  de-reddened $B-K$ colour for the SUDARE and the Local Universe galaxy samples are shown in Fig.~\ref{BmK}. 
 Galaxies in the SUDARE sample appear systematically bluer than those in the local sample in $B-K $ colour. This is not unexpected, since local galaxies are on the average older and characterised by a lower current SFR.  
Furthermore, our  intrinsic colours have been estimated  adopting the   \cite{calzetti:2000ht} dust attenuation law which, having been calibrated on starburst galaxies, could overcorrect the extinction for galaxies with low star formation activity.  In the Local Universe the intrinsic colours have been estimated and corrected for reddening with different methods.



\begin{figure}
\begin{center}
\includegraphics[width=8.5cm]{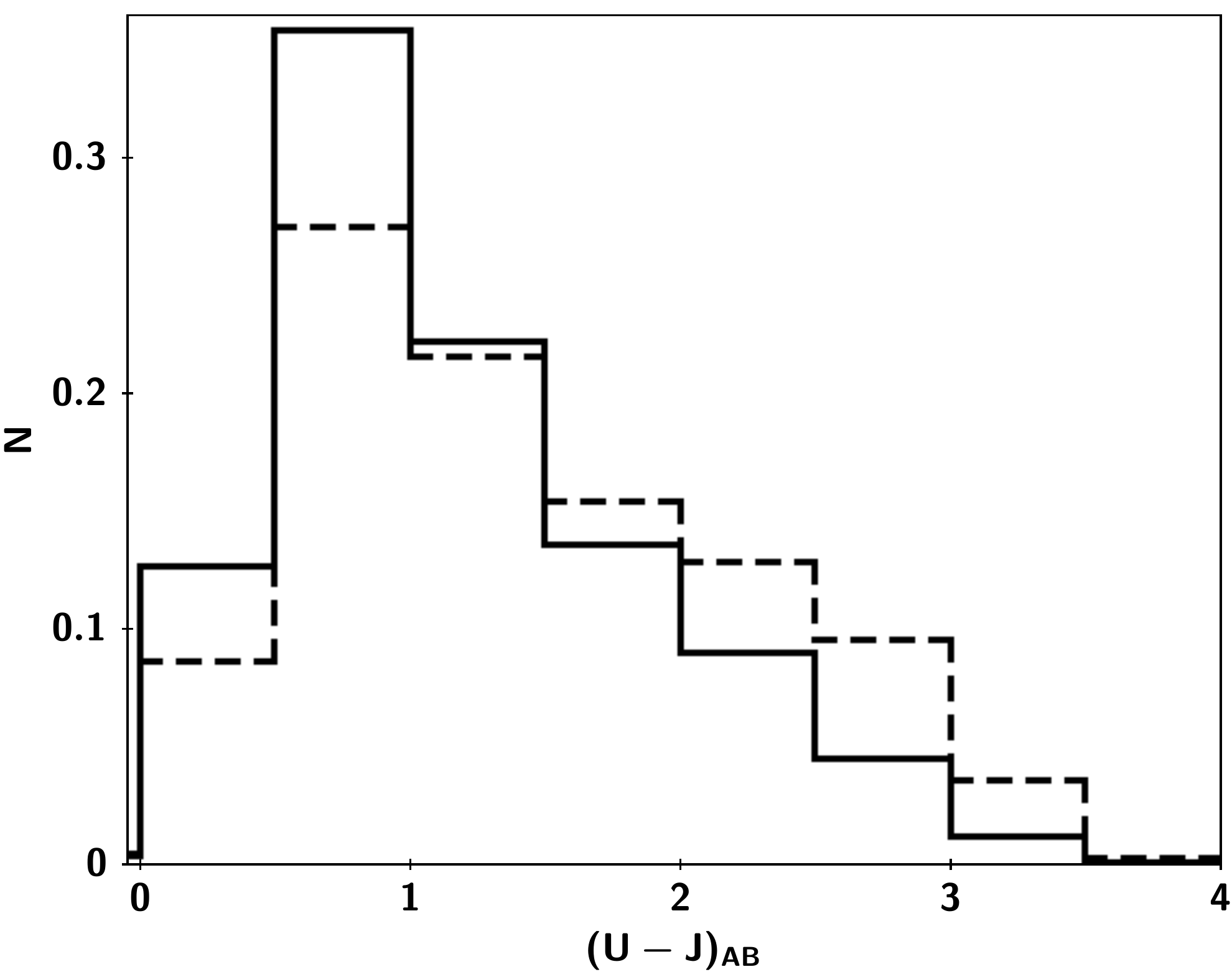}
\caption{The rest frame $U-J$ colour  in AB mag for the CDFS (solid line) and the COSMOS (dashed line) galaxy samples.}
\label{UmJ}
\end{center}
\end{figure}

In our analysis we also considered the intrinsic $U-J$ colour as SFH tracer for our galaxies (see Section \ref{RatevsUmJ}).   This wavelength combination well describes  the contrast between the mass in young and old component.  Moreover,  the star-forming galaxies identified as such as on the $U-V$ vs $V-J$ colour-colour diagram  have rest frame de-reddened $U-J$ colour bluer than the passive galaxies as shown Fig.~\ref{colvsparam}.
The distribution of the $U-J$ colours of galaxies in the CDFS and in the COSMOS samples, shown in Fig.~\ref{UmJ}, are quite consistent.

\section{SN sample}\label{sne} 
During the first two years of our programme we  discovered 117 SNe of which 57\% are Type Ia, 19\%  Type II, 9\%  Type IIn and 15\% Type Ib/c SNe.
A detailed description of the process of SN search, the criteria  and algorithm for SN typing and characteristics of the SN sample (coordinates, type, redshift, discovery epoch, best fit template and reddening) can be found in Paper\,$\textrm{I}$.    

The identification of the host galaxy for each SN in our sample has been obtained by measuring the separation of the SN from each candidate host galaxy in terms of  an elliptical radius (R)  defined by  {\sc SExtractor}  via the equation:
\begin{equation}
 CX(x_{\rm SN} -x_{c} )^2 + CY(y_{\rm SN} - y_{c} )^2 + CXY(x_{\rm SN} - x_{c} )(y_{\rm SN} -y_{c} ) = R^2 
\end{equation}
where $x_{\rm SN}$,  $y_{\rm SN}$  and $x_{c}$,  $y_{c}$ are the coordinates of the SN position  and of the galaxy centre, respectively. 
We assumed that  the  isophotal limit of a given galaxy corresponds to $R=3$.  
 The identification of the correct host galaxy, the  nearest in terms of R, has been verified through visual inspection and {verifying the consistency
of the host galaxy photometric redshift with the estimate  of SN  redshift from the photometric
classification (see Paper\,$\textrm{I}$).}

 The host galaxies of 19 SNe were not detected in the $K_{\rm s}$ stacks, the host galaxies of 4 SNe were fainter than our selection limit ($K_{\rm s}=23.5$\, mag), the hosts of 13 SNe had a poorly determined redshift ($Q_z>1$) and the hosts of 4 SNe had an unreasonable value of the SFR parameter (log(SFR)$ < -99$).
These 40 events have been discarded for the computation on the SN rates, since their hosts did not meet our selection criteria. 
 This reduced the SN sample to $13$ CC~SNe in the redshift range $0.15<z<0.35$  and  $36$ SNe Ia in the redshift range $0.15<z<0.75$.
 We point out that the same criteria that lead to the exclusion of the 40 events mentioned above are used to cull the whole galaxy sample, leading to a correspondent pruning of galaxies in which no SN was detected. 
 A fraction of SN candidates  (10\%) has an uncertain photometric classification, either because the light curve was not well sampled, or because affected by  large photometric
errors. These SNe have been classified and have been weighted by a factor  of 0.5 in the rate calculation.
The volumetric rates presented in Paper\,$\textrm{I}$  are based on a larger number of  events (26 CC~SNe and 53 SNe~Ia), since we selected all the SNe exploded within a given volume ($0.15<z<0.35$ and  $0.15<z<0.75$ for CC and Ia SNe, respectively)  without any constraint on the host galaxies.

\section{SN rate}\label{rate}
Ideally, we should measure detailed SFHs for the galaxies  monitored in our SN survey and relate the SNe directly to the stellar populations in which they originated \citep[e.g.][]{greggio:2010pd}.
  
  Since the probability of detecting a SN in one individual galaxy is low,  it is customary to add the data relative to  galaxies which are thought to share similar SFHs, in order to constrain the trend of the SN rate with the parent stellar population on a robust ground. The binning of galaxies can be performed in different ways, e.g. star-forming vs passive galaxies, or, in general, according to their intrinsic (rest frame and de-reddened) colour. The total SN rate of the bin will reflect the total SFH of the group,  that is the  mass and  the age distribution sampled in  the bin.

\subsection{SN rate and SFH}\label{ratesfh}

The SN rate in a given galaxy can be  expressed as :

\begin{equation}
R_{\rm SN}(t) =  \int_{\tau_{\rm min}}^{min(t,\tau_{\rm max})} \psi(t- \tau) k_{\alpha}(t- \tau) A_{\rm SN}(t- \tau) f_{\rm SN}(\tau) d{\tau}
\end{equation}
where $t$ is the time elapsed since the beginning of star formation in the considered galaxy, $\psi$ the SFR, $k_{\alpha}(t-\tau)$ is the number of stars per unit mass of the stellar generation born at epoch $(t-\tau)$, $A_{\rm SN}(t-\tau)$ is the number fraction of stars from this stellar generation that end up as SN, $f_{\rm SN}$ is  the DTD and $\tau_{\rm min}$ and $\tau_{\rm max}$ are  the minimum and maximum possible delay times, respectively. 
The factor $k_{\alpha}$ is given by:
\begin{equation}
k_{\alpha}(t-\tau)= \frac{\int_{m_l}^{m_u} \phi({m,t-\tau}) dm}{\int_{m_l}^{m_u} m\phi(m,t-\tau) dm }
\end{equation}
where $\phi$ is the IMF and  $m_l-m_u$ is the mass range of the IMF.  
The factor $A_{\rm SN}$ can be expressed by:
\begin{equation}
A_{\rm SN}(t-\tau)=P_{\rm SN}(t-\tau)\frac{\int_{m_{\rm SN,l}}^{m_{\rm SN,u}} \phi(m,t-\tau) dm}{\int_{m_l}^{m_u} \phi(m,t-\tau) dm}
\end{equation}
where $P_{\rm SN}$ is the probability that a star with suitable mass ($m_{\rm SN, u}-m_{\rm SN, l}$) to become a SN actually does it.  This probability depends on the SN progenitor models and on stellar evolution assumptions.
The  factors  $k_{\alpha}$ and $A_{\rm SN}$ can vary with galaxy evolution, for example due to the effects of higher metallicities and$/$or the possible evolution of IMF.  In the following we assume that $k_{\alpha}$ , $A_{\rm SN}$ and the IMF do not vary with time:
\begin{equation}
R_{\rm SN}(t) = k_{\alpha} A_{\rm SN} \int_{\tau_{\rm min}}^{min(t,\tau_{\rm max})} \psi(t- \tau) f_{\rm SN}(\tau) d{\tau}.
\end{equation}
We also consider  the DTD normalized to 1 over the total range of the delay times:
\begin{equation}
\int_{\tau_{min}}^{\tau_{max}} f_{\rm SN}(\tau) d{\tau}=1
\end{equation}

If we assume that all stars with suitable mass ($m_{\rm CC, u}-m_{\rm CC, l}$) become CC~SNe  with a negligible delay time , due to their short life time ($3-20$\,Myr), and that the SFR has remained constant over this timescale, we obtain that  the rate of  CC~SNe is simply proportional to the current SFR,  through a factor which is the number of CC progenitors per unit mass of the parent stellar population:
\begin{equation}\label{rcc}
R_{\rm CC}(t) = k_{\rm CC} \times  \psi(t) 
\end{equation}
with:
\begin{equation}\label{kcc}
k_{\rm CC} = \frac{\int_{m_{\rm CC,l}}^{m_{\rm CC,u}} \phi(m) dm}{\int_{m_l}^{m_u} m\phi(m) dm}
\end{equation}
where $\phi$ is the IMF and and  $m_l-m_u$ is the mass range of the IMF. 
On the other hand, the time elapsed between the birth of SNe~Ia progenitors and their explosions ranges from $\sim 30$\,Myr
to several billion years and controls the production rate of SNe~Ia. 
For SNe~Ia the observed rate includes progenitors born all along the past SFH: 
\begin{equation}\label{rateIa1}
R_{\rm Ia}(t) =  k_{\rm Ia}\int_{\tau_{\rm min}}^{min(t,\tau_{\rm max})} \psi(t- \tau) f_{\rm Ia}(\tau) d{\tau}
\end{equation}
where $k_{\rm Ia}=k_{\alpha} \times A_{\rm Ia}$, the productivity of SNe~Ia,  is the number of type Ia SNe produced per unit mass of the parent stellar population:
\begin{equation}
k_{\rm Ia}= P_{\rm SN}\frac{\int_{m_{\rm SN,l}}^{m_{\rm SN,u}} \phi(m) dm}{\int_{m_l}^{m_u} m\phi(m) dm }
\end{equation}

Both the productivity ( $k_{\rm Ia}$) and the DTD ( $f_{\rm Ia}$) of SNe~Ia are sensitive to the model for their progenitors and depend on many parameters including the distribution of binary separations and mass ratios,
the outcome of the mass exchange phases, and of the final accretion on top of the WD.

In the literature one can find different approaches at measuring the SN rates, which use different proxies for the galaxy SFH. For example the SN rates have been measured in galaxies of different Hubble types \citep[e.g.][]{van-den-Bergh:1990lr,ruiz-Lapuente:1995xy}, or $U-V$ and $B-K$ colours \citep[e.g.][]{cappellaro:1999dg,mannucci:2005mb,li:2011qf}. In this paper we measure the SN rates as function of $B-K$  colour, in passive and star-forming galaxies identified as such on $U-V$ vs $V-J$ colour-colour diagram and as a function of the sSFR estimated  through SED fitting technique.

\subsection{Computing the SN rate}\label{ratecalc}
Three basic ingredients are required to estimate the SN rate in a galaxy sample: the number and type of SNe discovered,
the time of effective surveillance of the sample (control time) in order to relate the detection frequency to the intrinsic SN rate, a physical parameter  to normalise the rate.

The control time (CT)  is defined as the time interval during which a SN occurring in a given galaxy could be detected in the search  \citep{zwicky:1942uq}. The CT of each single observation depends on SN light curve,  distance, reddening, efficiency of SN detection for that observation. The total  CT  is computed by  properly summing the contribution of individual observations.
The calculation of the CT is illustrated in detail in Paper\,$\textrm{I}$ (Sec. 7.1)
In the literature different  normalisation factors, proportional to the content  of stars that would eventually explode as SNe,  have been adopted to compute the SN rate  such as the  $B$-band luminosity  \citep[e.g.][]{cappellaro:1999dg},  $H$-band luminosity    \citep[e.g.][]{van-den-Bergh:1990lr}, the $K$-band luminosity   \citep[e.g.][]{della-Valle:1994uq,mannucci:2005mb,li:2011qf},  the far-infrared luminosity  \citep[e.g.][]{van-den-Bergh:1991uq, mannucci:2003wl}, the stellar mass derived from  the  $K$-band luminosity and the $B-K$ colour \citep[e.g.][]{mannucci:2005mb,li:2011qf}.
In this paper we normalise the SN rates to the galaxy stellar mass, with the latter derived from the SED fitting, as in \citealt{Sullivan:2006nq} and \citealt{smith:2012kx}.
The SN rate per unit mass is then computed as:
\begin{equation}
\label{massrate}
  r_{\mathrm{SN}}=\frac{N_{\rm SN}}{\sum_{i=1}^{\rm N_{\rm gal}}M_{i}\,{CT_{i, \rm SN}}} 
\end{equation}
\noindent where $N_{\rm SN}$ is the  total number of SNe of a given type, $N_{\rm gal}$ is the total number of galaxies, $M_{i}$, $CT_{i,\rm SN}$ are the stellar mass  and the control time of the $i$-th galaxy, respectively.
Galaxy masses in the literature are often referred to a  \citet{kroupa:2001jk}  IMF. To compare our results with those of other authors we need to convert the rates per unit mass, SFRs and galaxy masses, to our IMF of choice, that is a Salpeter IMF between $0.1$ and $100\, {\rm M}_\sun$. We do this by multiplying by a factor of $1.36$ the masses obtained with a Kroupa (or equivalent) IMF. This factor is the ratio between the total mass of SPSs with the same mass in stars heavier than $0.5\,{\rm M}_\sun$, but with a different content of low mass stars, as the two IMFs require.

\section{SN rates as a function of intrinsic colours}\label{ratecolor}
 \citet{mannucci:2005mb} analysing the local sample of  \citet{cappellaro:1999dg} demonstrated that the SN~Ia rates per unit  $K$ band luminosity and per unit stellar mass have a very strong dependence on the galaxy Hubble type and $B - K$ colour:  "blue" galaxies exhibit an SN~Ia rate larger by about a  factor of 30 than that of "red" galaxies and confirmed the earlier suggestions that a significant fraction of SNe~Ia  in late spirals/irregulars
originates in a relatively young stellar component. The evidence that  "blue" galaxies are more efficient in producing SN Ia events was interpreted as due to a DTD being more populated at short delay times ($<1$\,Gyr) already by
\citet{oemler:1979rf}  and \citet{greggio:1983gr}. 
A similar SN~Ia rate dependence on host galaxy colours was found  by \citet{li:2011qf} for the LOSS sample  that  suggested  an increasing  SN~Ia rate  from "red" to "blue" galaxies by a factor of $6.5$.

For CC~SNe  the dependence of the rate on galaxy colours is more enhanced  as already noted by \citet{cappellaro:1999dg} and confirmed by 
 \citet{mannucci:2005mb} and \citet{li:2011qf}.
The rate of the CC~SNe in the early type galaxies  is close to zero, it is  small in the reddest galaxies, and in general becomes progressively higher in bluer galaxies. 

\begin{figure}
\begin{center}
\includegraphics[width=10cm]{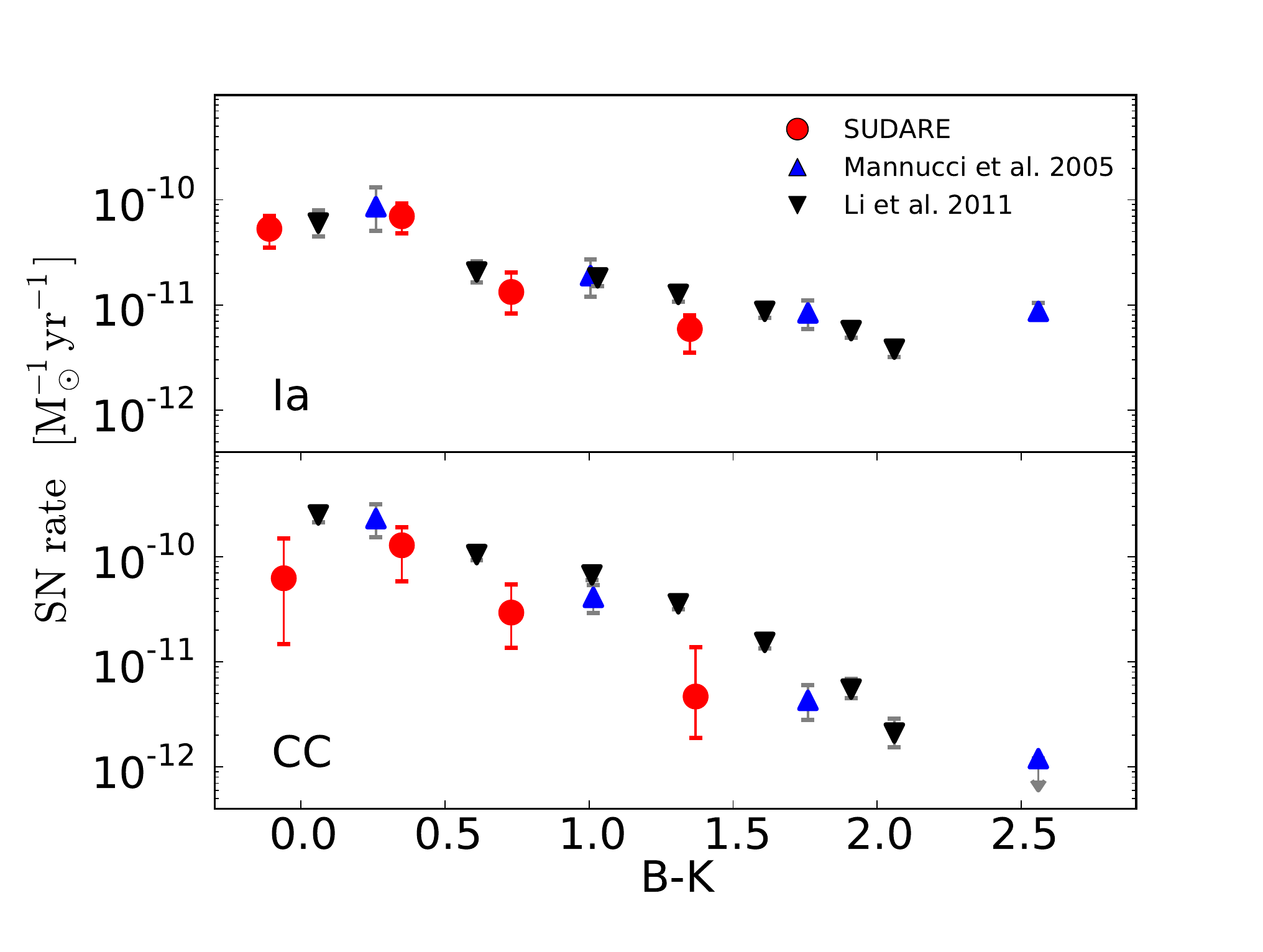}
\caption{SN rates  as a function of  $B-K$ galaxy colour  for  SNe~Ia (top panel) and CC~SNe  (bottom panel).  For comparison we report the same measurements obtained  in the Local Universe by \cite{mannucci:2005mb} and by \citet{li:2011qf}. The mass estimates for  the  galaxies by  \cite{mannucci:2005mb} and  \citet{li:2011qf} have been converted to a Salpeter IMF and $B-K$ colours are all in AB mag.} 
\label{rateBK}
\end{center}
\end{figure}

Our measurements of both Type Ia and CC~SN rates as  a function of  $B-K$  are shown in Fig.~\ref{rateBK} together with measurements in the Local Universe.
The  SN~Ia rate as a function of  $B-K$ colour  from SUDARE is in good agreement with local measurements, following the same trend. The  measurements of the CC~SN rate from SUDARE  are  lower than the  local ones in most colour bins, although consistent within the uncertainties.
Our results suggest that there is no evolution with cosmic time of the dependence on galaxy colours for both SN~CC and Ia rates.

\section{SN rates in star-forming and passive galaxies}\label{ratesfr}
The association between  a colour sequence and a sequence in the SFR is not straightforward, since the colours trace the global SFH in a galaxy, while the SFR refers to only the more recent past. The assumption of analytical expressions for the SFH in the SED fitting enforces a relation between the global colours and the recent SFR which may be inappropriate for the system.  Moreover,   the dust content and the presence of recent mergers can make extremely complex the relations between SFR and the galaxy colours. 
We showed in  Sec.~\ref{sed}  that separation of star-forming and quiescent galaxies in $U-V$, $V-J$ colour-colour space is reliable, thus to analyse the dependence of SN rates on SFR  we split our host galaxy sample in passive and star-forming subsamples  following the prescription by \citet{williams:2009fk} (Fig.~\ref{UmV_VmJ}).
Our results are reported in Table~\ref{ratecol} and show that  the Type Ia SN rate is about a factor of five higher in the star-forming sample with respect to the passive sample. This supports the notion of a DTD declining with increasing delay time.  There are not  CC~SNe discovered in the passive galaxies  and we can measure only an upper limit for their rate. This is expected given the short lifetimes of CC~SN progenitors which imply that the events die out soon after the end of the star formation activity. In star-forming galaxies we found a CC~SN rate  a factor two higher than the SN~Ia rate.

\begin{table}
\caption{SN rate per unit mass [$10^{-3}\,{\rm SNe}\,{\rm yr}^{-1}\,10^{-10}\,{\rm M}_\sun$] in "passive" and "star-forming" galaxies, identified as such on the $U-V$,$V-J$ diagram. The SN~Ia rate has been measured in the redshift range  $0.15<z<0.75$ while the CC~SN rate in the range $0.15<z<0.35$.}\label{ratecol}
\begin{tabular}{|c|cc|cc|}
\hline
 &         &       &               &\\
SN type & gal type&$N_{gal}$   &$N_{SN}$ & rate  \\ [6pt]

\hline
      &         &     &                 &\\
 Ia & passive&  12223    &    $4.0$  &  $0.5^{+0.2}_{-0.3}$    \\[4pt]  
     &star-forming & 68105&   $32.6$ &  $2.7^{+0.5}_{-0.4}$  \\[4pt] 
\hline
 &         &          &            &\\
 CC  &passive & 3013& $0.0$& $ < 0.1 $      \\[4pt]
       & star-forming& 14208& $13.6$ & $4.4^{+1.2}_{-1.3} $\\[4pt]
\hline
\end{tabular}
\end{table}

\section{SN rates as a function of sSFR}\label{ratessfr}
The sSFR is a proxy for the SFH and the evolutionary stage that the galaxy is in, in particular the inverse of the sSFR defines a timescale for the formation of the stellar population of a galaxy. 
To compare our results to previous works we adopt the same criteria to separate passive and star-forming galaxies as in   \citet{Sullivan:2006nq}, defining  three galaxy groups based on the nature of star formation: the first group of passive galaxies with a zero mean SFR; the second group of galaxies with a small or moderate sSFR ($-12.0 < \rm log(sSFR)<-9.5$); the third group of galaxies with a large sSFR ($ \rm log(sSFR)> -9.5$).
We stress that  the condition $\rm log(sSFR) > -9.5$ as division between highly and moderately star-forming is arbitrary.

\begin{figure*}
\begin{center}
$
\begin{array}{c@{\hspace{.1in}}c@{\hspace{.1in}}c}
\includegraphics[width=9.5cm]{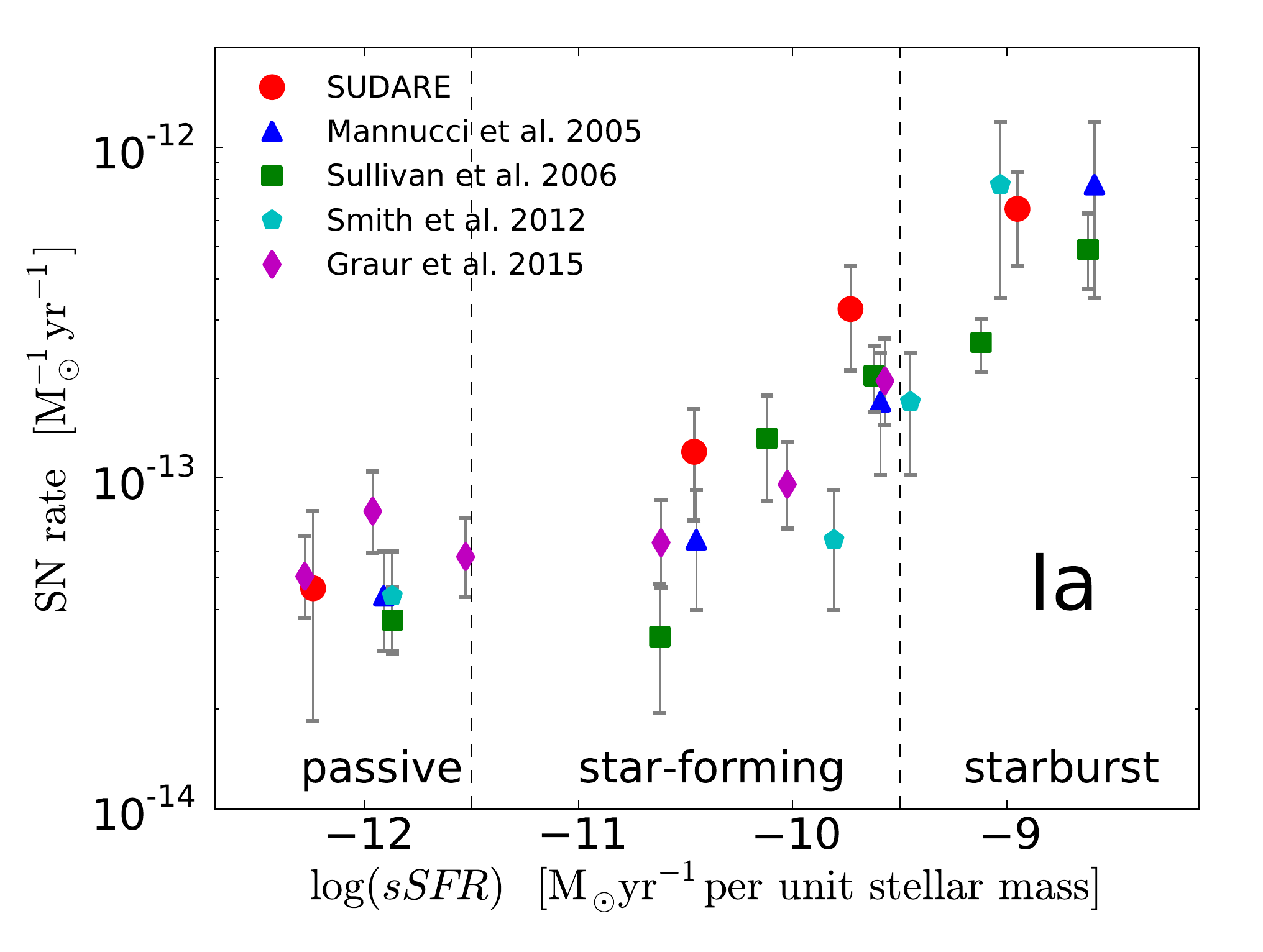} &
\includegraphics[width=9.5cm]{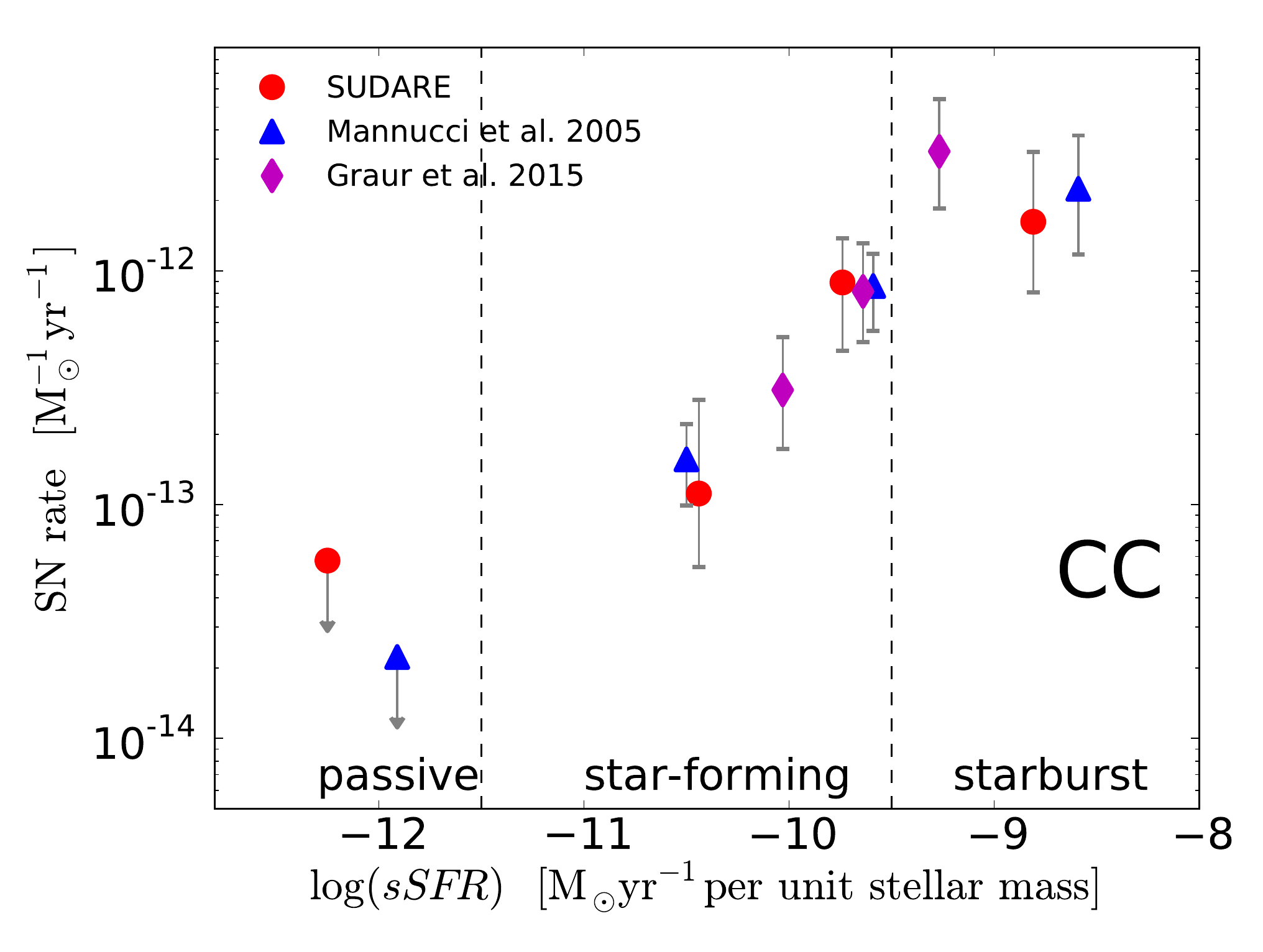}
\end{array}
$
\end{center}
\caption{SN rates in SNuM vs sSFR for Type Ia (left panel) and  CC~SNe  (right panel) in three different groups of galaxies based on  their sSFR: the first group of passive galaxies with a zero mean SFR; the second group of galaxies with $-12.0 < \rm log(sSFR)<-9.5$; the third group of galaxies with $ \rm log(sSFR)> -9.5$.  In the both panels the circles are from SUDARE, the triangles from \citet{mannucci:2005mb}, the squares from \citet{Sullivan:2006nq}, the pentagons are  from \citet{smith:2012kx} and the  diamonds from \citet{graur:2015lr}. The literature values have been corrected to a Salpeter IMF.}
\label{mass_ssfr}
\end{figure*}

Our results illustrated in Table~\ref{ssfr1} and  in Fig.~\ref{mass_ssfr}  confirm the increase in the rates per unit stellar mass with increasing galaxy sSFR with approximately the same trend in the Local Universe and at intermediate redshift.
The enhancement  of both  SN~Ia and CC~SN rate from star-forming to starburst galaxies is steeper for CC~SNe:  we found  an increase of  about a factor 5 for Type Ia  and of 15  for CC~SNe.

Fig.~\ref{mass_ssfr} shows that the results from previous surveys yield a 
consistent increase of the SN~Ia rate with increasing sSFR, with no evident dependence on redshift within the uncertainties.
The difference between the  SN~Ia rate in passive galaxies and starburst galaxies  is about a factor of  13 for our measurements, very close to \citet{Sullivan:2006nq}  estimates in an almost identical range of redshift.
A similar trend between the SN~Ia rate per unit mass and the sSFR  is reported by  \citet{graur:2015lr} in the redshift range $0.04 < z < 0.2$, while \citet{smith:2012kx} found  a steeper  increase  from star-forming to starburst galaxies in the same range of redshift ($0.05 < z < 0.25$). 
 We should mention that in the works of \citet{Sullivan:2006nq} and \citet{smith:2012kx} the galaxy masses were determined via SED fitting to PEGASE 2.2 models and that the sSFR is measured as the ratio between the mean SFR over the last 0.5\,Gyr and the current stellar mass of the galaxy.
In the Local Universe \citet{mannucci:2005mb}  adopted a galaxy classification based on the Hubble type and  \citet{Sullivan:2006nq}  converted  it in a sSFR classification, scaling  rates in E/S0   to match their rate in passive galaxies.

The trend of SN~Ia rate with the sSFR can be readily interpreted as shown in  \citet{greggio:2010pd} (Fig. 12), and reflects the fact that the higher the SFR over the past 0.5\,Gyr the higher the SN~Ia rate. The similar scaling of the SN~Ia rate with sSFR found at the different redshifts supports the notion that  the ability of the stellar populations to produce SN~Ia events does not vary  with cosmic time.

The dependence of the  CC~SN rate per unit mass  on the sSFR is expected to be  linear since the CC~SN rate is proportional to the recent SFR and the trend from SUDARE data is very similar to that observed  in the Local Universe \citep{mannucci:2005mb}  and  in the redshift range $0.04 < z < 0.2$   \citep{graur:2015lr}.

\begin{table*}
\begin{center}
\caption{SN rate per unit mass [$10^{-3}\,{\rm SNe}\,{\rm yr}^{-1}\,10^{-10}\,{\rm M}_\sun$] in different bins of sSFR and galaxy mass as measured by  {\sc FAST}. The Type Ia SN rate has been measured in the redshift range  $0.15<z<0.75$ while the CC SN rate in the range $0.15<z<0.35$. The SN rates as a function of stellar mass have been estimated only for star-forming galaxies.}\label{ssfr1}
\begin{tabular}{|c|cccc|cccc|}
\hline
SN type &$N_{\rm gal}$ & log$(<\rm sSFR>)$  & $N_{\rm SN}$ & rate & $N_{\rm gal}$ &$\rm log <M> $& $N_{\rm SN}$ & rate  \\[6pt]
\hline
& & & & & & & &  \\
 &$8718$ &$-12.2$ & $3.5$ & $0.50^{+0.28}_{-0.33}$ & 16861&$8.8$ & $3.6$  &$14.8^{+9.1}_{-9.7}$  \\[4pt]
Ia&$21595$&$-10.5$ &  $9.8$ & $1.2^{+0.4}_{-0.5}$ & 19468& $9.3$&  $5.7$ & $8.0^{+3.8}_{-4.0}$ \\[4pt]
&$22094$ &$-9.7$ & $10.6$ & $3.2^{+1.1}_{-1.1}$ &14948& $9.8 $& $8.9$ & $5.8^{+2.0}_{-2.4}$ \\[4pt]
&$ 24101$&$-9.0$ &$12.8$ & $6.5^{+1.9}_{-2.2}$ &16513& $10.9$& $15.0$  &$1.4^{+0.5}_{-0.4}$ \\[4pt]
\hline
& & & & & & & &  \\
 &$2198$ & $-12.2$ & $0.0$ & $<0.6$ & 6415 & $8.8$& $1.9$ &  $26^{+15}_{-19}$   \\[4pt]
CC &$5130$&$-10.5$ & $2.9$ & $1.1^{+0.6}_{-1}$ &3092 & $9.3$& $2.3$ & $11 ^{+7}_{-20}$ \\[4pt]
& $5580$ &$-9.7$ &$5.9$& $8.9^{+4.4}_{-6.9}$&2106 & $9.8$ &$3.9$ &$13 ^{+7}_{-10}$  \\[4pt]
&$3119$ &$-8.8$ &$4.2$ & $16.2^{+8}_{-16}$ &2216& $10.9$&$5.1$& $2.0^{+1.0}_{-1.5}$ \\[4pt]
 \hline
\end{tabular}
\end{center}
\end{table*}

\section{SN rates as a function of galaxy mass}\label{ratemass}
In the recent literature several papers suggest that the rate of both SNe~Ia and  CC per unit stellar mass scales with the mass of the parent galaxy  \citep{Sullivan:2006nq,li:2011qf,graur:2015lr}, but with some controversy regarding the trend in passive and star-forming galaxies.  \citet{smith:2012kx} suggest that the different results may be related to the different redshift range targeted in the various surveys.
 
\begin{figure*}
\begin{center}
$
\begin{array}{c@{\hspace{.1in}}c@{\hspace{.1in}}c}
\includegraphics[width=9.5cm]{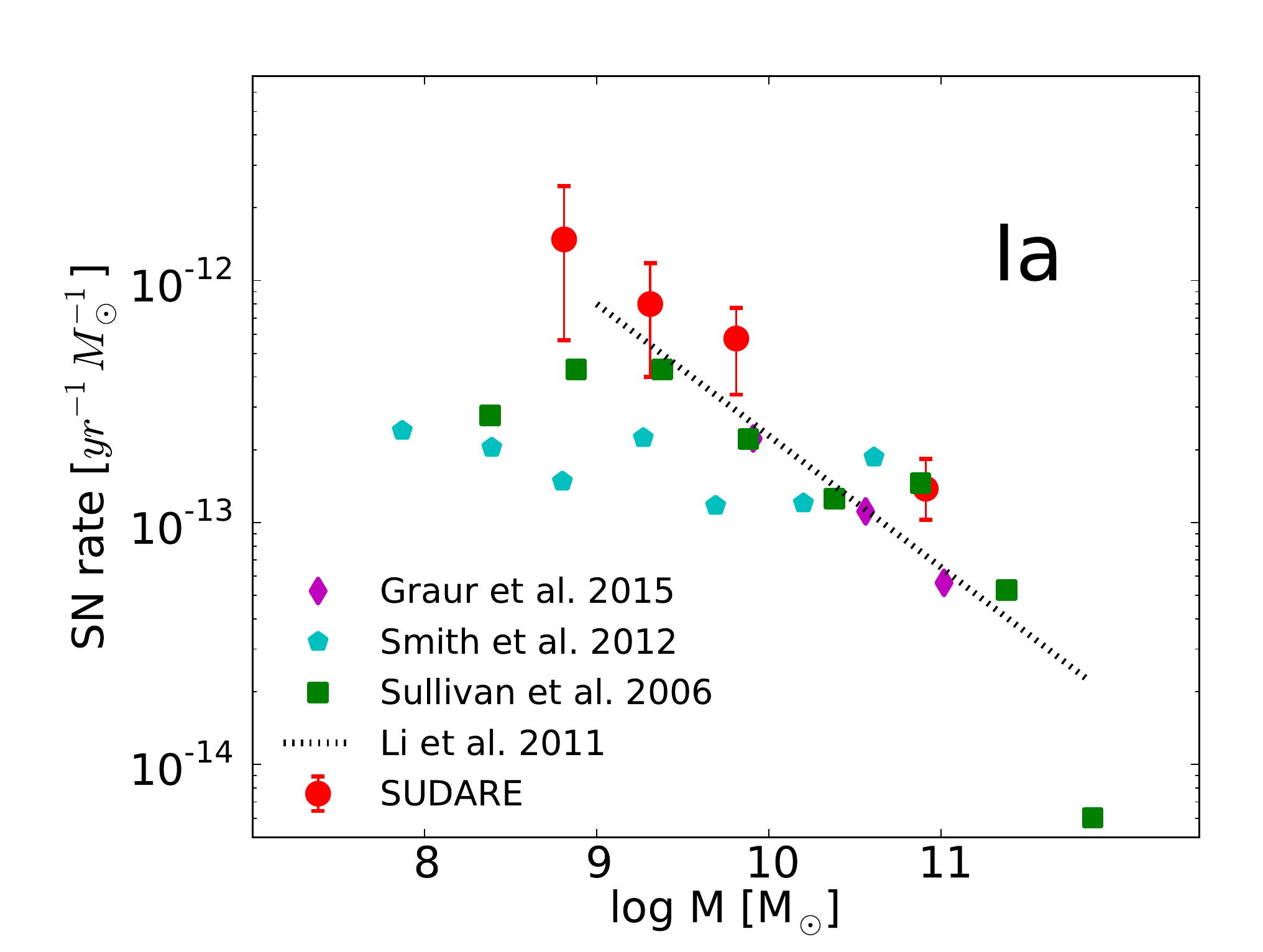} &
\includegraphics[width=9.5cm]{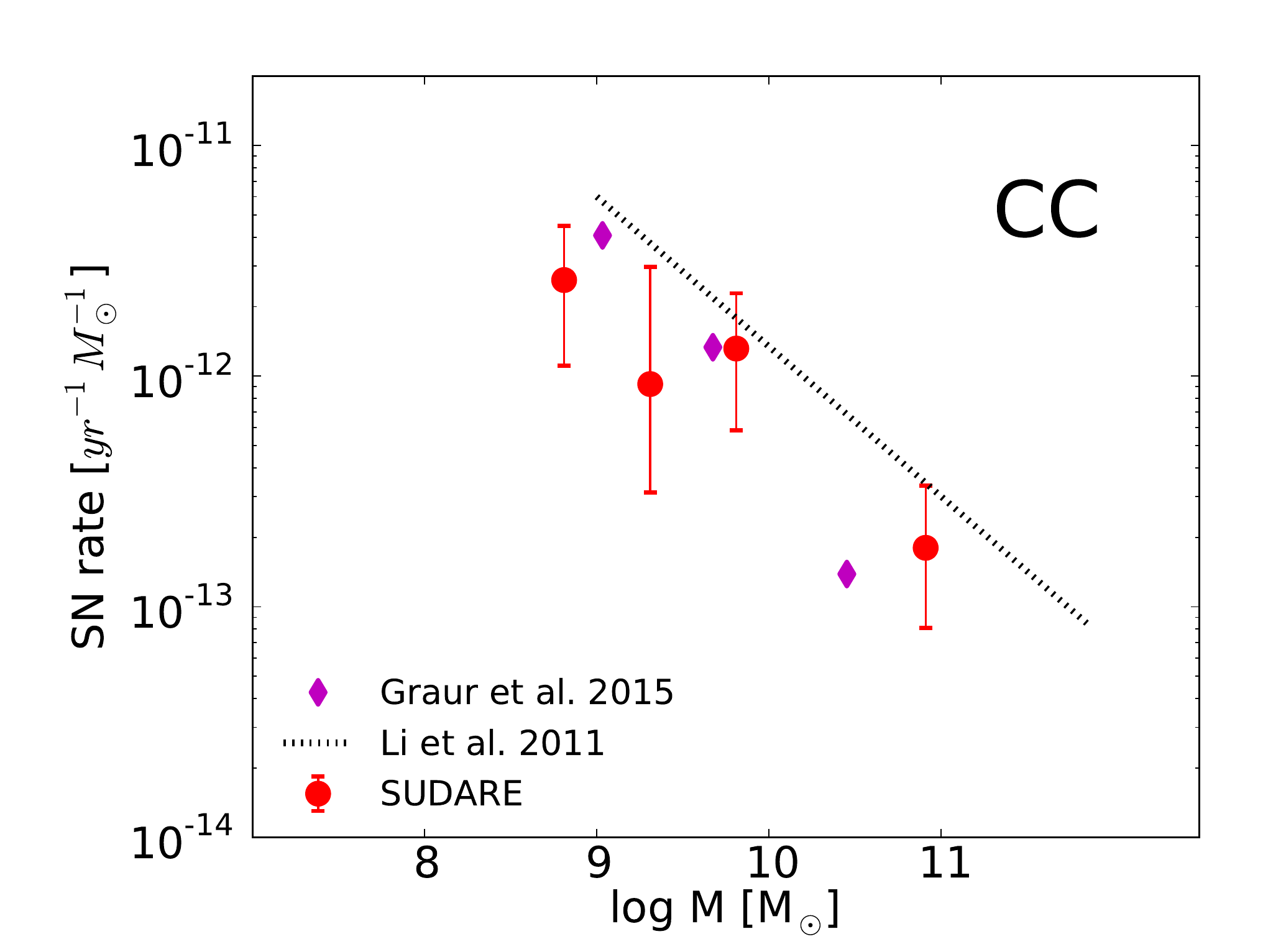}
\end{array}
$
\end{center}
\caption{SN rates per unit stellar mass [$10^{-3}\,{\rm SNe}\,{\rm yr}^{-1}\,10^{-10}\,{\rm M}_\sun$] as a function of the mass of the parent galaxy  in the star-forming subsample for Type Ia (left panel) and  CC~SNe  (right panel).  In both panels the circles are from SUDARE, the triangles from \citet{mannucci:2005mb}, the squares from \citet{Sullivan:2006nq}, the pentagons  from \citet{smith:2012kx} and the  diamonds from \citet{graur:2015lr}. SN rates and masses have been rescaled to a Salpeter IMF.}
\label{rate_mass}
\end{figure*}

The rate of CC~SNe per unit mass is  proportional to the sSFR and, since  the latter is found to be lower in massive star-forming galaxies with respect to less massive ones \citep[e.g.][] {schiminovich:2007lr}, it is expected that the CC~SN rate per unit mass decreases for an increasing mass of the parent galaxy. The case for the rate of SN~Ia is different.  The SN~Ia rate per unit mass of formed stars can be expressed as (see Eq.~(\ref{rateIa1})):
\begin{equation}\label{rateIa2} 
r_{\rm Ia}(t) = k_{\rm Ia} <f_{\rm Ia}>_{\psi(t)}  
\end{equation}
where $< f_{\rm Ia}> $ is the average of the DTD weighted by the SFH over the whole galaxy lifetime. We point out that in Eq.~(\ref{rateIa2})  the rate is normalised to the integral of the SFR over the total galaxy lifetime, which is greater than the current mass in stars because of the mass return from stellar populations. From Eq.~(\ref{rateIa2}), a dependence of  $r_{\rm Ia}$ on the mass of the parent galaxy may reflect a trend of either the SN~Ia productivity ($k_{\rm Ia}$), the shape of the DTD ($f_{\rm Ia}$), or the SFH. Since the latter is known to be more populated at old ages  in more massive galaxies (i.e. the \textit {downsizing} phenomenon, e.g. \citealt{thomas:2005kx}), it is expected that the SN~Ia rate per unit  stellar mass is smaller in the more massive galaxies because these are older, and at long delay times the DTD is less populated  than at short delays \citep{kistler:2013lr}. 
Fig.~\ref{rate_mass} and Table~\ref{ssfr1} show the results of our survey for the  star-forming galaxy sample, which are similar to those in the literature for both CC and Type Ia SNe. We  find a trend of the SN~Ia rate per unit  stellar mass with the mass of the parent galaxy in the star-forming sample along the same slope as in \citet{li:2011qf}. However, we note that  our measurement in galaxies with low mass ( ${\rm log(M)}  <9$) seems to suggest a steeper trend with respect to that  from the measurements  by \citet{Sullivan:2006nq} and by  \citet{smith:2012kx}.  The trend for CC~SN rate per unit  stellar mass  observed from SUDARE data is  in good agreement with that from \citet{graur:2015lr}  but not as steep as  in \citet{li:2011qf}.
All in all, however, the trend of the rate with the 
galaxy mass for star-forming galaxies does not seem to vary with redshift, similar to what found for the scaling with colour and sSFR of the parent galaxy.

\section{Comparison with models}\label{tmodel}
We compare the observed SN rates with model predictions to derive constraints on the SN progenitors.
The number of  SNe ($n_{\rm SN}$) of a given type which we expect to detect in a galaxy of our sample  can be expressed as:
\begin{equation}\label{eqrate}
N_{\rm SN} = r_{\rm SN} \, CT_{\rm gal} \, M_{\rm gal}
\end{equation}
\noindent where  $r_{\rm SN}$ is the SN rate per unit mass,  the control time ($CT_{\rm gal}$) for the specific galaxy is computed as described in Sec.~\ref{ratecalc} and the galaxy mass ($M_{\rm gal}$) is obtained from the SED fitting by {\sc FAST}.

\subsection{SNe Ia}
In a galaxy that includes stars with age up to $AGE$  the  SN~Ia rate ($r_{\rm Ia}$) per unit mass of formed stars  can be written as:
\begin{equation}\label{rateage}
r_{\rm Ia} = k_{\rm Ia} \frac {\int_0^{AGE} \psi(A) f_{\rm Ia}(\tau= A) dA} {\int_0^{AGE}\psi(A) dA}  
\end{equation}
\noindent where the variable $A$ is the age of the stellar generations present in the galaxy, and $\psi(A)$ is the SFH, expressed as the age distribution by mass in the galaxy. 
Thus, the SN~Ia rate per unit mass is given by the the productivity $k_{\rm Ia}$, which we assume constant in time and the same in all galaxies, multiplied by the average DTD weighted by the SFH, or age distribution. The difference of the rate in different galaxies is totally ascribed to this second factor, and specifically to the age distribution, or SFH.
The rate is the convolution between the age distribution and the DTD; therefore knowing the former, one may solve
directly for the latter, as for example in  \citet{maoz:2011lr} and in  \citet{childress:2014vn}. 
 However, we think that the age distribution recovered from the SED fitting is not accurate enough to proceed along this line, and we prefer to test some physically motivated DTD models versus the observations. 
\begin{figure}
\begin{center}
\includegraphics[width=8cm]{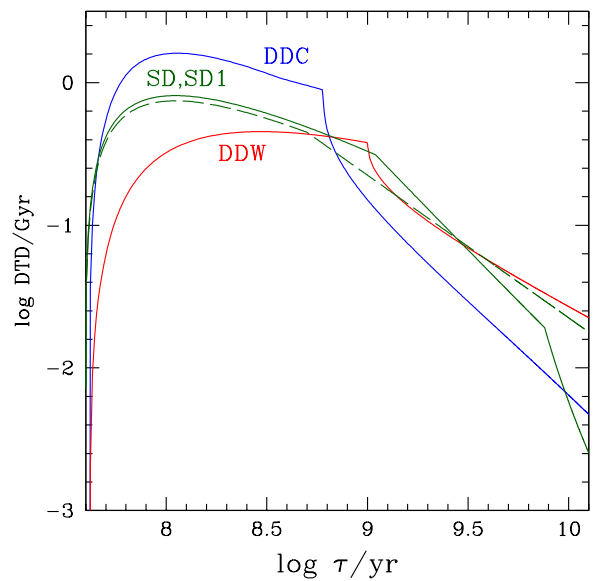} 
\caption{Distributions of the delay times adopted in our modeling, selected from the family in \citet{greggio:2005ph}. SD (green solid): a single degenerate model which assumes Chandrasekhar mass explosions; SD1(green dashed): the same as SD up to a delay time of $0.5$\,Gyr, proportional to the inverse of the delay time onward; DDC (blue): a DD CLOSE model with parameters $\beta_a=-0.9,\tau_{n,x}=0.6$; DDW (red): a DD WIDE model with parameters   $\beta_a =0., \tau_{n,x}=1$. See \citet{greggio:2005ph} for a description of the models. The DTD models are normalized to 1 over the range of delay times from 0 to $13$\,Gyr.}\label{dtd}
\end{center}
\end{figure}
As in Paper\,$\textrm{I}$, we select three DTD models from the
realizations in \citet{greggio:2005ph}, showed in Fig.~\ref{dtd}, specifically one SD model which assumes Chandrasekhar mass explosions, one very steep and one very flat DD model (hereafter DDC and DDW, respectively). 
The selected models correspond to a very different time evolution following a burst of star formation:  50\% of the explosions occur within the first $0.45$, $1.0$, and $1.6$\,Gyr, for DDC, SD, and DDW models, respectively, while the late epoch decline scales as $ t^{-1.3}$ and  $t^{-0.8}$ for the DDC and DDW models, respectively.
We add a fourth, DTD model (hereafter SD1) which is equal to the SD distribution for delay times shorter than $0.5$\,Gyr and falls off as $ t^{-1}$ at delay times longer than $0.5$\,Gyr. This DTD is similar to what was adopted in \citet{rodney:2014eu}. Compared to the SD model, the SD1 DTD maintains a relatively high  rate in old stellar populations, with a late epoch decline.  
 We perform the comparison of models with the observations in two ways: one relying on the SFH from {\sc FAST}  to predict the events in each galaxy, the other analyzing the  trend of the rate with the $U-J$ colour  of the parent galaxy, assumed as a tracer of the age distribution.

\subsubsection{Fit of the SNe Ia observed in the galaxies of the sample}\label{expIa}
We compute $r_{\rm Ia}/k_{\rm Ia}$ adopting  for $f_{\rm Ia}$ the  DTD models  in Fig.~\ref{dtd}, and for the SFH the best fit parameters ($AGE$ and $\tau$) from {\sc FAST}. For each galaxy of the sample we then derive  the
quantity $n_{\rm Ia}/k_{\rm Ia}$, and by summing them, we obtain the value of the productivity $k_{\rm Ia}$ required to reproduce the number of observed events. 




We test the  four DTD models above mentioned and two SFHs, namely an exponentially decreasing function ( $\psi_{\rm exp} \propto  \exp(\frac{-t}{\tau_{\rm SFR}})$ ),  and a delayed SFH ($\psi_{\rm del} \propto  t \times \exp(\frac{-t}{\tau_{\rm SFR}})$ ), and  for each combination we derive the SN~Ia productivity ($k_{\rm Ia}$). 
Table~\ref{expIa} also lists the expected number of SNe~Ia 
in  the passive and star-forming galaxy samples for the eight  combinations of DTD model and SFH, having adopted the productivity normalized to the total number of events.
In the SUDARE  SN sample we count 
$4^{+3.1}_{-1.9}$ and  $32^{+6.1}_{-6.2}$  SN Ia in passive  and star-forming galaxies, respectively. All errors are  $1\sigma$ and were based on the confidence limits for small numbers according to \citet{gehrels:1986lr}. Comparing these figures with the expected events in Tab.~\ref{expIa}, it seems that all combinations are similarly adequate, but the DDW model seems slightly disfavoured.
      The productivity results in the vicinity of $k_{\rm Ia}=0.001 \msun^{-1}$, a value which is greater than we obtained in Paper\,$\textrm{I}$  analyzing the cosmic SN Ia rates as function of redshift. 
In fact, assuming the cosmic SFH by \citet{madau:2014uf}, we found  $k_{\rm Ia}=7.5 \times 10^{-4} \msun^{-1}$ for the SD and the DDW model and $k_{\rm Ia}=8.5 \times 10^{-4}\, \msun^{-1}$ for the DDC model. This discrepancy is not large, and we postpone the assessment of its significance until this survey is
complete, when more robust statistics will be available.

\subsubsection{Rates as a function of galaxy colour}\label{RatevsUmJ}
The method described in the previous section strongly relies  on the SFH determined from the fit of the SED of each galaxy in the sample, and takes full advantage of the multi-band dataset. On the other hand, one can constrain the DTD through the analysis of the trend of the rate with the colour of the parent galaxy \citep[e.g.][]{greggio:2005ph}. 
 In this approach, galaxies are stacked in colour bins, so that each bin samples a large stellar mass range, and the value of the rate benefits from a high statistical significance. At the same time,  it not necessary that the fitted SFH is a precise  representation for each galaxy; rather it is assumed that the average colour describes the global feature of the SFH of the bin, and the uncertainties on the individual SFHs are averaged out. 
  We choose the $U-J$ colour as representative of the global SFH: the $U$ band is very sensitive to the young stellar populations, while the $J$ band is a good tracer of the total mass.

\begin{table}
\caption{Expected number of SNe~Ia in the  redshift range  $0.15 \le z< 0.75$ for the four DTD models
in Fig.~\ref{dtd} and for two different  analytical forms of the SFH: an exponentially declining function ($\psi_{\rm exp}$)  and a  delayed exponentially declining function  ($\psi_{\rm del}$).  The SN~Ia productivity  $k_ {\rm Ia}$ is in unit of $10^{-3} \, \msun^{-1}$. Observed SNe~Ia from SUDARE are $4^{+3.1}_{-1.9}$  in passive galaxies and  $32.6^{+6.1}_{-6.2}$ in star-forming galaxies}.\label{expIa}
\begin{tabular}{|c|c|cccc|}
\hline
SFH& gal. type& SD  & SD1& DDW &DDC\\[6pt]
\hline
& passive &9.7 & 9.7&  11.5& 4.9 \\[4pt]
$\psi_{\rm exp}$ & star-forming & 30.3&30.4& 28.6& 35.2\\[4pt]
& $k_ {\rm Ia}$ & 1.2&1.5&1.5&1.2 \\[4pt]
\hline
 & passive &10.16 & 10.14&  11.9& 5.3 \\[4pt]
$\psi_{\rm del}$ & star-forming & 29.9& 29.9&28.2& 34.8\\[4pt]
& $k_ {\rm Ia}$ & 1.2&1.4 &1.4&1.1 \\[4pt]
\hline
\end{tabular}
\end{table}

\begin{figure*}
\begin{center}
\includegraphics[width=18cm]{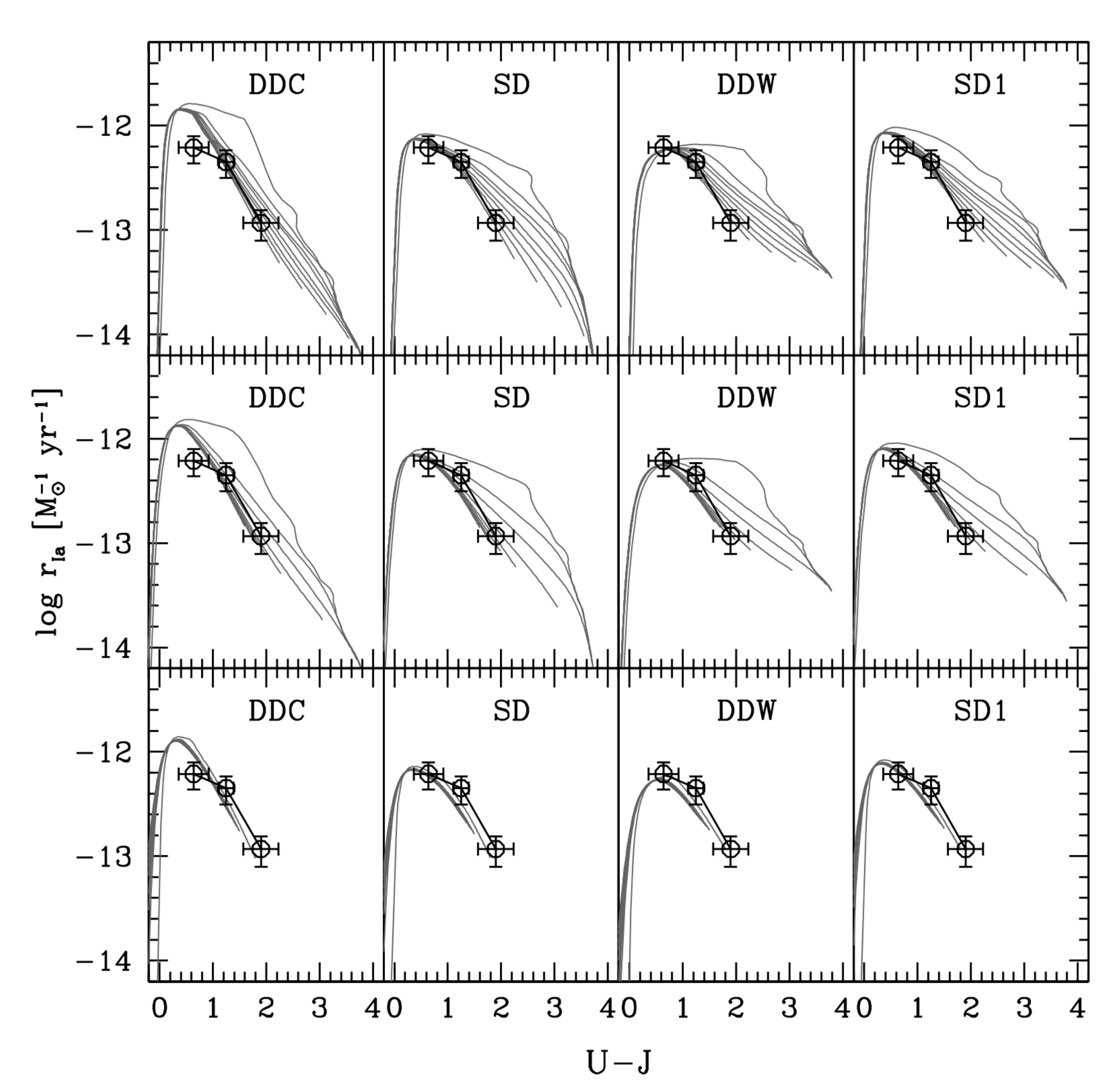}\\
\caption{Models of the SN~Ia rate per unit mass as a function of the $U-J$ colour of composite stellar populations obtained with different
kinds of SFH and the four DTDs shown in Fig.~\ref{dtd}. Along each line the parameter $AGE$ of the composite stellar population varies from $0.05$ (bluest point in $U-J$) to $13$\,Gyr (reddest point in $U-J$). In the top panels the assumed SFH is an exponentially declining law with e-folding times of $(8, 4.5, 3.0,2.0, 1.5, 1.0, 0.8, 0.5, 0.1)$\,Gyr (left to right). In the central panels the assumed SFH is a delayed exponential model peaked at $(20, 10, 7.5, 5, 3.5, 2.0, 1.0, 0.5, 0.1)$\,Gyr (left to right). In the bottom panels the adopted SFH is a power law $\psi(t) \propto t^{-p}$ with p ranging from $0$ to $2.8$: all these models describe the same locus, and do not account for populations redder than $U-J \sim 2$. The adopted values of the productivity are different for the different DTDs, and equal to
$k_ {\rm Ia}= (1.2, 1.2, 1.5, 1.5) 10^{-3}$ M$_\odot^{-1}$ for the DDC, SD, DDW and SD1 model respectively.  The circles show the observed rates, colours and error bars for star-forming galaxies reported in Table~\ref{ratecolter}.
}
\label{modelsfr}
\end{center}
\end{figure*}

\begin{figure*}
\begin{center}
\includegraphics[width=18cm]{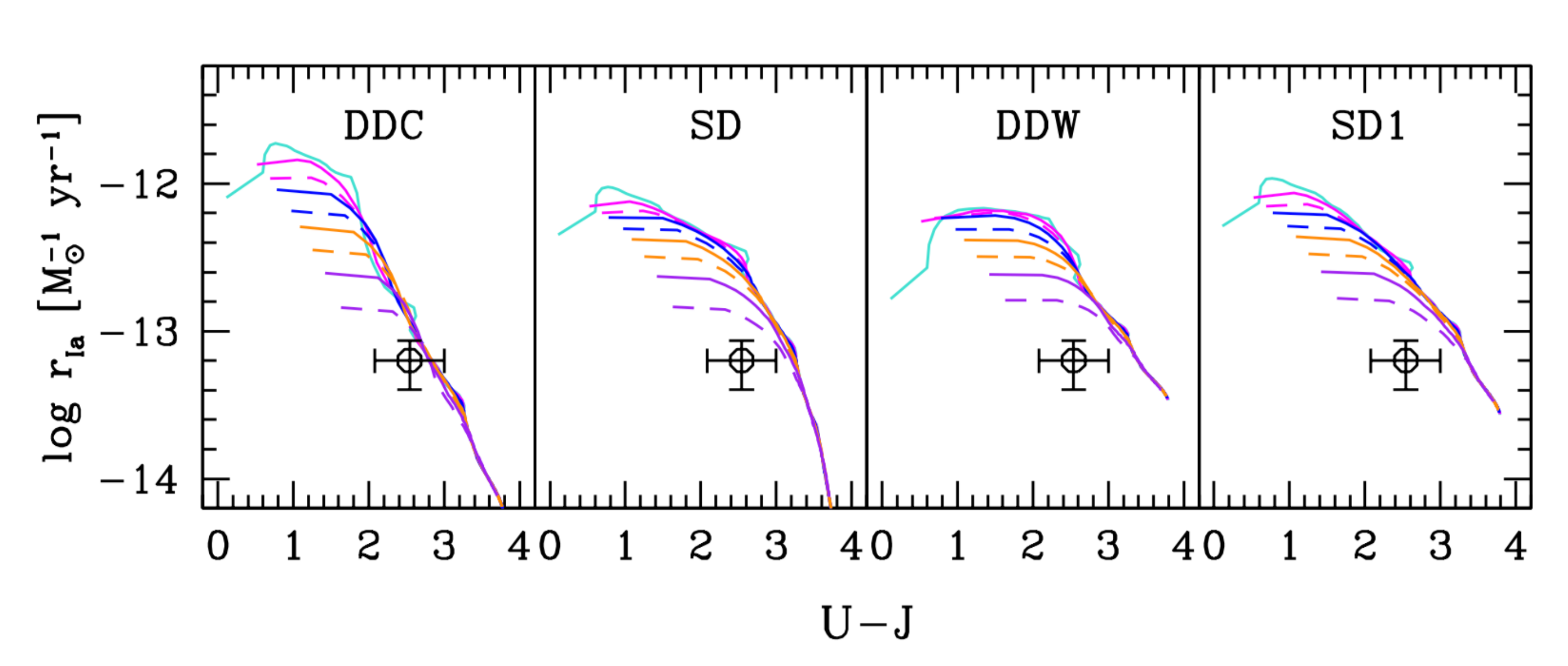}
\caption{
Models of the SN~Ia rate per unit mass as a function of the $U-J$ colour for passive galaxies, as described by the evolution past a bust of star formation. Each curve refers to different durations of the bursts, namely $(0.1, 0.5, 0.8, 1, 1.5, 2, 3, 4.5, 8)$\,Gyr (top to bottom). Different panels show the results for different DTD models as labelled, and adopt the values for $k_{\rm Ia}$ as in Fig.\ref{modelsfr}, namely $k_Ia = (1.2,1.2,1.5,1.5) 10^{-3}$M$_\odot^{-1}$. The circle shows the observational determination for the passive galaxies in our sample, and its error bar, as given in Table~\ref{ratecolter}.
}
\label{modelpass}
\end{center}
\end{figure*}

The lines in Fig.~\ref{modelsfr} show the  trend of the SN~Ia rate per unit mass with the colour of the parent galaxy as the parameter $AGE$ increases, for different descriptions of the SFH, and the four selected DTD models.
The productivity ($k_{\rm Ia}$) for each DTD model has been chosen so as to reproduce the total number of events in the sample (see Table\ref{expIa}), while the colours have been computed with  a SPS model based on   \citet{Bruzual2003}, with solar metallicity, as adopted for the SED fitting with {\sc FAST}. 
In this way, the colours from the population synthesis agree with the rest frame de-reddened colours of galaxies in our sample, output of the {\sc EAZY} and {\sc FAST}  codes.
Open circles in Fig.~\ref{modelsfr} show the observed rates  reported in Table~\ref{ratecolter} in star-forming galaxies, 
since  the models SFH assume a currently active SFR. 
As in previous works \citep[e.g.][]{mannucci:2005mb,greggio:2005ph} the SN~Ia rate per unit mass decreases as the colour becomes redder, because the DTD models are more populated at short delay times, which implies a higher efficiency of SN~Ia production in younger stellar populations.
The slope of the trend depends on both the shape of the DTD and on the SFH, so that the 
diagnostic power of this kind of correlation is relatively poor, and all four DTD models appear equally 
compatible with the data.
 
 \begin{table}
\caption{SN Ia rate per unit mass [$10^{-13}\,{\rm SNe}\,{\rm yr}^{-1}\,{\rm M}_\sun^{-1}$] as a function of $U-J$ colour for galaxies in the redshift range $ 0.15 \le z< 0.75$}\label{ratecolter}
\begin{tabular}{|c|cc|cc|}
\hline
gal. type &$N_{\rm gal}$& $<U-J> $& $N_{\rm SN}$ & rate  \\ [6pt]
 \hline
 & & & & \\
passive& 12221& $2.5  \pm 0.5$    &    $7.4 $    & $0.63^{+0.23}_{-0.23}$    \\ [6pt]
 \hline
  & & & & \\
&  18644 &$1.9 \pm0.3$  &    $ 9.5$ & $1.17^{+0.38}_{-0.38}$   \\   [4pt]
star forming &21705 & $1.3\pm 0.1$   &  $11.2$  & $4.48^{+1.3}_{-1.3}$   \\  [4pt] 
&27740&  $0.6 \pm0.3$  &  $11.9$ & $6.14^{+1.8}_{-1.8}$   \\   [4pt] 
\hline  
\end{tabular}
\end{table}

 Considering now the rate in passive galaxies, we derive one important indication. Fig.~\ref{modelpass} shows the predicted trend of the SN~Ia rate per unit mass as a function of $U-J$ colour for a family of models meant to describe passive galaxies, which assume an initial burst of star formation, with a constant rate over a time interval $\Delta t$, and which then drops to zero. The lines in the figure show the evolution 
of the rate and the colour of the produced stellar population from the end of the star formation episode onward, and different lines refer to different durations of this episode. The values adopted for $k_{\rm Ia}$ are the same as in Fig.~\ref{modelsfr}.
It is interesting to note that  in passive galaxies after an initial stage in which the SN~Ia rate stays constant as the color of the stellar population gets redder, all models converge to the same locus. This behaviour offers a diagnostic tool which does not seem to depend on the details of the SFH.  
The circle in Fig.~\ref{modelpass} shows the rate and average color of the passive galaxies. 
Only the DDC model meets at the same time the level of the rate and the colour of the passive galaxies. Agreement between models and data could be found also for the other DTDs, but with a productivity of SN~Ia higher than that appropriate for star-forming galaxies. 
Notice that our formalism does not allow for different values in $k_{\rm Ia}$ in different types of galaxies, while the trend of the rate with the properties of the parent galaxy should trace the different SFH, modulo the DTD. 
We also considered mixed models, with a contribution from both SD and DD channels, in the same fashion as \citet{greggio:2010pd}. Again we found that the combinations which include the DDC model provide a better fit to the star-forming and passive galaxies, while those with the DDW DTD tend to be too shallow to describe at the same time the rates in the star forming and in the passive galaxies.

\subsubsection{Correlation of the rate with the galaxy mass} 
\begin{figure}
\begin{center}
\includegraphics[width=9cm]{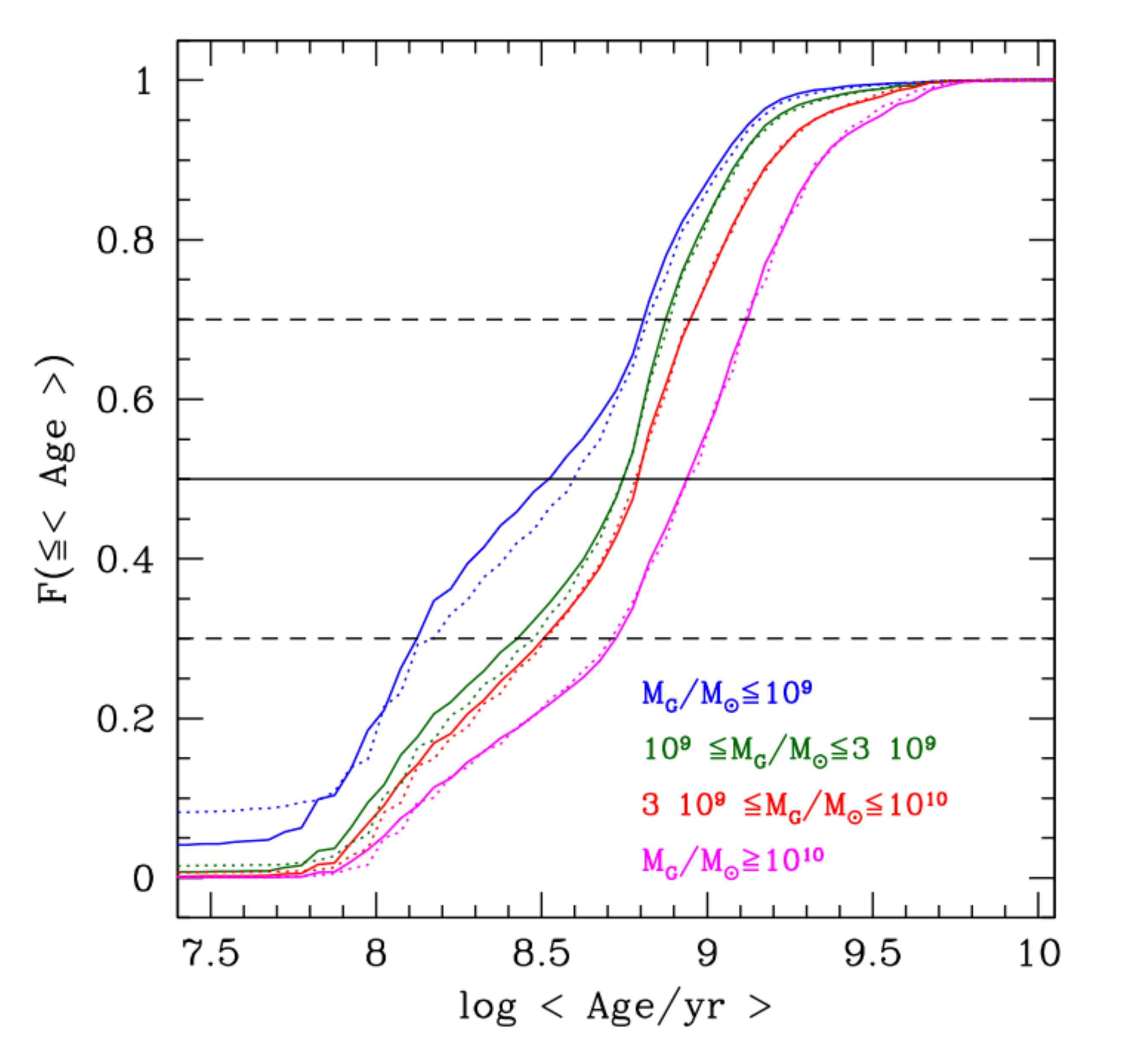} 
\caption{Cumulative distributions of the average mass-weighted age of star-forming galaxies, in four bins of galaxy mass as labelled. The solid lines show the average ages relative to fits with the exponentially declining SFH; the dotted lines refer instead to fits which adopt the delayed SFH. The horizontal lines intercept the curves at the median (solid) average age and at the 30-th and 70-th percentiles (dashed).} \label{aagedist}
\end{center}
\end{figure}

As pointed out previously, the SN~Ia rate per unit mass in star-forming galaxies appears to decrease 
for increasing mass of the parent galaxy (see Fig.~\ref{rate_mass}), an effect which could be ascribed to the  \textit{downsizing} phenomenon (see Section~\ref{ratemass}). But is downsizing present in our galaxy sample? To answer this question we binned our star-forming  galaxies in four mass ranges, and within each range we computed the  distribution of the galaxy mass-weighted average ages 
($<Age> $),  as given by:  
\begin{equation}\label{aveage}
<Age> = \frac {\int_0^{AGE} A \psi(A)  dA} {\int_0^{AGE}\psi(A) dA} ,
\end{equation}
\noindent and calculated by adopting  the best fit parameters from {\sc FAST} to define $\psi(A)$.  Fig.~\ref{aagedist} shows the cumulative distribution of the average ages of the star forming galaxies for the combined COSMOS and CDFS sample, for the two options considered here to describe the SFH law. 

The \textit{downsizing} effect is indeed present in our sample, with the lower mass bin (blue line) hosting a larger fraction of galaxies with  young average age. The solid horizontal line intercepts the distributions at the median average age of the four mass ranges, which gets older and older going from the least to the most massive bin, similar to what was found in \citet{gallazzi:2005kq}. The average age distributions appear almost insensitive to the law adopted to describe the SFH. The dashed lines are drawn at the 30-th and 70-th percentiles in order to illustrate the width of the distributions of $<Age>$ in the four mass bins. 

\begin{figure}
\begin{center}
\includegraphics[width=9cm]{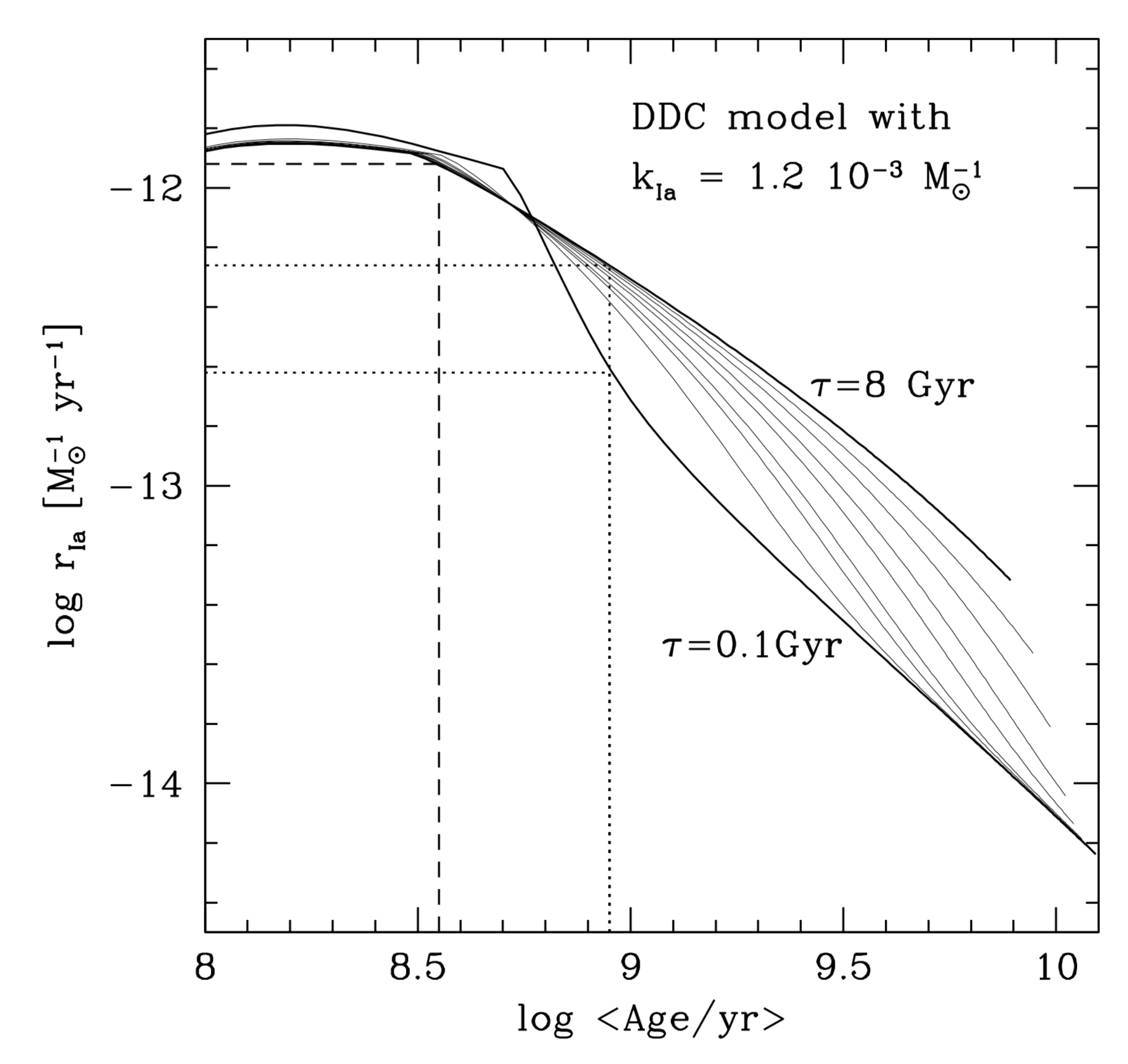} 
\caption{Relation between the SN~Ia rate model  and the average age of the stars in the parent galaxy,  for exponentially declining SFHs and a DDC model description of the DTD. The different solid lines are obtained for different values of the e-folding SF timescale, ranging from $0.1$\,Gyr (lowermost curve) to $8$\,Gyr (uppermost curve). Along each line the parameter AGE increases from $0.05$ to $12.5$\,Gyr. The dashed and dotted lines are drawn at the median average ages respectively of the least and the most massive bin in Fig.~\ref{aagedist}.}\label{agerc1}
\end{center}
\end{figure}

We model the effect of  \textit{downsizing} in our sample on the measured SN~Ia rates by considering the correlation between the parameter $<Age>$ and the rates in the family of models described in the previous section.
Fig.~\ref{agerc1} shows such correlation for the exponentially declining SFH models in the case of the DDC description of the delay times distribution. Similar plots can be drawn for other DTD models. In general, the SN~Ia rate per unit mass decreases as the average age of the parent galaxy increases, except for the systems with $<Age>$ younger than $\simeq 0.5$\,Gyr , for which it appears virtually constant. This characteristics reflects  the wide peak of the  DTD models at the short delay times. At average ages older than $0.5$\,Gyr, the rate declines with a slope sensitive to the e-folding timescale of the SFR, so that, at given $<Age>$ the model rate spans a wide range, depending on the parameter $\tau$. We find that the correlation between the SN~Ia rate and the average age of the parent galaxy is very similar for the two SFHs considered here, while it turns out sensitive to the DTD, with the DDW delay time distribution  providing a quite shallow correlation.   

\begin{figure}
 \begin{center}
\includegraphics[width=9.5cm]{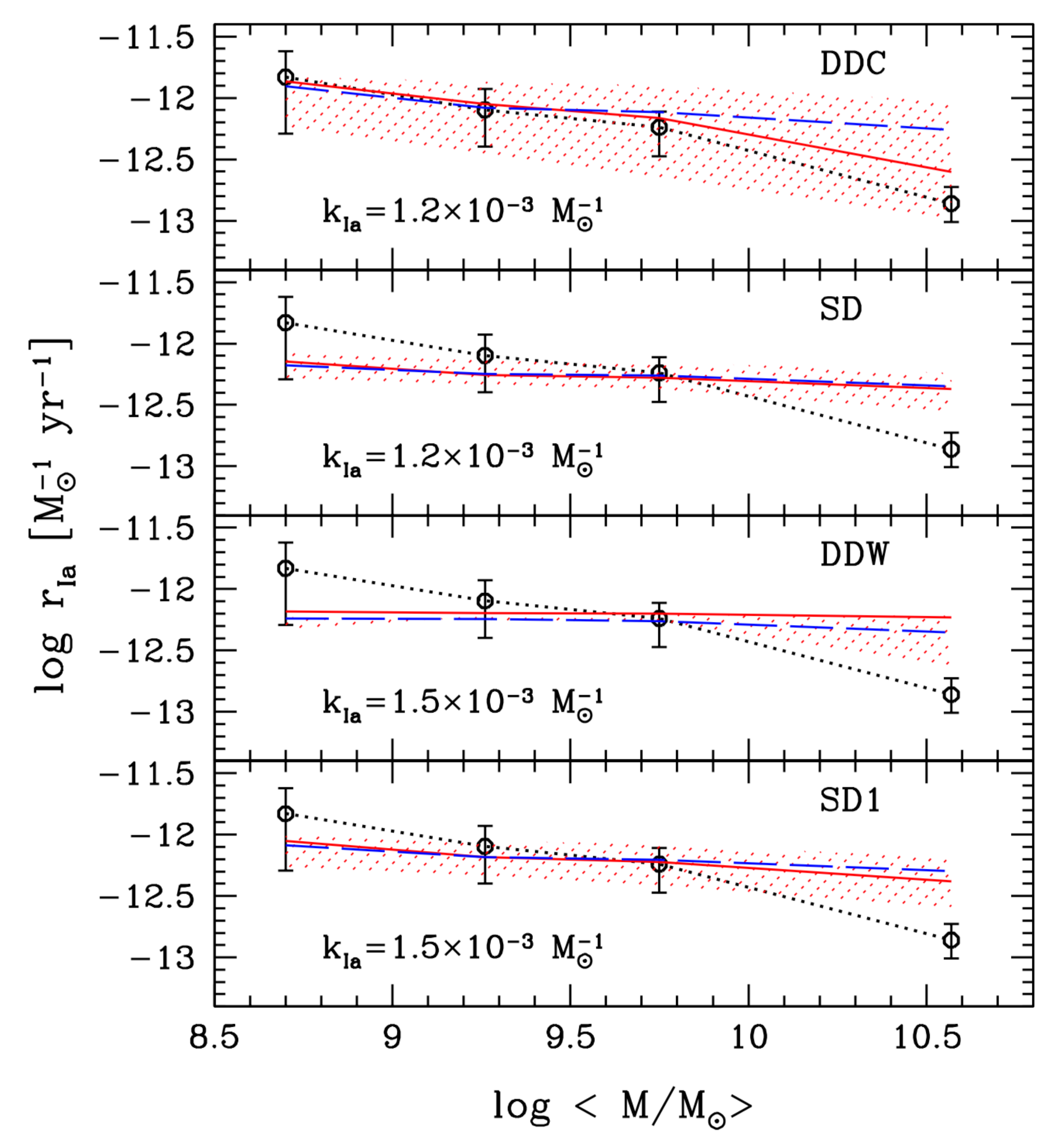} 
\caption{SN~Ia rate as a function of galaxy mass. Black circles are the observed rates in the four mass bins, reported in Table~\ref {ssfr1}, while lines connect the theoretical rates at the median $<Age>$ of the distributions in the four mass bins. Solid (red) and dashed (blue) lines refer to exponentially declining SFH models with $\tau = 0.1$ and $8$\,Gyr, respectively. The shaded areas show the range covered  when using the 30-th or 70-th percentiles of the $<Age> $ distribution instead of the median $<Age>$, for the  exponentially declining SFH models with $\tau = 0.1$\,Gyr. }\label{hodor}
\end{center}
\end{figure}

We now turn to consider the expected correlation between the SN Ia rate and the parent galaxy mass by combining the relation between $<Age>$ and galaxy mass  bin, and that between $<Age>$ and SN~Ia rate.  Rather then computing the expected rate for each  galaxy in the sample, and examining the distribution of the results in the four mass bins,  as in \citet{graur:2015lr}, we
compute the expected rate at some characteristic $<Age>$ of the four mass bins, in particular at the intercepts of the horizontal lines with the four distributions in Fig.~\ref{aagedist}.  Given the relatively poor statistics of SN~Ia events collected to far, we aim at characterizing the expected correlation and exploring its variation with the DTD and the SFH.  We will follow a more sophisticated approach once the
survey has been completed.
Fig.~\ref{hodor} shows  how the SN~Ia rate is expected to vary in the four mass bins whose $<Age>$ distributions are shown in Fig.~\ref{aagedist}.
Solid (red), and dashed (blue) lines connect the model rates at the median $<Age>$ of the
galaxies in the four mass bins, when  adopting exponentially declining SFH models with e-folding times of $0.1$ and $8$\,Gyr, respectively. At a given <Age>, models with longer SFH timescales yield a higher rate, as depicted in 
Fig.~\ref{agerc1}. The four panels of Fig.~\ref{agerc1} show the results for the four selected DTD models. Notice that for each DTD model the value of $k_{Ia}$ comes from the fit described in Sect.~\ref{expIa}. 

The shaded areas illustrate the range of variation of the SN~Ia rate model for the $\tau=0.1$\,Gyr SFH models, corresponding to the 30-th and 70-th percentiles of the distribution of $<Age>$ in the four bins of galaxy mass (see 
Fig.~\ref{aagedist}).  Finally the open circles are the observed rates.

The observed correlation is best reproduced with the combination of the DDC distribution of the delay times coupled to an exponentially decreasing SFH with short e-folding timescale. The models predict a relatively shallow correlation when the other 3 DTDs are considered, as well as when SFHs with long e-folding timescales are adopted.  There is however ample room to accomodate either less steep DTDs and/or flatter SFHs if one considers the width of the distributions of the parameter $<Age>$ within the four mass bins, as the shaded areas indicate. Similar considerations hold when the delayed SFH models are considered. As a cautionary note, we point out that these conclusions rely on the output SFHs from the {\sc FAST} fit, and in particular on the distribution of the parameter $<Age>$, or quantitive \textit{downsizing} relation. In this respect we notice that the average ages of galaxies in \citet{gallazzi:2005kq} are systematically {\bf older than} those found here by  a factor which ranges from $2$ to $5$, at the same average galaxy mass. This may be due to the different age tracers used in the two studies,  and /or to the different galaxy samples. A thorough study of the SFH in the individual galaxies  is required for a robust interpretation of the correlation between SN~Ia rate and the mass of the parent galaxies. 

\subsection{CC SNe}\label{expCC}
\begin{figure}
\begin{center}
\includegraphics[width=9cm]{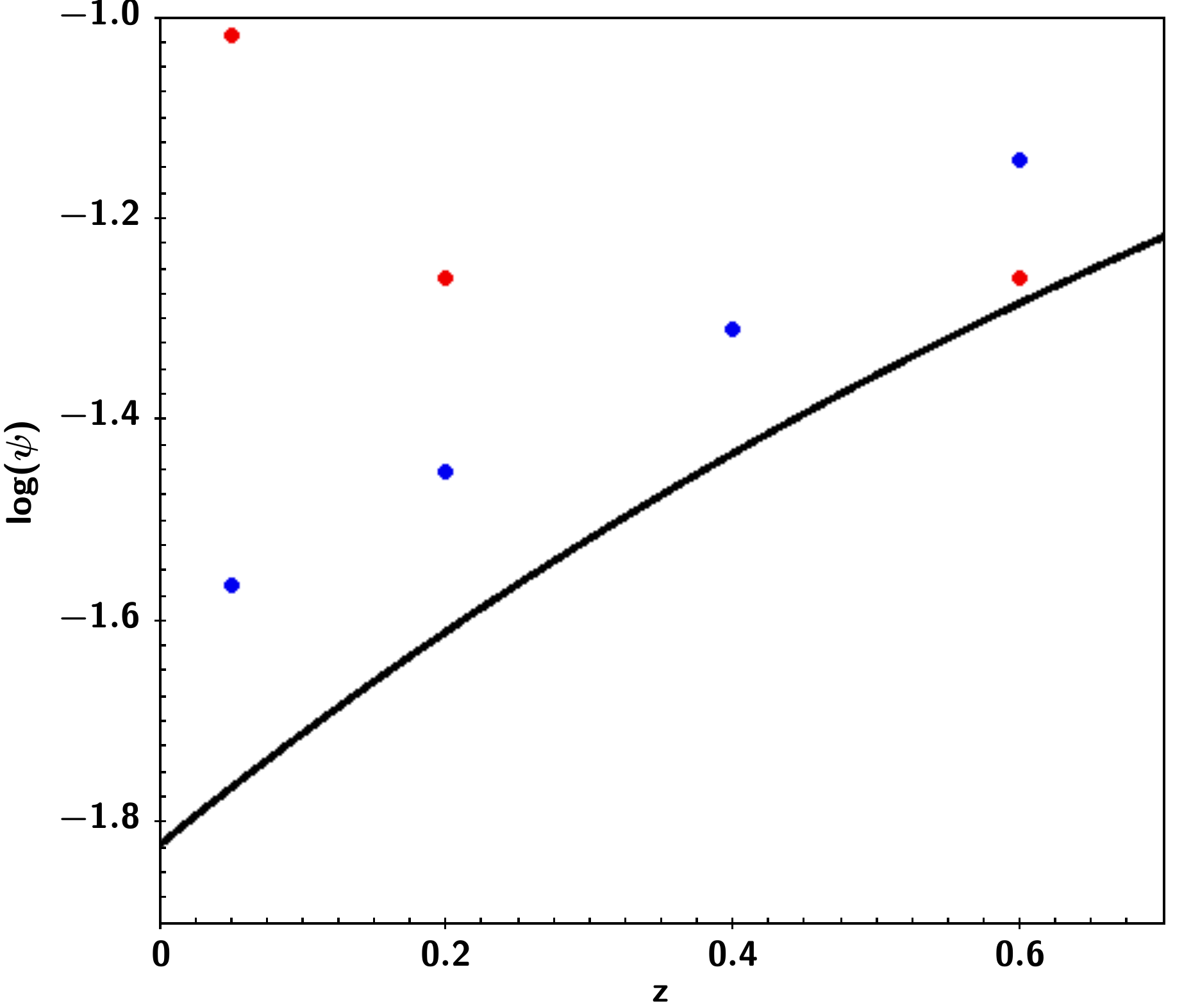} 
\caption{SFR density as a function of  redshift. The black line is  the analytical function from  \citet{madau:2014uf}. The red circles are measurements from SFRs in the COSMOS galaxy sample estimated via {\sc FAST} SED fitting, while the blue points are measurements via  UV and NIR luminosity by  \citet{muzzin:2013fj}. }\label{sfrcomp}
\end{center}
\end{figure}

For the CC SNe, we use Eq.~(\ref{eqrate}) with the rate ($r_{\rm CC}$) given by Eq.~(\ref{rcc}). We adopt the 
the SFR ($\psi$) from the {\sc FAST}  best fit  and $ k_{\rm CC}= 0.0067\, {\rm M}_{\odot}^{-1}$, which corresponds to a progenitor mass range of $8-40\, {\rm M}_{\odot}$ for a Salpeter IMF. 
To estimate the $CT_{\rm gal}$  for CC SNe   we combined the CTs for  each CC SN subtype assuming the fraction of different subtypes reported in Paper\,$\textrm{I}$ (56\% II 40\% Ibc and 4\% IIn).  
We computed the expected number of CC~SNe  for each star-forming galaxy of the sample and the total number of expected CC~SNe ($n_{\rm CC}$)  by summing the contribution of each galaxy in the redshift range $0.15 < z <0.35$.
The expected number of CC~SNe is $21$ in our galaxy sample  while the observed number is $13 ^{+4.1}_ {-4.2}$(about 60\%).
At face value, this discrepancy suggests  a higher limit for the minimum progenitor mass, in the vicinity of  $\sim 10\, {\rm M}_\odot$. 
In  Paper\,$\textrm{I}$ comparing the volumetric CC~SN rates as function of redshift, we found agreement between the observed rates and the expectations from the cosmic SFH by \citet{madau:2014uf}  in combination with $k_{\rm CC}= 0.0067\,  {\rm M}_\odot^{-1}$ (see Sect. 8.1 in Paper\,$\textrm{I}$). The expected number of CC~SNe in the  sampled cosmic volume   is  33, close to the number of discovered CC~SNe, 26 events (about 80\%). 
We note that the selection criteria on the galaxies properties (see Sect.\ref{sne}) have reduced  the CC SN sample by a factor of two with respect that exploited in Paper\,$\textrm{I}$.
Therefore, we attribute part of the discrepancy between the expected and observed CC~SNe to a fluctuation due to the relatively poor statistics, which is to be verified once the full data-set will be analyzed. 
   It is also interesting to compare the total volumetric  SFRs in different redshift bins for the COSMOS galaxy sample to the volumetric  SFRs by  \citet{madau:2014uf}. 
Fig. \ref{sfrcomp} shows the SFR  density in different redshift bins  estimated  by summing the SFRs from the {\sc FAST} code (and the SFRs from the galaxy UV and NIR luminosity by  \citet{muzzin:2013fj} for the  galaxy sample in the COSMOS field.
In both cases we  found higher values than those estimated  assuming the SFH by   \citet{madau:2014uf}. 
From this comparison we conclude that the cosmic SFH is not known with sufficient accuracy to allow us to pinpoint the value of $k_{\rm CC}$ with an accuracy better than $\sim 50\%$.

\section{Conclusions}\label{conclusions}

The aim of the SUDARE survey is to constrain the SN progenitors  analysing the dependence of the SN rate on the properties of the parent stellar population.
The uncertainties in both SN rate and SFH measurements prevent an analysis based on individual galaxies therefore we averaged over a population of galaxies with different ages  and SFR following two different approaches. 

In  Paper\,$\textrm{I}$  we measured the SN rates in a large volume as a function of the cosmic time, while in this work we measured the SN rates in a sample of galaxies as a function of the galaxy intrinsic colours or sSFR ---both tracers of galaxy age.

The galaxy sample (about $130000$ galaxies) has been obtained selecting  in the CDFS and COSMOS fields all galaxies with $i)$ $K_{\rm s}$ band magnitude  $\le 23.5$\, mag, $ii)$ redshift  within the range  $0.15 \le z \le 0.75$  and   $iii)$ a reliable redshift estimate ($Q_z < 1$).
We performed the SED fitting for each galaxy with the {\sc EAZY} code to estimate the photometric redshift and rest frame colours. 
The galaxy mass, SFR and sSFR have been estimated by  the {\sc FAST} code, with redshifts fixed to the values derived by  {\sc EAZY}, adopting the \cite{Bruzual2003} SPS model library, a Salpeter IMF, solar metallicity, an exponentially declining SFH  and the  \cite{calzetti:2000ht} dust attenuation law.
 
 The separation of  star-forming and quiescent galaxies  has been obtained  exploiting both the $U-V$ vs $V-J$ colour-colour diagram, following the prescription by \citet{williams:2009fk}, and the sSFR best fit values from {\sc FAST}, adopting the same criteria as in   \citet{Sullivan:2006nq} (the passive galaxies with a zero mean SFR, the star-forming galaxies with $-12.0 < \rm log(sSFR)<-9.5$, the star burst galaxies with $\rm log(sSFR)> -9.5$).
 
In this galaxy sample we discovered   $13$ CC~SNe in the redshift range $0.15<z<0.35$  and  $36$ SNe Ia in the redshift range $0.15<z<0.75$. 
The criteria and algorithm for SN search and photometric classification have been illustrated  in Paper\,$\textrm{I}$.
The SN rates have been measured with the method of the control time  and normalized per unit stellar mass.
 
 Our principal findings are as follows.
 
 \begin{enumerate}
 
  \item The  trend of the SN rates  as a function of the $B-K$ colour of the parent galaxy from SUDARE data is consistent with that observed in the Local Universe. Both CC and Type Ia SN rates become progressively higher for bluer galaxies but CC~SNe with a steeper slope.
 
  \item  The SN~Ia rate per unit mass is about a factor of five higher in the star-forming with respect to the passive galaxies identified as such on the $U-V$ vs $V-J$ colour-colour diagram. 
Only  a lower limit for CC~SN rate has been estimated in  passive galaxies. This is expected given the short lifetimes of CC~SN progenitors which imply that the CC~SN rate is proportional to the recent SFR.
 
   \item The trend of the  SN~Ia rate as a function of the sSFR  that we observed is very similar to previous results \citep{mannucci:2005mb,Sullivan:2006nq, smith:2012kx,graur:2015lr}.  Our results confirm that the higher the sSFR the higher the SN~Ia rate per unit mass suggesting a DTD declining with delay time.
 The fact that this trend is similar at the different redshifts suggests that  the ability of the stellar populations to produce SN~Ia events does not vary  with cosmic time. 
The trend of the CC~SN rate as a function of the sSFR is very similar to that observed  in the Local Universe \citep{mannucci:2005mb}  and  in the redshift range $ 0.04 < z < 0.2$   \citep{graur:2015lr}.  This dependence  is expected to be  linear since the CC~SN rate  per unit mass is proportional to the sSFR.

 \item  Our measurements of  both CC and Type Ia SN rate  per unit mass in the star-forming galaxy sample decrease  for an increasing mass of the parent galaxy. The trend observed from SUDARE data is similar  to that in \citet{li:2011qf} for both CC and Ia SNe.   This trend is expected for CC~SN rate per unit mass which is proportional to the sSFR, since  the latter  is found  to be lower  in massive  star-forming galaxies  with respect to less massive ones. The dependence of Type Ia SN rate on galaxy mass may reflect a trend of either the SN~Ia productivity, the shape of the DTD  or the SFH. Since the latter is known to be  skewed towards  old ages  in more massive galaxies, it is expected that the SN~Ia rate per unit mass is smaller in the more massive galaxies because these are older, and at long delay times the DTD is less populated  than at short delays ($< 1$\,Gyr).

\item The  number of expected type Ia events for each galaxy has been determined by convolving  the galaxy SFH  from  the {\sc FAST} SED fitting with DTD models. The expected number of SNe~Ia both in  the passive and star-forming galaxy sample is  in agreement with the observed one  for all  DTD models considered although the DDC model seems to be slightly favoured. 

\item  We derive the SN~Ia productivity  by comparing the total number of SN~Ia observed in our sample to the expected number of events obtained convolving the SFH in each galaxy with DTD models, and taking into account the appropriate control times. This normalization made on the total sample which includes star-forming and passive galaxies, yields similar values for $k_{\rm Ia}$ for the four DTD models considered, $k_{\rm Ia}=1.2 \times 10^{-3} \msun^{-1}$ for SD, $k_{\rm Ia}=1.5 \times 10^{-3} \msun^{-1}$ for DDW and $k_{\rm Ia}=1.2 \times 10^{-3} \msun^{-1}$ for DDC model.
In  Paper\,$\textrm{I}$ we derive a complementary constraint on $k_{\rm Ia}$ by fitting the cosmic evolution of the SN Ia rate adopting the same DTD models as here, and he analytical formulation  by \cite{madau:2014uf} to describe the cosmic SFH. These $k_{\rm Ia}$ values are typically smaller that those derived here ($k_{\rm Ia}=7.5 \times 10^{-4} {\rm M}_\sun^{-1}$ per SD and DDW model and $k_{\rm Ia}=8.5 \times 10^{-4} \msun^{-1}$ for DDC model).
Given the different systematics affecting the two determinations we cannot
assess with confidence the significance of this discrepancy.

\item  The comparison between the observed and expected trend of the SN~Ia rate with the intrinsic $U-J$ colour  of the parent galaxy in star forming sample is acceptable for 
 all four DTD models. In the passive galaxies  only the DDC model meets at the same time the level of the rate and the colour of the passive galaxies.

\item We  modeled the effect of  {\it downsizing}  in our galaxy sample on the SN~Ia rate and find that the observed correlation between the SN~Ia rate and the mass of the parent galaxy is best reproduced with a DDC model.

\item  The number of expected CC~SN events for each  star-forming  galaxy has been estimated  by assuming  the SFR from the {\sc FAST}  SED fitting and a mass range of $8-40\, {\rm M}_{\odot}$ for CC~SN progenitors which corresponds to $ k_{\rm CC}= 0.0067\, {\rm M}_{\odot}^{-1}$ for a Salpeter IMF.  The expected number of CC~SNe is higher than the observed one by a factor of about 40\%. In Paper\,$\textrm{I}$  the expected number of CC SN events has been determined by assuming the SFH by \cite{madau:2014uf} and the same value for the CC~SN productivity and it is higher than the observed one by a factor of 20\%.
  We attribute part of this discrepancy to a fluctuation due to the relatively poor statistics of CC~SN events part to the difference between cosmic SFR and SFR best fit values from the {\sc FAST} code.

 \end{enumerate}

We conclude from the observed correlations between the SN rates and properties of the parent galaxies that,  according to the expectations of stellar evolution, the minimum mass for CC SN progenitors is between $8-10 \, \msun$ and  the DTD of SN Ia progenitors  is skewed at the short delay times  ($<1$\,Gyr).
However, the uncertainties on both  CC and Type Ia SN rates are still too high to constrain the mass range of SN CC progenitors with sufficient accuracy and to discriminate between SD and DD progenitor scenarios.

\begin{acknowledgements}
We thank the referee for useful comments and suggestions.
This work is partially supported by the PRIN-INAF 2014 with the project
"Transient Universe: unveiling new types of stellar explosions with PESSTO" (PI A. Pastorello).
G.P. acknowledges support by the Proyecto Regular FONDECYT 1140352 and by  Millennium Institute of Astrophysics (MAS) through grant IC120009 of the Programa Iniciativa Cientifica Milenio del Ministerio de Economia, Fomento y Turismo de Chile.
MV acknowledges support from the Square Kilometre Array South Africa project,
the South African National Research Foundation and Department of Science and
Technology (DST/CON 0134/2014), the European Commission Research Executive 
Agency (FP7-SPACE-2013-1 GA 607254) and the Italian Ministry for Foreign Affairs
and International Cooperation (PGR GA ZA14GR02).

\end{acknowledgements}


\bibliographystyle{aa}
\bibliography{sudareII}


\begin{appendix}

\section{CC SN subtypes}\label{CCsubtypes}

\begin{figure}
\begin{center}
\includegraphics[width=10cm]{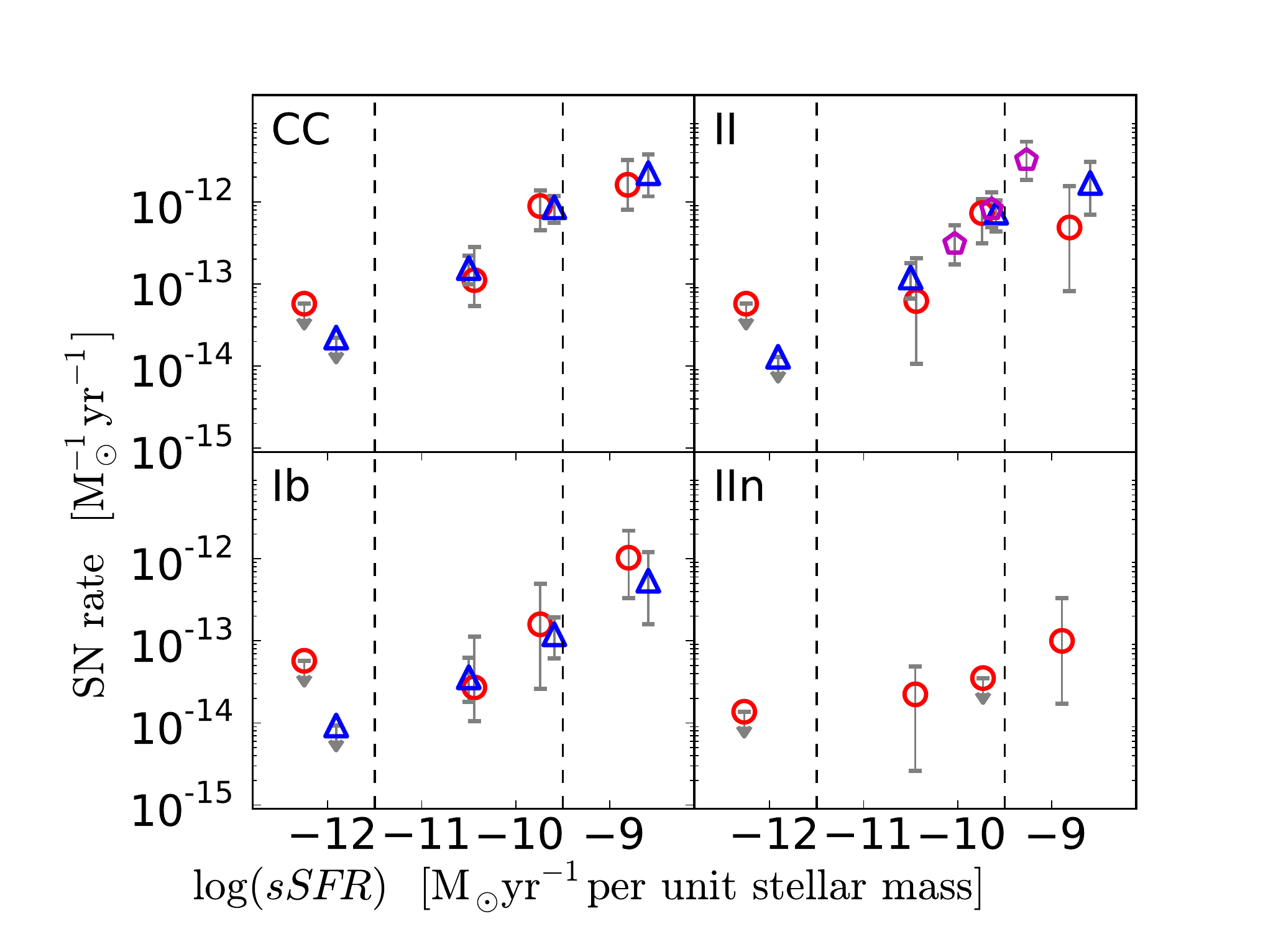}
\end{center}
\caption{SN rates per unit stellar mass vs sSFR for different CC~SN types in three different groups of galaxies based on  their sSFR: the first group of passive galaxies with a zero mean SFR; the second group of galaxies with
 $-12.0 < \rm log(sSFR)<-9.5$; the third group of galaxies with $ \rm log(sSFR)> -9.5$.  The circles are from SUDARE, the triangles from \citet{mannucci:2005mb} and the  pentagons from \citet{graur:2015lr}. 
 }
\label{mass_ssfr_sntypes}
\end{figure}

We estimated the rates for different CC~SN sub-types (II, Ibc and IIn)  as a function of both the sSFR and the stellar mass (Table~\ref{ssfr3} and Fig.~\ref{mass_ssfr}).  
Although our SN sample is still small, it is adequate to show a good match with the SN rates in the Local Universe  \citep{mannucci:2005mb} and at intermediate redshift  \citep{graur:2015lr}. Particularly our measurement of the Type II SN rate in starburst galaxies  is  not significantly lower than the measurement from \citet{graur:2015lr} in the redshift range $0.1<z<0.4$. The trend of the rate as a function of stellar mass for both Type II and Type Ib/c SNe  is similar to that obtained from \citet{li:2011qf}. The scanty statistic prevents us to confirm the suggestion  by  \citet{li:2011qf} that  the Type Ib/c  SN rates are deficient relative to the Type II rates in less-massive galaxies since we found  very similar values for them.

\begin{table*}
\begin{center}
\caption{SN rate per unit mass [$10^{-3}\,{\rm SNe}\,{\rm yr}^{-1}10^{-10}\,{\rm M}_\sun^{-1}$] for different bin of sSFR}\label{ssfr3}
\begin{tabular}{|c|ccc|ccc|}
\hline
type &  $\rm log <sSFR> $ & $N_{SN}$ & rate     &  $\rm log <M> $ & $N_{SN}$& rate     \\[6pt]
\hline
 &  &  &  &  & &     \\
& $-12.2$ &  $5.5$ & $ < 0.7$ & $8.7$& $0.95$ & $14_{-11}^{+13}$      \\[4pt]
II & $-10.5$ &  $1.0$ & $0.8_{-0.6}^{+1.8} $ & $9.3$&$1.0$  & $7.3_{-6 }^{+17}$  \\[4pt]
& $-9.7$ &  $3.8$ & $7.1_{-4.6}^{+3.8}$  & $9.8$ & $2.5$ & $8.5_{-6}^{+7}$   \\[4pt]
& $-8.8$ &  $2.0$ & $11.9_{-7.8}^{+15.3}$ & $10.6$ & $2.4$ & $1.2_{-0.9} ^{+1}$ \\[4pt]
\hline
 &  &  &  &  & &     \\
  &$-12.2$ &  $0.0$ & $< 0.7$ & $8.7$& $0.88$ & $13_{-10}^{+14}$   \\[4pt]
Ibc &$-10.5$ &  $0.9$  & $0.7_{-0.6}^{+0.7}$  & $9.3$& $0.26$ & $1.9_{-0.7}^{+11}$  \\[4pt]
&$-9.7$  & $0.6$ & $1.1_{-0.8}^{+2.3}$  & $9.8$ &$1.38$  & $4.7_{4}^{+6}$ \\  [4pt]
&$-8.8$ &  $2.2$  &$12.8_{-8.7}^{+14.5}$    &$10.6$ & $1.16$ & $0.6^{+0.5}_{0.2}$   \\[4pt]
\hline
 &  &  &  &  & &       \\
&$-12.2$ &  0.0 &  $< 0.2  $ & $8.7$ & $0.0$  & $< 4  $    \\[4pt]
 IIn & $-10.5 $&  $1.5$ & $0.3$ &  $9.3$ & $1.0$ &$ 1.7_{ 1.4}^{+4} $  \\[4pt]
& $-9.7$ &  $0.0$ & $<0.4$ & $9.8$ & $0.0$ &$< 0.8  $    \\[4pt]
&$-8.8$  &  $1.0$ & $1.2_{-1}^{+0.27} $  & $10.6$& $1.5$ & $0.2_{-0.16}  ^{+0.2}$  \\[4pt]
 \hline
\end{tabular}
\end{center}
\end{table*}

\section{Systematic errors}\label{systerrors}
Rest frame colours, mass and sSFR, as determined from broad band photometry, suffer from various uncertainties related to incomplete coverage of photometric bands, errors in photometric redshifts, incorrect assumptions in SED fitting  and the systematic interplay between these parameters.

Several studies have recently presented a comprehensive analysis of the accuracy of the  galaxy properties estimated through SED fitting  as a function of the various parameters (SFH, metallicity, age grid, IMF, reddening law, wavelength coverage and filter setup), both  in star-forming and passive galaxies and in a wide redshift range \citep{pforr:2012yu,pforr:2013kx,mobasher:2015qv}.
In the following we discuss the impact of systematics, affecting galaxies samples, on our results.  
\\

\begin{figure*}
\begin{center}
$
\begin{array}{c@{\hspace{.1in}}c@{\hspace{.1in}}c}
\includegraphics[width=6cm]{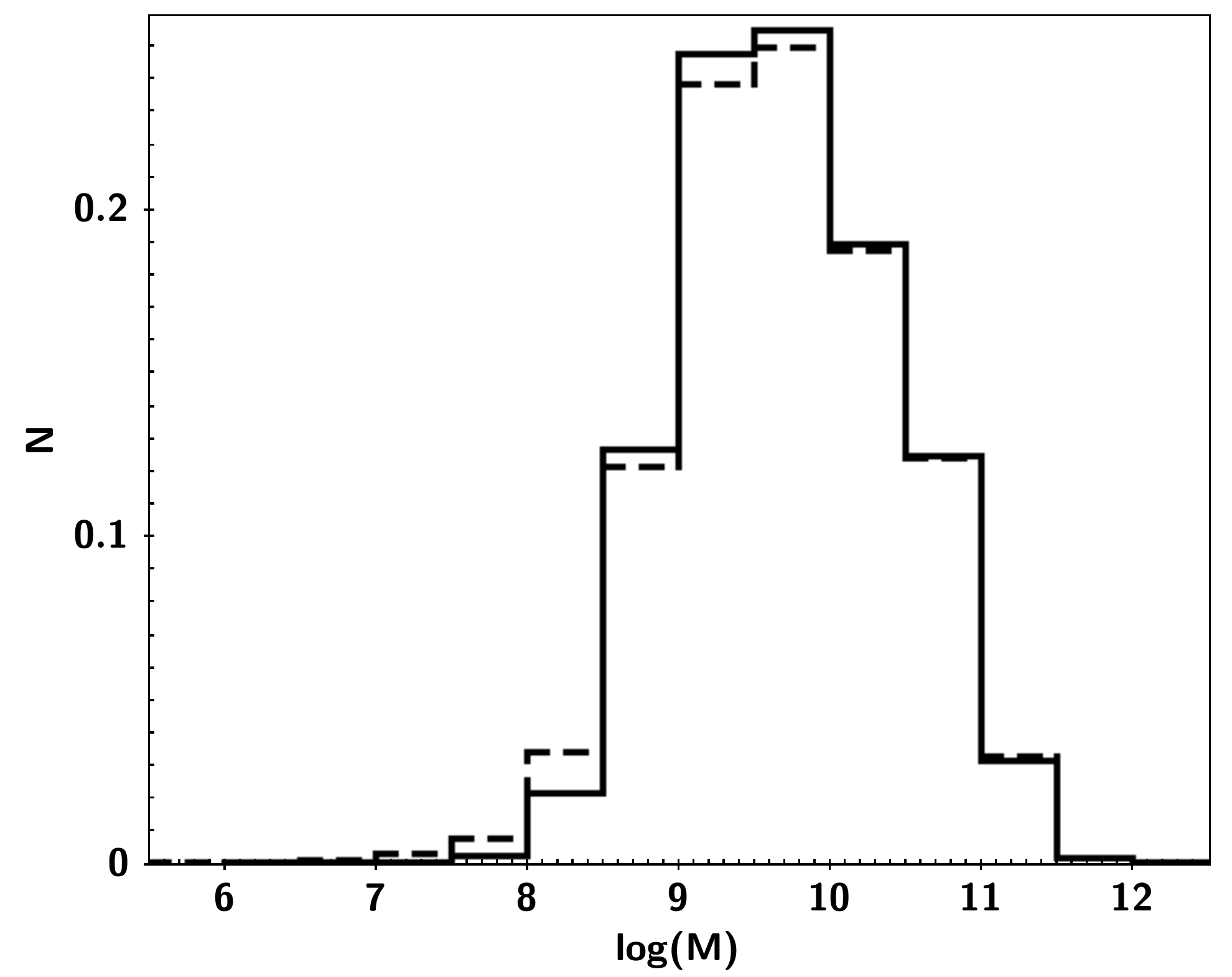} &
\includegraphics[width=6cm]{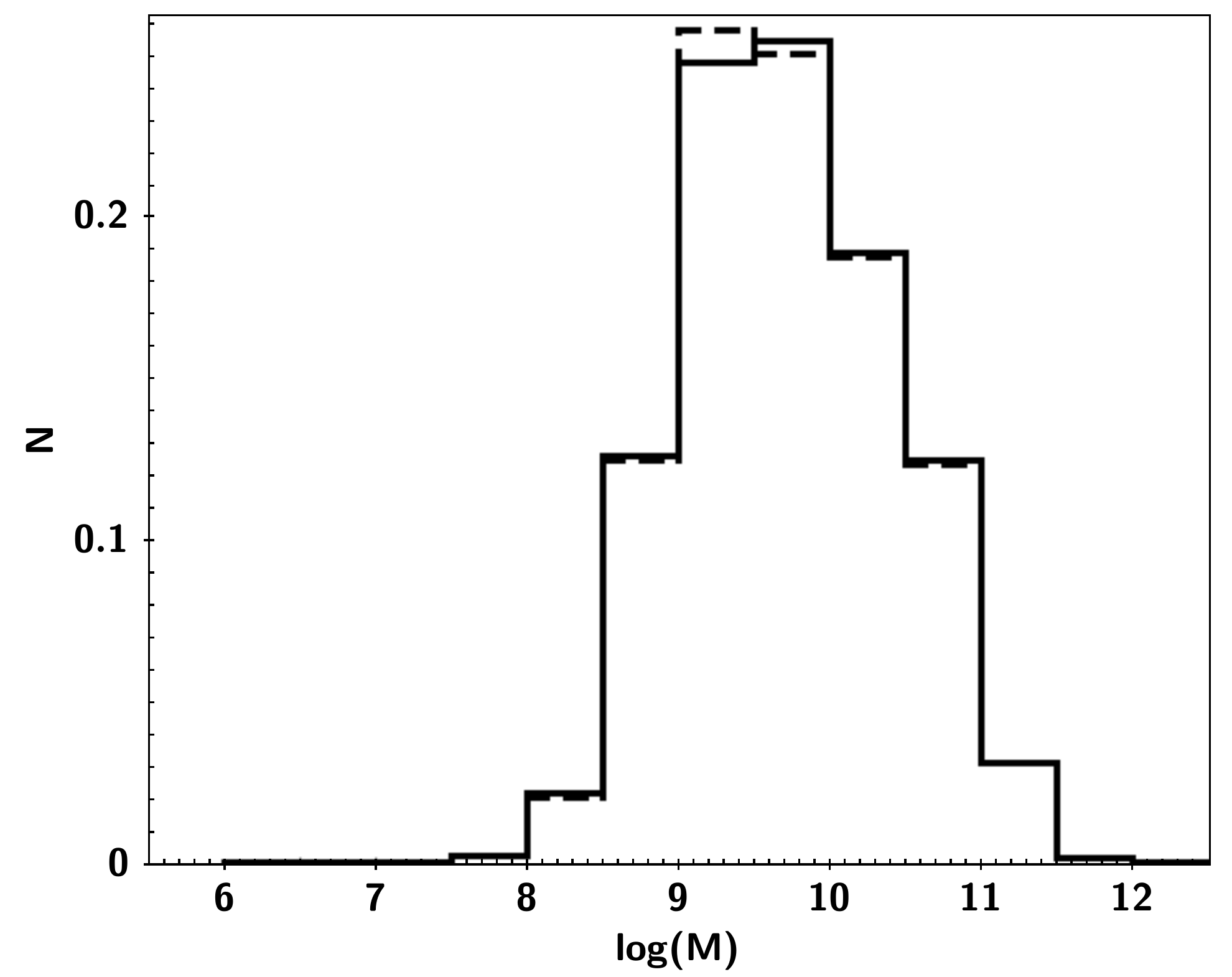}&
\includegraphics[width=6cm]{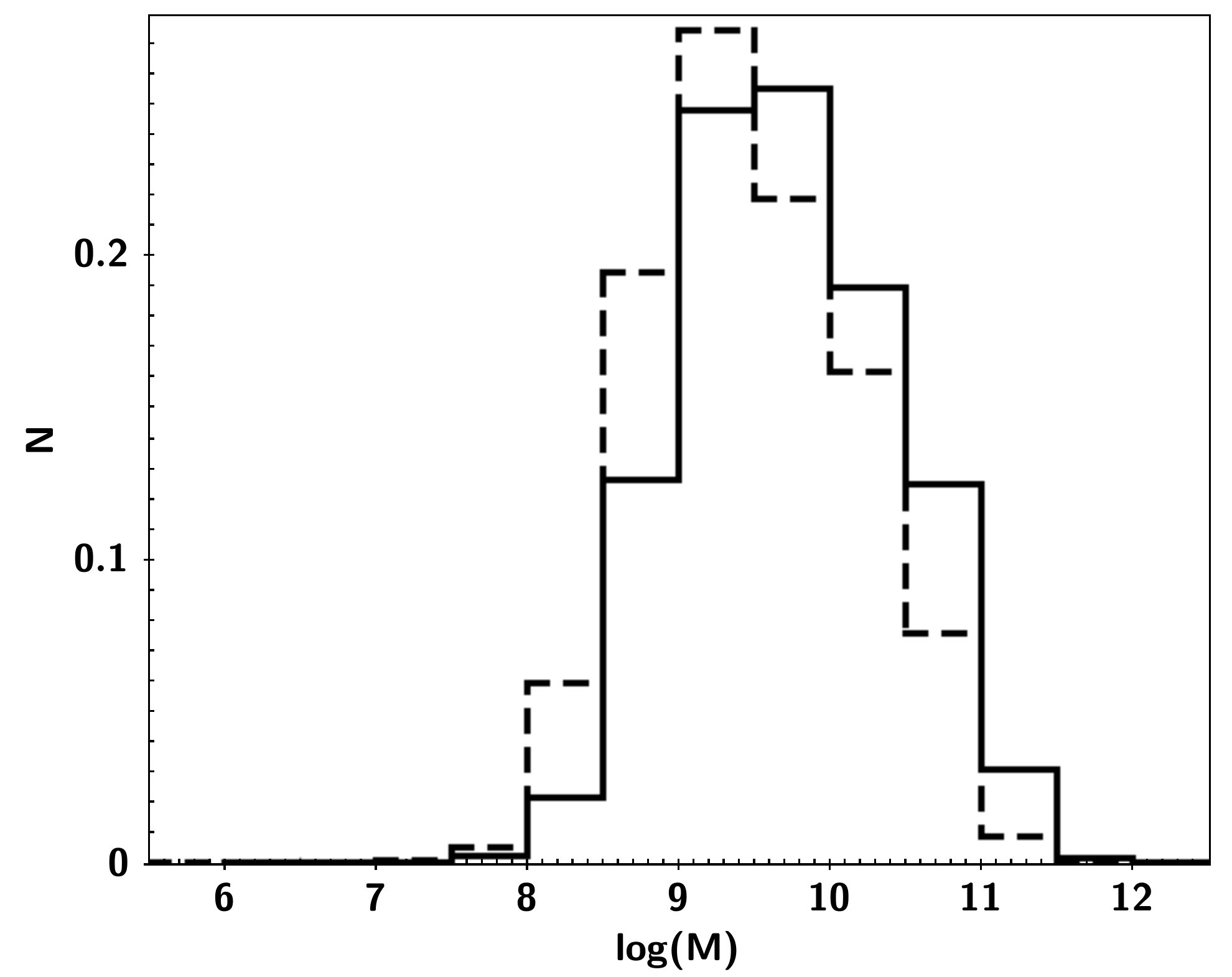} \\
\includegraphics[width=6cm]{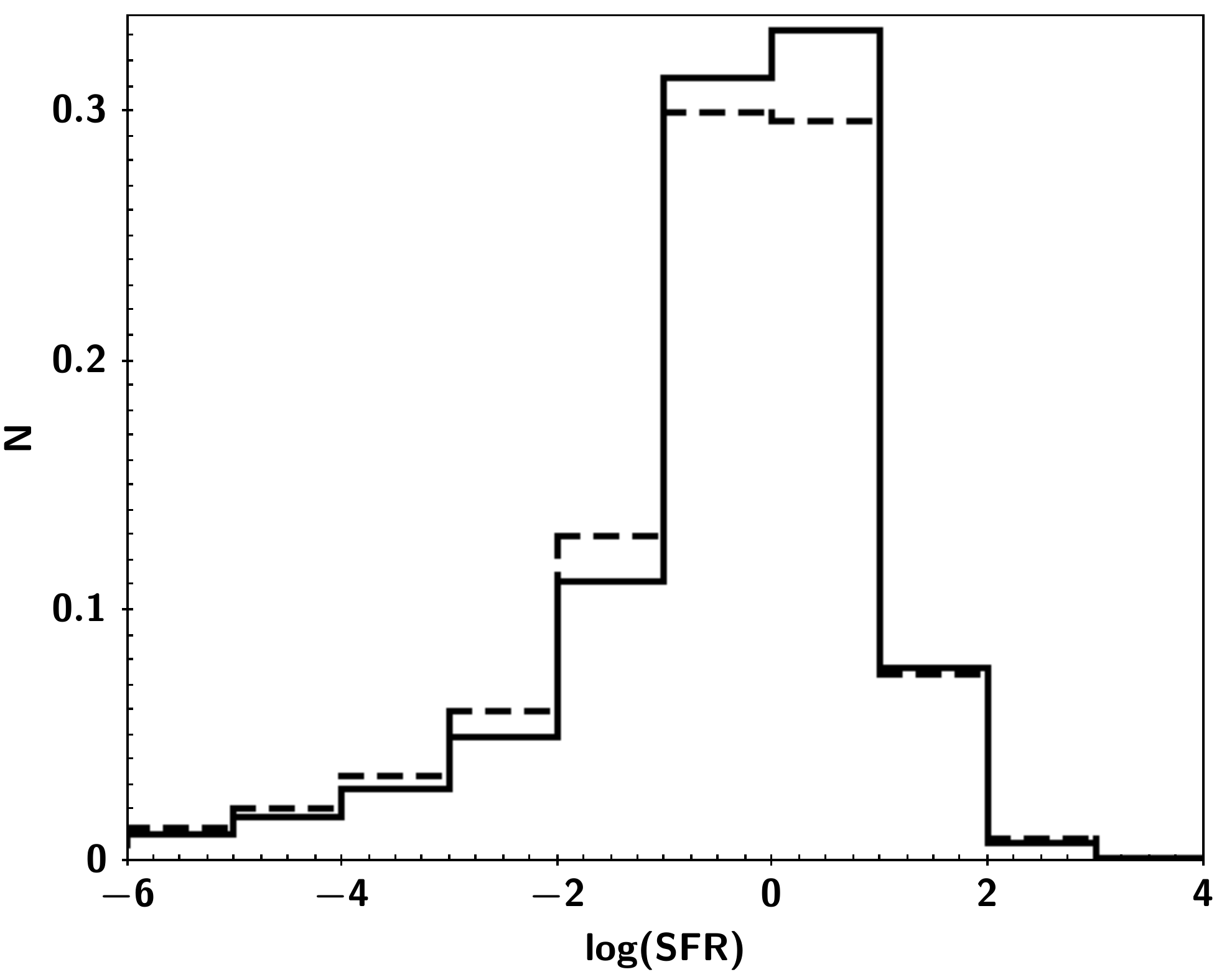} &
\includegraphics[width=6cm]{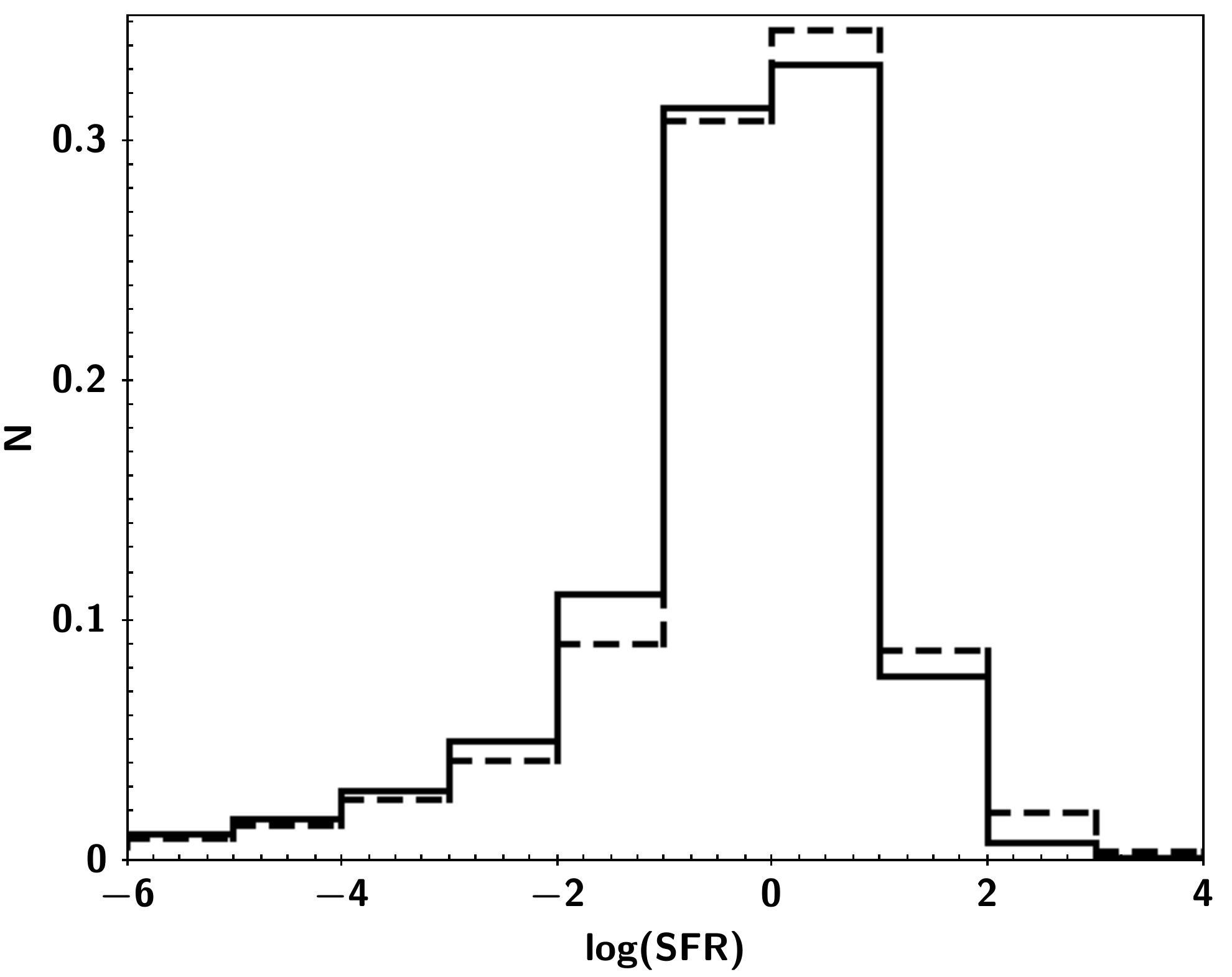}&
\includegraphics[width=6cm]{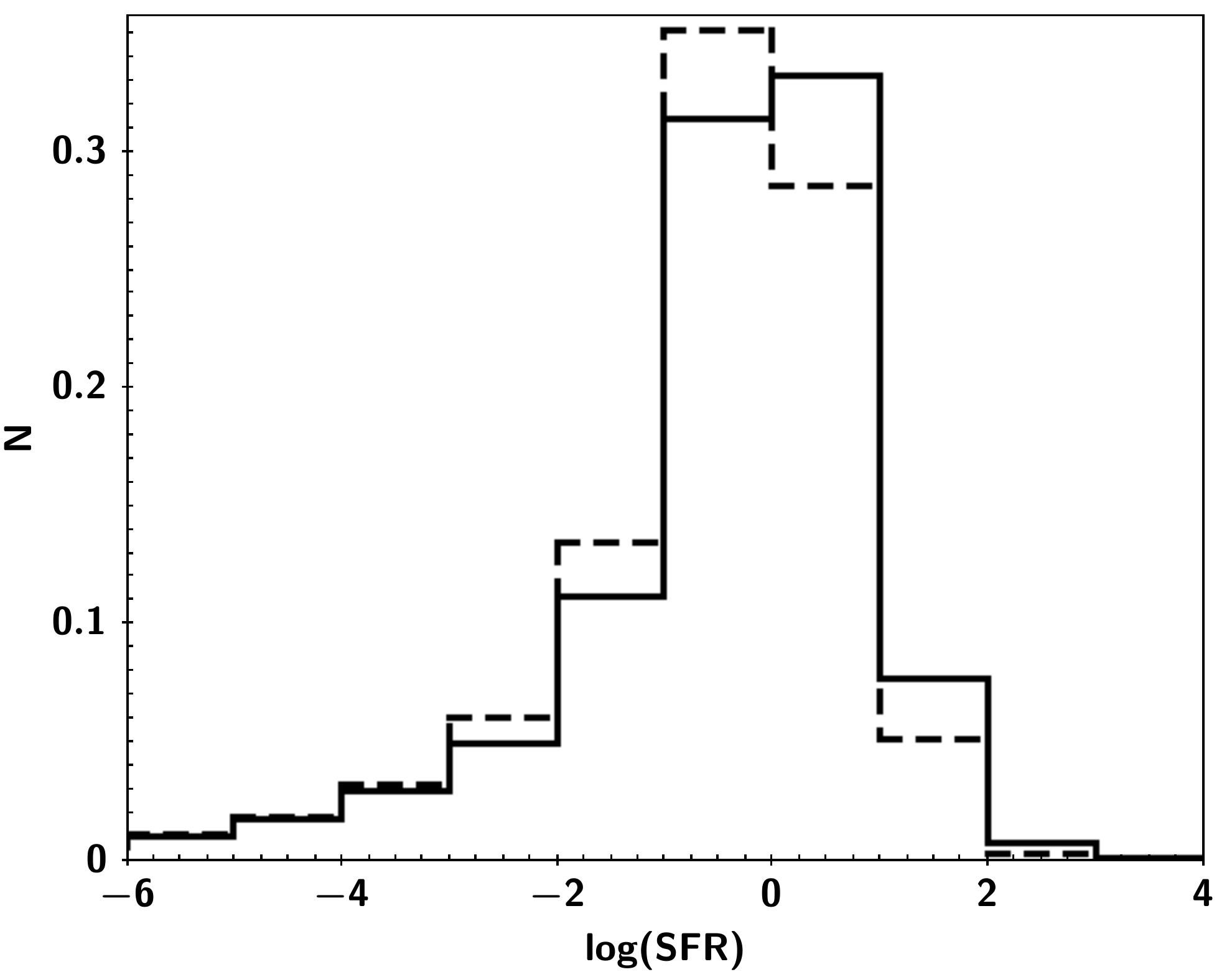}\\
\includegraphics[width=6cm]{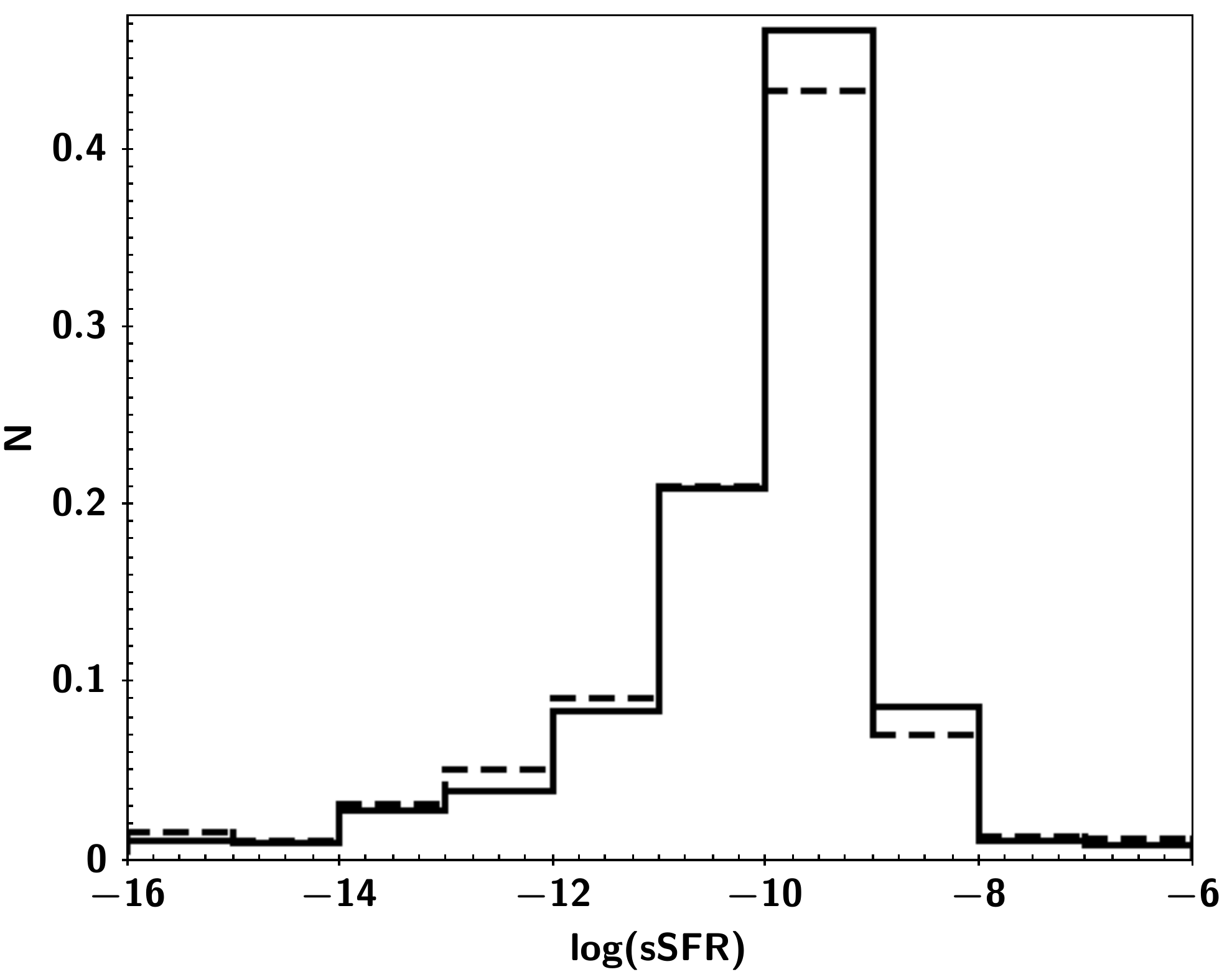} &
\includegraphics[width=6cm]{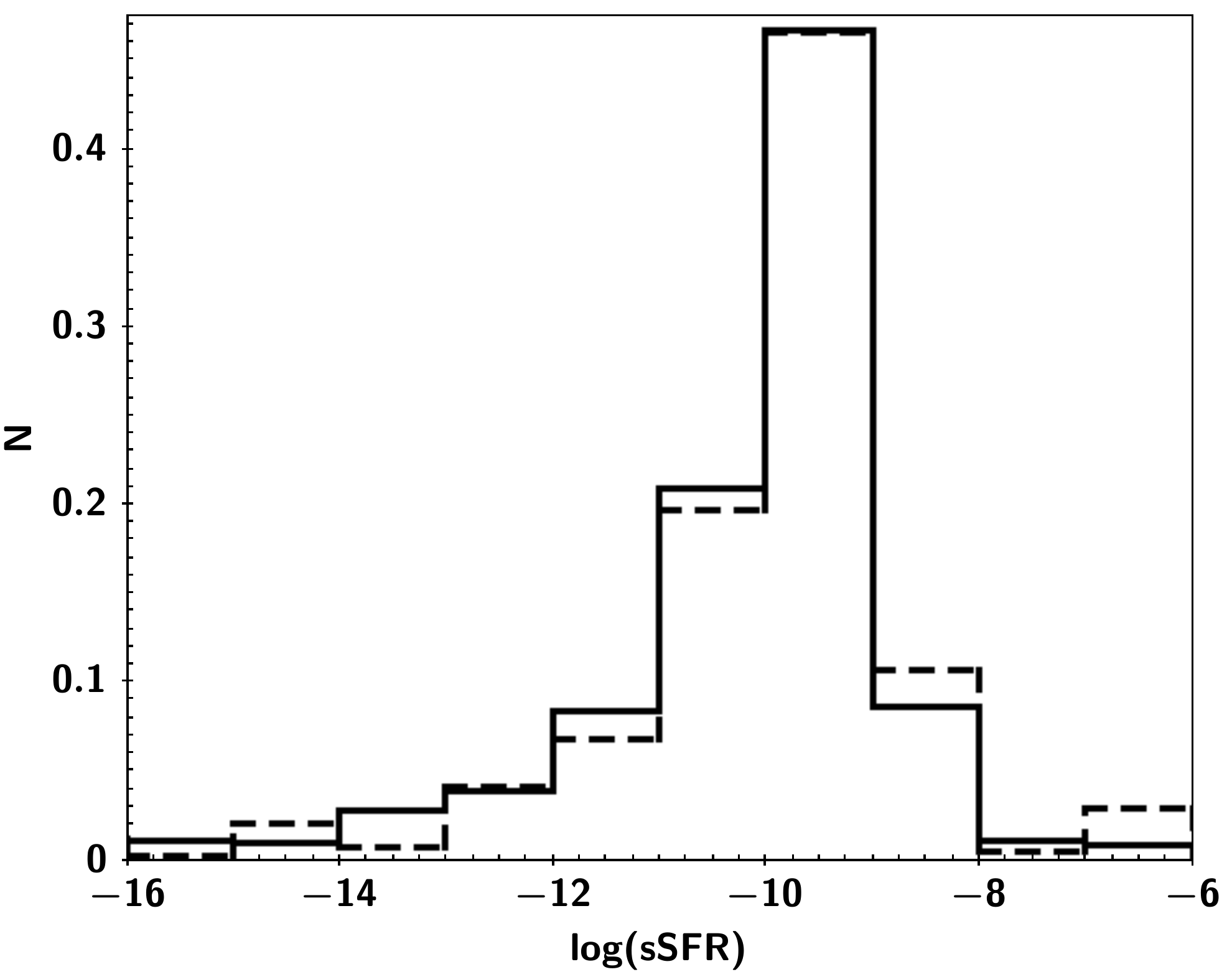}&
\includegraphics[width=6cm]{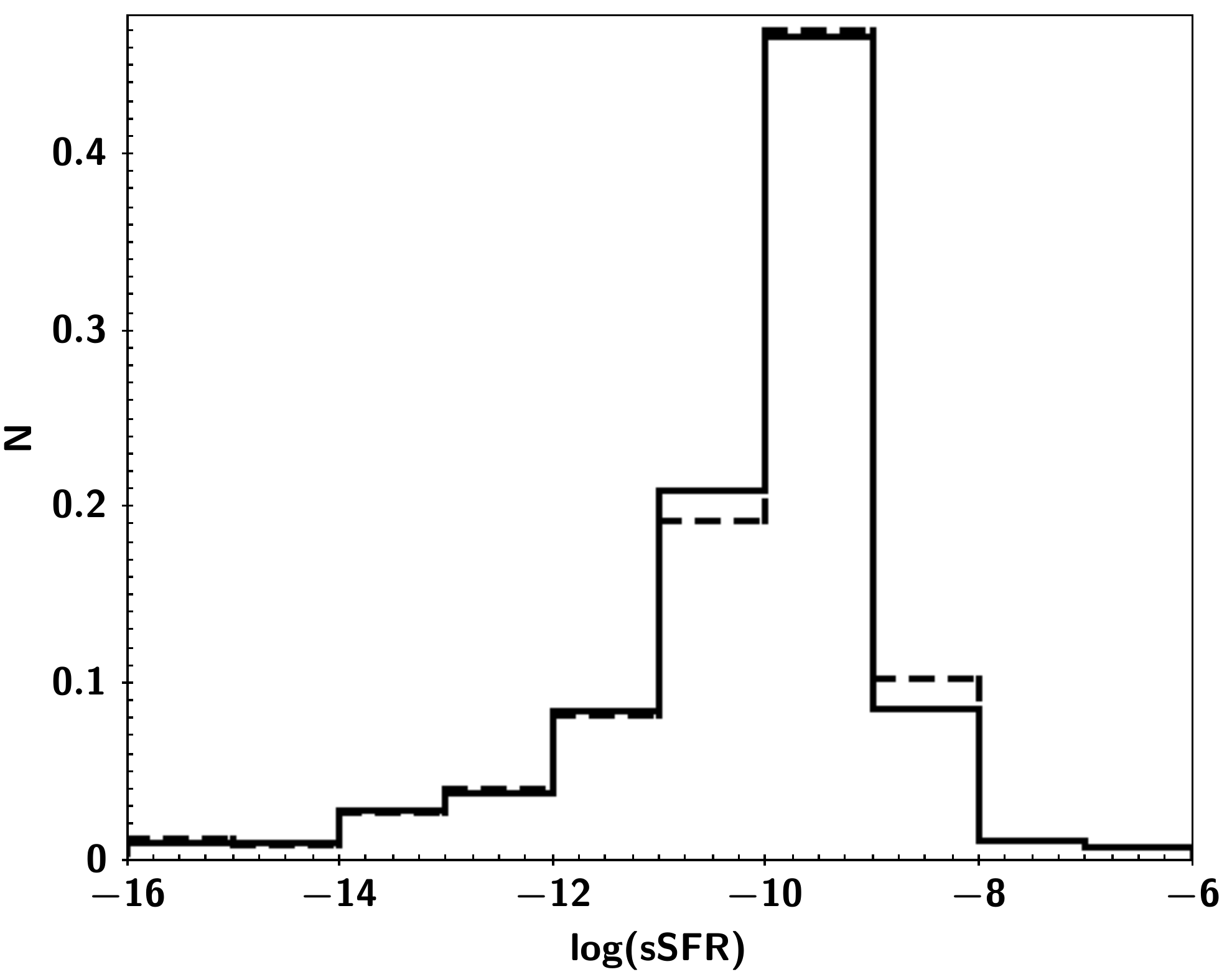} \\
\end{array}
$
\caption{Top panel:  the different distributions of  stellar masses  for galaxies in CDFS obtained from the SED fitting assuming, in the left panel,  $z_{\rm peak}$ (black line)  and $z_{\rm MC}$ (dotted line), in the middle panel  $\psi \propto  \exp(-t/\tau$  (black line)  and  $\psi_{\rm del}\propto t \times \exp(-t/\tau$  (dotted line), in the right  panel a  Salpeter (black line) and  Chabrier (dotted line)  IMF.  Middle panel: the different distributions of SFRs for galaxies in CDFS obtained from the SED fitting assuming, in the left panel,  $z_{\rm peak}$ (black line)  and $z_{\rm MC}$ (dotted line), in the middle panel  $\psi \propto  \exp(-t/\tau$  (black line)  and  $\psi_{\rm del}\propto t \times \exp(-t/\tau$  (dotted line), in the right  panel a  Salpeter (black line) and  Chabrier (dotted line)  IMF.
Bottom panel: the different distributions of  sSFR for galaxies in CDFS obtained from the SED fitting assuming, in the left panel,  $z_{\rm peak}$ (black line)  and $z_{\rm MC}$ (dotted line), in the middle panel  $\psi \propto  \exp(-t/\tau$  (black line)  and  $\psi_{\rm del}\propto t \times \exp(-t/\tau$  (dotted line), in the right  panel a  Salpeter (black line) and  Chabrier (dotted line)  IMF. }
\label{fastdist}
\end{center}
\end{figure*}

\noindent{\it Photometric redshifts}\\
\noindent We explored different methods to assess the quality of photometric redshifts and discussed  them  in detail in Paper\,$\textrm{I}$.
Here we only  analyse the effects on SED fitting of choosing a redshift estimator different  from $z_{\rm peak}$\footnote{We denoted   by $z_{\rm peak}$ the estimator which finds discrete peaks in the redshift probability function
and returns the peak with the largest integrated probability. }.
 In this test we adopted $z_{\rm MC}$  which  is drawn randomly from  the redshift probability distribution and  incorporates the redshift uncertainties \citep{Wittman:2009rt}.  
The comparison between the distributions of the photometric redshifts  obtained by adopting alternatively $z_{\rm MC}$ or  $z_{\rm peak}$  is analysed in Paper\,$\textrm{I}$ for  both CDFS and COSMOS fields.  
We find that the distributions of mass, SFR and sSFR obtained with $z_{\rm MC}$ and $z_{\rm peak}$  in input to  {\sc FAST} are very similar  for both CDFS and COSMOS sample, as shown in Fig.~\ref{fastdist}.
The SN rates per unit stellar mass as a function of  sSFR  measured by adopting  $z_{\rm MC}$ in the SED fitting are reported in Table~\ref{errsist}.
 The errors  due to photometric redshift  uncertainty  are negligible with respect to the statistical errors for both Type Ia and CC~SN rate. We note that  the errors estimated for CC~SN rates are higher than those for SN Ia rates.\\

\noindent{\it SFH}\\
\noindent \citet{pforr:2012yu} after analysing the accuracy of galaxy properties  as a function of the various parameters of the SED fitting  found that the most important parameter  in recovering stellar mass is the SFH, in agreement with \citet{maraston:2010fp}. 
\citet{mobasher:2015qv}  performed a  similar study to quantify the differences between the stellar masses estimated using different SED templates and different fitting techniques. 
In agreement with previous studies, they found that the lack of knowledge of the correct SFH, combined with inherent degeneracy between age, dust and metallicity, are the main reasons for uncertainties in the stellar masses. 
We analysed the distribution of some galaxy properties obtained assuming a different SFH, a delayed exponentially declining function ($\psi_{\rm del}\propto t \times \exp(-t/\tau)$ where $t$ is the time since the onset of SF  and $\tau$ is a parameter, ranging from $10^{8}-10^{10}$ Gyr).
The distributions of galaxy mass, SFR and sSFR for the two adopted SFH laws  are shown in Fig.~\ref{fastdist} for the CDFS galaxy sample. We note that the distributions of galaxy masses are very similar, while the distributions of SFR and sSFR show more galaxies with higher SFR and sSFR for the delayed SFH assumption.
The  SN rate measurements obtained by assuming $\psi_{\rm del}$ are reported in Table~\ref{errsist}. The errors due to  the uncertainty in the SFH assumption are negligible with respect to statistical errors for both  SN~CC and Ia  rates. \\ \\
\noindent{\it IMF}\\
\noindent We also analysed the sensitivity of the SED fitting to the IMF choice.  
Recent studies showed that different choices on the IMF produce different estimates of stellar mass and age, especially for top-heavy IMFs.
We checked how the parameters from our SED fitting  vary when the template setups are compiled using Salpeter and Chabrier IMFs \citep{salpeter:1955ky,chabrier:2003lr}.
We found that the sSFRs are very similar  but   
the masses and SFRs obtained assuming a  \citet{chabrier:2003lr}  IMF  are systematically lower by  $\sim 0.2$ dex. 
This is expected since these two IMFs essentially differ for the mass fraction in stars more massive than 1 M$_\odot$, which amounts to $f_{\rm high}= 0.61$ and $0.39$  for Chabrier and Salpeter IMF, respectively. Since these stars contribute almost the total light of the stellar population, the mass to light ratio scales as $1/f_{\rm high}$. Therefore, at given total luminosity, the Chabrier IMF yields a total mass which is a factor of $\sim 1.6$ lower than obtained with Salpeter IMF.  The SN rates become higher by a factor of two  in the bins with higher sSFR, when we adopt a  \citet{chabrier:2003lr}  IMF.  The errors estimated assuming a different IMF are larger than those in the previous cases.


\begin{table*}
\begin{center}
\caption{SN rates per unit mass [$10^{-3}\,{\rm yr}^{-1}\,10^{-10}\,{\rm M}_\sun^{-1}$] in different bins of sSFR  obtained by adopting in the SED fitting a different redshift estimator ($z_{\rm MC}$), a  delayed exponentially declining SFH ($SFH_{\rm del}$) and  a different IMF (Chabrier).   The results presented in Table~\ref{ssfr1} (ref), obtained assuming the redshift estimator $z_{\rm peak}$, an exponentially declining SFH and a Saltpeter IMF,  have been reported for comparison.
The Type Ia SN rate has been measured in the redshift range  $0.15<z<0.75$ while the CC SN rate in the range $0.15<z<0.35$.}\label{errsist}
\begin{tabular}{|c|cc|cc|cc|cc|cc|}
\hline
               & &        &  &           &                                                            &                   &                      &           &   & \\
SN type & $N_{\rm gal}$& $\rm log <sSFR> $&  \multicolumn{2}{c|}{ref}   &  \multicolumn{2}{c|}{$z_{\rm MC}$}    & \multicolumn{2}{c|}{$SFH_{\rm del}$} & \multicolumn{2}{c|}{$IMF_{\rm Chabrier}$}\\[6pt]
\hline
 &    &          & $N_{\rm SN}$ & rate  & $N_{\rm SN}$ & rate  &  $N_{\rm SN}$ & rate  &  $N_{\rm SN}$ & rate   \\[6pt]
  &  8718 &$-12.2$ & $3.5$ & $0.5^{+0.28}_{-0.33}$&         4.9          &     $0.9^{+0.4}_{-0.5}$      &          4.4           &   $0.6^{+0.4}_{-0.3}$        &         2.5             & $0.6^{+0.5}_{-0.4}$  \\[4pt]
 Ia& 21595  &$-10.5$ &   $9.8$ & $1.2^{+0.4}_{-0.5}$&      8.7             &   $1.4^{+0.5}_{-0.6}$         &     8.1                &    $1.0^{+0.5}_{-0.4}$       &       7.7              & $1.7^{+0.7}_{-0.7}$   \\[4pt]
    &22094 &$-9.7$  & $10.6$ & $3.2^{+1.1}_{-1.1}$ &      11.5              &    $4.5^{+1.5}_{-1.6}$         &     16.2               &    $5.0^{+1.3}_{-1.5}$       &      11.1                & $6.0^{+2.3}_{-1.8}$   \\[4pt]
   & 24101 &$-9.0$  &$12.8$ & $6.5^{+1.9}_{-2.2}$&      12.5             &     $8.0^{+2.6}_{-2.5}$        &       9.9             &  $5.0^{+1.6}_{-1.9}$         &        15.3              &  $14^{+4.1}_{-3.7}$   \\[4pt]
\hline
&  &&                  &     & &      &                      &           &   & \\
 & 2198 &$-12.2 $ & $0.0$ & $<0.6$ &      0.5             &   $0.4^{+0.1}_{-0.2}$         &      0.1              &    $0.05^{+1}_{-0.05}$       &      0.0                & $<1$     \\[4pt]
CC &5130 &$-10.5$ & $2.9$ & $1^{+0.6}_{-1}$&     3.8            &   $2^{+0.9}_{-1.5}$        &        2.6               &   $0.7^{+}_{-}$      & 3.4  &  $2.5^{+2.8}_{-1.3}$    \\[4pt]
      &5580 & $-9.7$ &$5.9$& $9^{+4.4}_{-6.9}$  &      4.7           &  $9^{+4.5}_{-6.0}$         &          7.0            &     $11^{+}_{-}$       &  4.3 & $12^{+7}_{-8}$    \\[4pt]
       & 3119&$-9.0$ &$4.2$ & $16^{+8}_{-16}$ &    4.1             &    $23^{+}_{-}$        &             3.4         &     $12^{+}_{-}$       &  5.3  &  $38^{+22}_{-20}$   \\[4pt]
\hline
\end{tabular}
\end{center}
\end{table*}

\end{appendix}


\end{document}